# Fundamental Principles in Bacterial Physiology - History, Recent progress, and the Future with Focus on Cell Size Control: A Review

**Suckjoon Jun**

Department of Physics, University of California San Diego, 9500 Gilman Dr, La Jolla, CA 92093, USA

Section of Molecular Biology, Division of Biology, University of California San Diego, 9500 Gilman Dr, La Jolla, CA 92093, USA

E-mail: `suckjoon.jun@gmail.com`

**Fangwei Si**

Department of Physics, University of California San Diego, 9500 Gilman Dr, La Jolla, CA 92093, USA

E-mail: `fsi@physics.ucsd.edu`

**Rami Pugatch**

Department of Industrial Engineering and Management, Ben-Gurion University of the Negev, Beer-Sheva 8410501, Israel

E-mail: `rpugatch@bgu.ac.il`

**Matthew Scott**

Department of Applied Mathematics, University of Waterloo, 200 University Ave West, Waterloo, Ontario N2L 3G1, Canada

E-mail: `mscott@uwaterloo.ca`

**Abstract.** Bacterial physiology is a branch of biology that aims to understand overarching principles of cellular reproduction. Many important issues in bacterial physiology are inherently quantitative, and major contributors to the field have often brought together tools and ways of thinking from multiple disciplines. This article presents a comprehensive overview of major ideas and approaches developed since the early 20th century for anyone who is interested in the fundamental problems in bacterial physiology. This article is divided into two parts. In the first part (Sections 1 to 3), we review the first 'golden era' of bacterial physiology from the 1940s to early 1970s and provide a complete list of major references from that period. In the second part (Sections 4 to 7), we explain how the pioneering work from the first golden era has influenced various rediscoveries of general quantitative principles and significant further development in modern bacterial physiology. Specifically, Section 4 presents the history and current progress of the 'adder' principle of cell size homeostasis. Section 5 discusses the implications of coarse-graining the cellular protein composition, and how the coarse-grained proteome 'sectors' re-balance under different growth conditions. Section 6 focuses on physiological invariants, and explains how they are the key to understanding the coordination between growth and the cell cycle underlying cell size control in steady-state growth. Section 7 overviews how the temporal organization of all the internal processes enables balanced growth. In the final Section 8, we conclude by discussing the remaining challenges for the future in the field.

# Contents

# 1. Introduction

## 1.1. Prologue

Figure 1A shows Trueba and Woldringh's classic photograph of the bacterium *Eschericia coli*, arguably the most well-studied model organism in biology [1]. We see two groups of cells, one bigger/fatter and the other smaller/skinnier. These cells are isogenic, *i.e.*, they have exactly the same genetic information. They are different in their size because they were grown under different growth conditions; the larger cells were grown in nutrient 'rich' medium, whereas the smaller cells were grown in nutrient 'poor' medium.

In the 1950s, the biologist Ole Maaløe and his group carefully measured physiological parameters of growing bacteria, emphasizing reproducibility of quantitative data [2–5]. In particular, Schaechter, Maaløe, and Kjeldgaard found that the average size of a *Salmonella* bacterium has a robust exponential dependence on the nutrient-imposed growth rate (Figure 1B and C) [2]. Importantly, their results were independent of the chemical composition of the growth media. Because molecular details – 'prefactors,' in addition to 'exponents' in the language of physics – are also often important in biology, this exponential relationship represents a rare example of a *general*, quantitative law in biology. We will refer to the results by Schaechter, Maaløe, and Kjeldgaard the 'nutrient growth law' ‡.

In our view, there are parallels between the development of bacterial physiology in the latter half of the 20th century and the development of physics in the 16th and 17th centuries. By collecting significantly better data, Brahe led Kepler to conclude that planetary orbits were ellipses and not circles (with or without epicycles). Kepler's elliptical model said nothing about the physical origins of ellipses, but his kinematic modeling was an essential starting point for Newton's work on dynamics 50 years later.

Like Kepler's laws, the nutrient growth law is 'kinematic' insofar as it allows prediction of cell size without understanding the underlying mechanism ('dynamics'). For example, if we were to pick one *E. coli* cell in Figure 1A and grow it in a growth medium with an unknown chemical composition, we would be able to predict the average cell size in the new medium just by measuring the growth curve. This is the predictive power of the phenomenology that the nutrient growth law represents. Of course, we do not know whether biology as a whole is following the footsteps of the history of physics. Bacterial physiology, however, has been transforming rapidly in the past several years so that there is hope that we might one day have a universal 'dynamical' view of bacterial growth.

This review provides a detailed account of the development of major ideas in the field of bacterial physiology during its first golden era (from the late 1940s to about the early 1970s), followed by remarkable recent advances. We set two internal rules: first, we will use language accessible to a general audience in physical and mathematical sciences, yet have endeavored to keep the content as informative as possible for biologists interested in the field. Second, we have tried to provide as comprehensive a list of references as possible reaching back to the beginning of the 20th century. There are several important topics we were unable to cover (*e.g.*, molecular and cellular biology); where these omissions arise, we have suggested other reviews in hope of covering the gaps.

The road map of the review is loosely conveyed by the historical flow chart in Figure 3, as elaborated in the next section.

## 1.2. Major questions in bacterial physiology

The major quest in microbial physiology is to understand the fundamental principles underlying the regulation and coordination of biosynthesis in a given growth environment. Physiological parameters (whatever these might be) must be measured with sufficient precision that causal relationships can be inferred.

### 1.2.1. Growth, cell division, and their distributions.
Consider one *E. coli* cell transferred to transparent liquid growth medium in a flask (Figure 2A). With good shaking for aeration at $37\,^\circ$C, the medium gradually becomes turbid due to cellular growth. The change in turbidity can be quantified by measuring the optical density (OD) of the cell culture using a spectrometer, which measures the growth of total cell mass in the culture (Jacques Monod was the first who put the turbidimetry to greatest use to measure the life cycle of cells in batch culture [6]). Plotted against time, the total cell mass exhibits a sigmoidal curve in a typical growth experiment, which is called growth curve; when plotted on a semi-log scale, the growth curve exhibits an 'exponential phase' during which the total cell mass in the culture increases exponentially. Preceding exponential phase is "lag phase" as it takes takes time for cells to readapt to the new growth environment. Once growing cells have consumed the nutrient in the growth media during exponential phase, they gradually transition to "stationary phase" and the optical density becomes stationary (Figure 2A) [7, 8].

Cells can be kept in the exponential phase of growth for many generation via serial dilution. In this state, extensive properties of the cell culture such as the total number of cells, the total mass of protein, total mass of various metabolites, all increase exponentially; whereas intensive properties such as the average cell size or the average amount of DNA per cell remains invariant with time. As such, after several generations in exponential growth, the culture reaches 'steady state'.

‡ We will discuss a number of 'growth laws' in this review, as summarized in Box 6 of Section 6

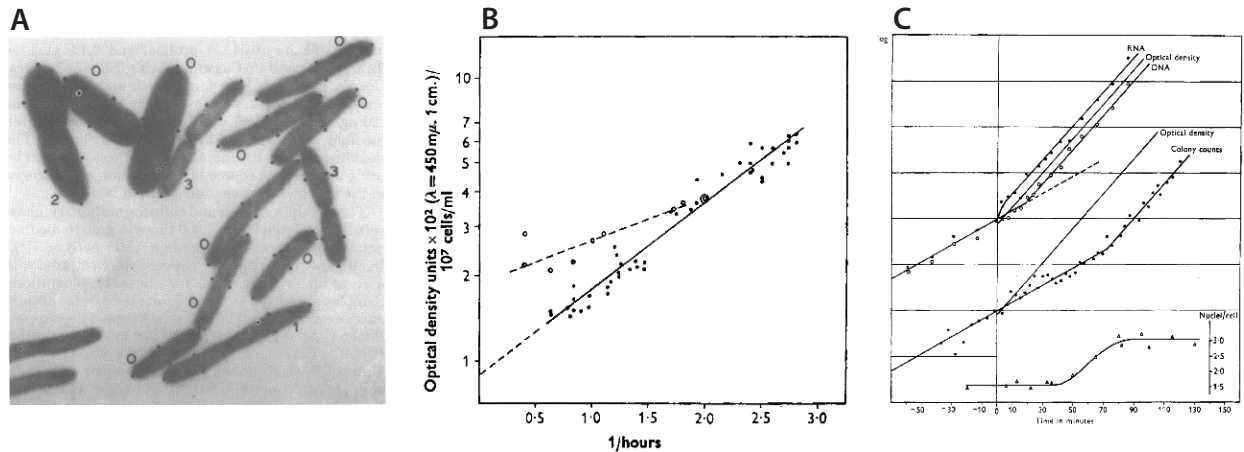

Figure 1: **E. coli cell size is different under different growth conditions. A.** Electron microscopic picture of *E. coli* cells grown in different nutrient conditions, adapted from [1]. **B.** The exponential relationship between cell size and nutrient-imposed growth rate, by Schaechter, Maaløe and Kjeldgaard in 1958 (figure adapted from [2]). The shorter dashed line is the relation obtained from continuously cultured cells. The Y axis shows the logarithm of optical density which measures the total mass of the cell culture, plotted against growth rate on X axis (see definitions in Section 1.2.1). **C.** The transitions of cell size and cellular composition when growth medium is changed from nutrient poor to nutrient rich (figure adapted from Kjeldgaard, Maaløe and Schaechter [3]).

Several physiological parameters can be measured experimentally at the population level, and still others inferred from these measurements (see, for example, [10]). First, the growth rate can be directly measured based on the rate of change in the cell number density (Figure 2A). Second, the distribution of cell size (and therefore the average size) can be measured by microscopy (Figure 2B). Third, at the molecular level, the total amount of proteins, nucleic acids and other biomolecules can be measured with the help of biochemistry (Figure 2B). Because the total cell number per unit culture volume is known, the per-cell average of a given biomolecule can be estimated.

There are, however, parameters that cannot be directly measured in flask growth. An obvious example is the 'age' of the cell, *i.e.* the time elapsed since birth (Figure 2D and E). For an ideal case where all cells divide precisely in the middle when they reach the same size, the general age distribution $\varphi(a)$ of a steady-state population can be derived analytically (Section 2.2.1). If we assume specific growth dynamics of individual cells (*e.g.*, exponential or linear in time, which itself has been a major subject of debates in the past, see Section 2.3.1), the measured cell length $l$ can be converted to cell age $a$. Conversely, the cell size distribution $\rho_l(L)$ can be analytically calculated from the age distribution $\varphi(a)$ if the growth dynamics are known.

An important realization from the mid 20th century is the stochasticity of growth and cell division dynamics. This came from the comparisons between the theoretical and the experimental size distributions. The experimental data exhibits smooth tails for the size distribution $\rho_l(L)$ at both lower and upper ends (Figure 2E; green curves), whereas the idealized theoretical distribution predicts sharp cut-offs (Figure 2E; red curves). The presence of smooth tails indicates that the coefficient-of-variation (CV) of the dividing cell size distribution is non-zero.

The stochasticity of cell division was directly confirmed by a pioneering single-cell time-lapse data from the 1950-1960s [11–18]. Both the division size and the generation time showed significant cell-to-cell variability, typically with coefficient-of-variation between 10% and 30%. These numbers are in good agreement with microfluidic based high-throughput microscopy measurements in the 2010s. As we will review in Section 3 and Section 4, some of the most intense research efforts in the field have focused on the biological origins of the cell-to-cell variability in a variety of physiological parameters, and how they are quantitatively related to one another [9, 19–30].

*1.2.2. Coupling between growth and the cell cycle, and consequences on cell size.* The cell cycle is one of the most basic controls underlying cellular reproduction, and heart of the cell cycle is replication of the chromosome. The chromosome of the model organism *E. coli* is arranged as a circular loop of

**A  Typical growth of cell culture**

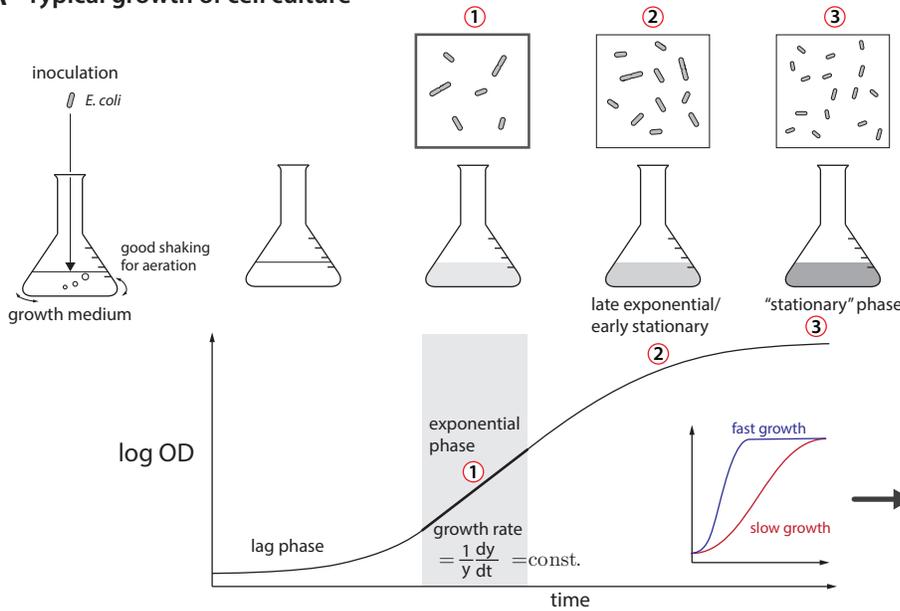

**B  What is measurable**

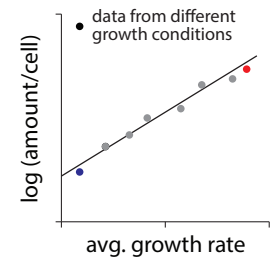

**C  "(nutrient) growth law"**

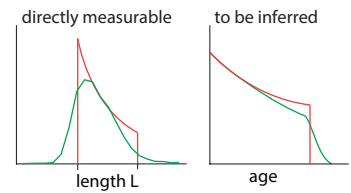

**D  Growth and division of a single cell (*E. coli*) and their variability**

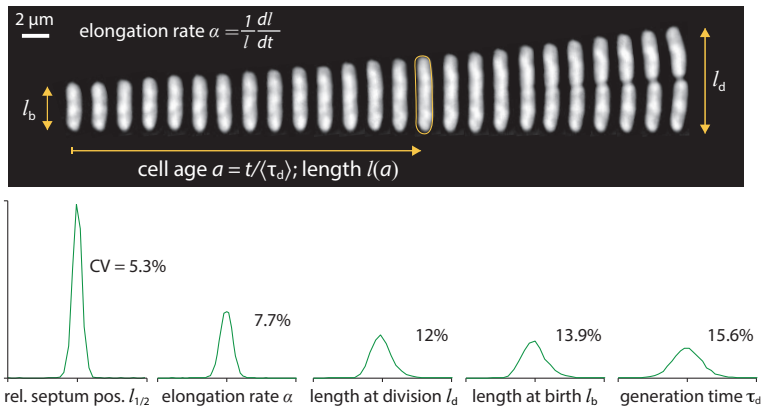

**E  Canonical (deterministic) vs. variable (stochastic) distributions**

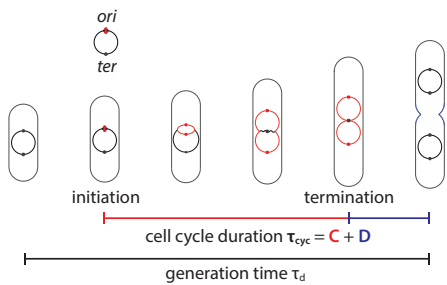

**F  Replication cycle of the *E. coli* chromosome**
(case of non-overlapping cell cycles)

**G  Coarse-grained partitioning of cellular resources (e.g., proteins)**

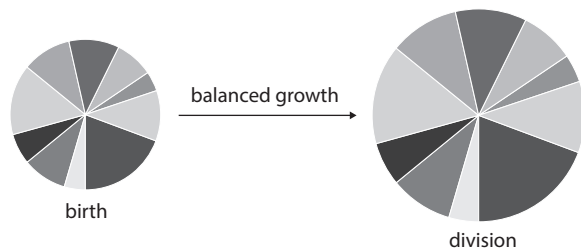

Figure 2: **Schematic diagrams of basic definitions in bacterial physiology. A.** Growth curve and growth phases of cell culture. (Cell death is not considered here.) **B.** The measurable properties from an exponentially growing population during balanced growth. **C.** The exponential relationship between cell size and growth rate (the nutrient growth law). Blue marks the slowest growth and red the fastest. The cell image is adapted from [9] with permission, and distributions are calculated from experimental data in [9]. **D.** The measurable properties of individual cell during one generation from cell birth to division, and example data of distributions of each property. **E.** The deterministic versus stochastic distributions of cell length and age of an exponentially growing population. **F.** Diagram showing one cell cycle in a slowly growing cell. Here the cell cycle parameters are defined. The generation time $\tau_{\mathrm{d}}$ is the period from cell birth to division. The cell cycle duration $\tau_{\mathrm{cyc}}$ is defined as the time period between replication initiation and cell division, which consists of **C** period (or replication period, from initiation to termination) and **D** period (from termination to cell division). **G.** The partitioning of cellular resources during balanced growth.

approximately $4.5 \times 10^6$ base-pairs of DNA, which replicates bidirectionally starting from a well-defined origin of replication (called *ori*). The average replication speed is approximately the same along both chromosome arms and the two replication forks meet at the opposite side of the chromosome from the *ori* (in a region called the terminus, or *ter*). One of the most fundamental questions in *E. coli* physiology is what ensures that one, and only one, replication cycle starts for every division cycle under all growth conditions? Considering the intrinsic stochasticity, the coupling between the replication cycle and the division cycle poses both conceptual and technical challenges. We will cover these issues throughout the review, especially in Section 6.

*1.2.3. Coarse-graining cellular resources.* Typical *E. coli* cells contain $O(10^5)$ to $O(10^6)$ proteins expressed from about 4000 genes encoded in the chromosome, and on average these proteins double their numbers in each generation. While all the proteins are present in the cell for a reason, it would be neither efficient nor necessary to study gene expression and the synthesis of each and every protein in the cell. In physics, kinematics has been successful in understanding a general phenomenon with predictive power. Examples include Kepler's laws (vs. Newton's dynamics), thermodynamics (vs. statistical mechanics), and the Landau theory of phase transitions. For such an approach to be useful in bacterial physiology, the key is to pay close attention to the biological functions such as cell cycle, cell envelope synthesis and so on, and to carefully choose appropriate 'state variables.' For example, proteins can be grouped depending how they respond to specific growth inhibition (Figure 2G and Section 5), and biomolecules and their synthesis can be grouped and connected in a graph by their roles and function (Section 7). These approaches are relatively new (2010s), and how they may be integrated into the study of cell size control remains an open question (Section 8).

## 2. The first golden era of bacterial physiology (late 1940s - early 1970s)

The period between late 1940s and early 1970s represents the first golden age of bacterial physiology. A typical progression during this period was first a new technology allowed novel experiments that were not possible before, followed by modeling efforts to explain the data. Bacterial physiologists in this period were comfortable with both biology and mathematics. This pattern is refreshingly modern, similar to how physics advances. The timeline is elaborated in Figure 3 and throughout the remainder of this section. In Box 1, we list some of the founding figures and their main contributions to the field.

### 2.1. Part I: Key technology development and experiments

*2.1.1. Carlsberg pipette and colorimetric assays: quantification of growth in the Copenhagen school.* Inspired by Max Delbrück's quantitative studies on phage dynamics, Ole Maaløe [then at the State Serum Institute in Copenhagen, Denmark] developed a similarly quantitative approach to the study of bacterial physiology. We will discuss the conceptual significance of their works in Section 2.2; here, we review the technological innovations the Copenhagen school developed and refined which set the stage for the golden era of bacterial physiology. Maaløe's lab set rigorous standards for the reproducibility of their experiments. One example is their pipetting technique. By that time, even before the invention of modern pipette (the first Schnitger's pipetter was invented in 1958), Maaløe's lab was able to transfer down to ten microliters of liquid by using their handcrafted Carlsberg pipettes. This transfer technique enabled reliable serial dilution and plate counting, with direct bearing on the accuracy of cell-number measurements.

One major technical hurdle in bacterial physiology at that time was to measure the cell composition. Moselio 'Elio' Schaechter, who was then a post-doc in Maaløe's lab, first adapted colorimetric assays to reliably quantify the macromolecular composition of bacterial cell in different growth conditions. By using RNA stained with orcinol and DNA with diphenylamine, Schaechter was able to accurately quantify the nucleic-acid cell content across different growth conditions, which proved to be an essential characterization method in modern bacterial physiology [4].

Note that before their work in 1958, the growth curve as shown in Figure 2A was thought to be the 'obligatory life cycle' of the bacterial cell, but little attention was paid to either the steady state of growth or balanced growth (Figure 2; also see Box 2) [5]. The Copenhagen school overturned this standard point of view. In 1958, Maaløe's lab published two back-to-back papers: the first focused on the steady-state growth of *Salmonella typhimurium* by modulating the quality of the nutrient of growth medium [2], the second studied the transition between physiological states by shifting the growth medium[3].

The idea of the nutrient-limitation experiment is to quantify the physiological state of cell (*e.g.*, cell size, macromolecular composition and so on), and to examine its dependence on growth rate as modulated by changes in the nutrient quality of the medium. To maintain steady-state growth for tens of generations, all nutrients in the medium are available in saturating amounts such that their concentration does not change appreciably during the experiment. What does change is the rate at which the bacteria can metabolize the nutrients. For example, *E. coli* requires fewer enzymes and less time to metabolize glucose as a carbon source as opposed to succinate; *E. coli* can synthesize all amino acids, but will grow more rapidly if amino acids are supplied in the medium. In a similar fashion, changing the nitrogen source, the carbon source, adding amino acids, nucleotides and vitamins, the growth rate can be modulated over a wide range. In their study, Maaløe and colleagues grew *Salmonella* in over 20 different media where they fine-tuned the chemical composition to yield different growth rates at steady state [2]. We will review their results in Section 2.2. Interested readers can find a more detailed (and personal) review in [4, 5].

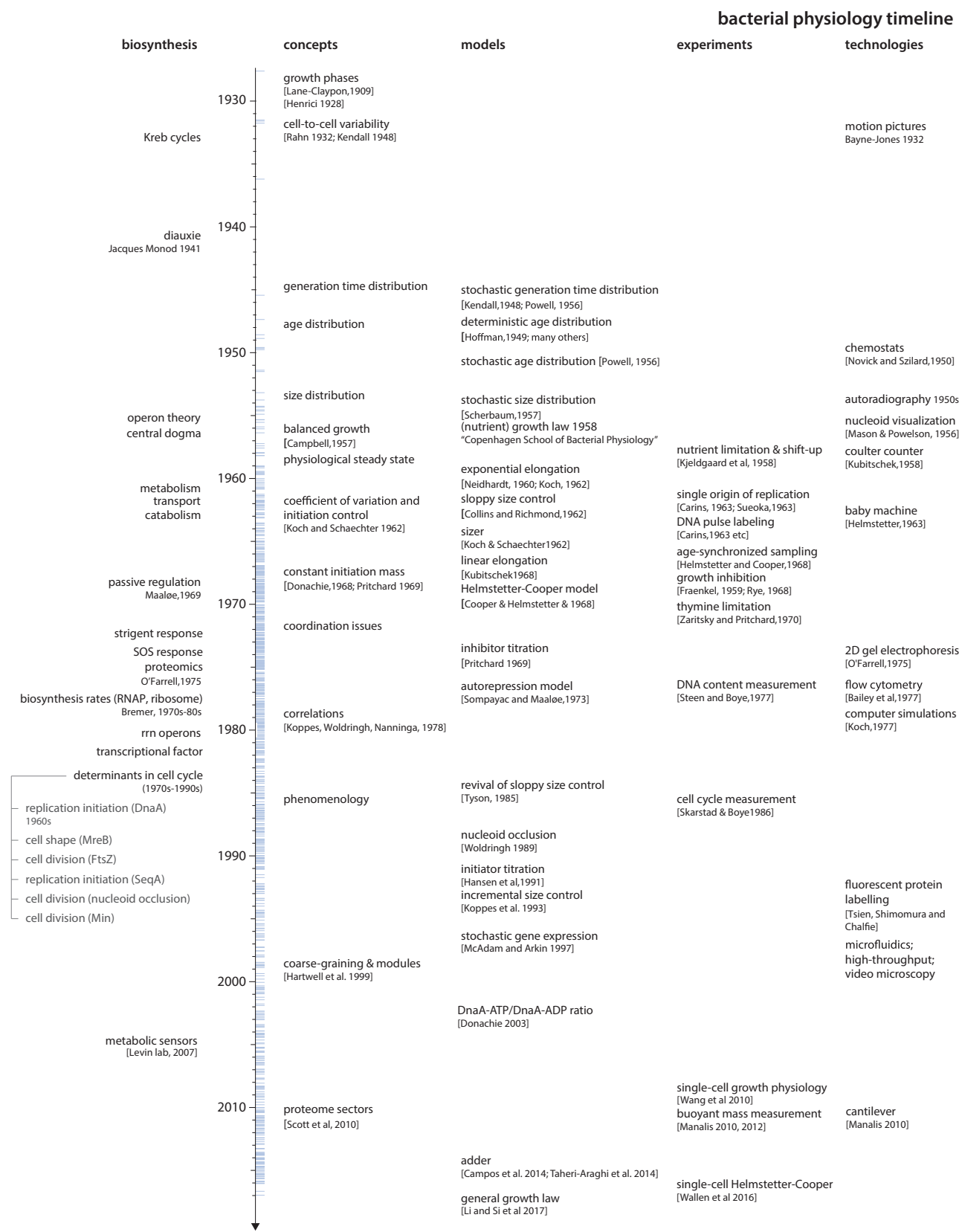

Figure 3: **Timeline of bacterial physiology (1900 - present).** Shown on the right hand side of the time axis are the major technological developments, experiments, models, and conceptual advancements. Each blue tick on the time axis represents one publication at that time (analyzed from the bibliography of this review). Shown on the left hand side are the major advancements in molecular biology of biosynthesis for those interested (which is beyond the scope of this review). Representative researchers and papers are shown beneath each keyword.

In their second paper, they examined the transition between different physiological steady states, by adding nutrients to a poor medium where the cells have already reached a steady-state of growth. This type of experiment is called 'nutrient shift-up' or simply 'shift-up' [3]. The motivation of doing a shift-up experiment is to reveal transition patterns of cell size and molecular composition, which in turn helps better understand how the cell coordinates its biosynthesis with growth. Again we will review the results in detail in Section 2.2. Subsequently, shift-up or shift-down experiments were done by other groups to study bacterial growth [31, 32], cell cycle [33–36], cell division [37], DNA synthesis [38], RNA [39–44] and protein sysnthesis [45, 46], gene regulation [47–49], metabolism [50], morphogenesis [51–53] and many other aspects of microbiology [54].

## Box 1 – Key players in bacterial physiology and bacterial cell cycle

We highlight some of the founders of bacterial physiologists.

### Ole Maaløe

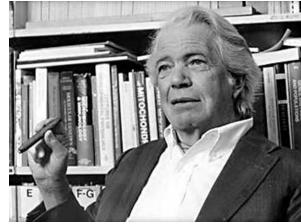

Maaløe was an influential leader. His two-part series on steady-state and transitional growth of *Salmonella* [2, 3] played a foundational role in establishing the Copenhagen school of Bacterial Physiology (Section 2.2.2 ). Maaløe and Kjeldgaard's book, *Control of Macromolecular Synthesis: A Study of DNA, RNA, and Protein Synthesis in Bacteria* [55], summarizes the state-of-the-art during this golden age. With Sompayrac, Maaløe proposed an autorepressor model for DNA replication initiation control [56] (Section 2.2.7).

### Moselio Schaechter

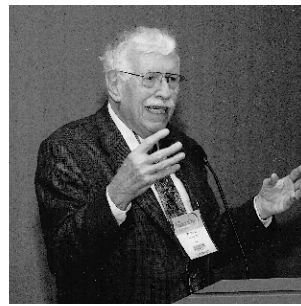

Schaechter joined Maaløe's lab in 1956 as a post-doc. He led the discovery of the first "growth law" in bacterial physiology, together with Maaløe and Kjeldgaard [2]. He also contributed to our understanding of bacterial cell division and cell size control, chromosome replication and segregation, and co-authored the textbook *Physiology of the Bacterial Cell: A Molecular Approach* [57]. His notable personal weblog *Small Things Considered* provides invaluable original essays about microorganisms to both general and specific audience.

### Niels Ole Kjeldgaard

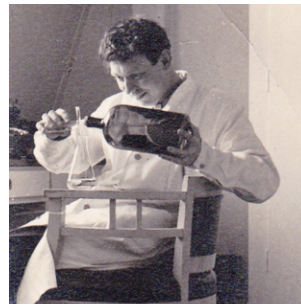

Kjeldgaard's major contributions to bacteriology include the UV light induction of bacteriophage during his PhD with André Lwoff in Paris, and the study of bacterial growth physiology during his post-doc in Maaløe's lab.

In 1968, as professor in Aarhus University, he founded the first institute of molecular biology in Denmark and had led the molecular biology research in the country.

**Fred Neidhardt**

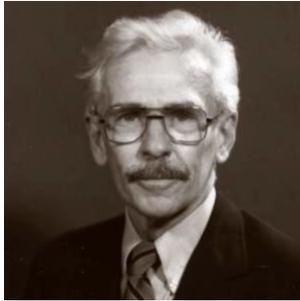

In addition to important work on the role of ribosomes in protein synthesis [58–62] (Section 2.2.3), Neidhardt edited the landmark reference book *Escherichia coli and Salmonella: Cellular and Molecular Biology* [63] and co-authored the textbook *Physiology of the Bacterial Cell: A Molecular Approach* [57]. Neidhardt and Pedersen (see below) were among the first to recognize the potential of proteomic studies using 2D gel electrophoresis [64, 65]. For more personal perspective by Neidhardt, we recommend his writing in [66].

**Steen Pedersen**

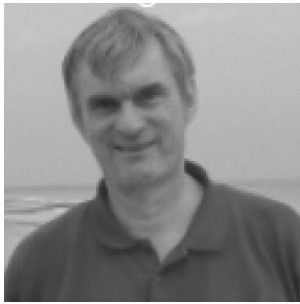

Pedersen was a student of Maaløe, and continued his study in bacterial physiology as a post-doc in Neidhardt's lab, where he was the one who adopted the 2D gel electrophoresis method that was being developed by Patrick O'Farrell, a graduate student at the University of Colorado at Boulder [67]. This method finally allowed global analysis of protein composition in a physiology dependent manner starting in the 1970s. Pedersen was a true master of pulse-chasing experiments, and much of our current understanding of the kinetics of protein synthesis and degradation is due to his rigorous and precise measurements.

**Arthur L. Koch**

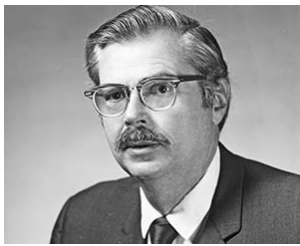

Koch was a Renaissance man; an experimentalist and theorist of broad scope and depth. He devoted his insight and quantitative skills to explaining the physical and biochemical basis of bacterial growth and form in the broadest sense. His surface stress theory is an extraordinary example of characterizing the complex nature of cell shape control using simple physical concepts [68]. He authored the text *Bacterial Growth and Form* [69] among others, summarizing his original thinking and approach to bacterial physiology.

**Charles Helmstetter and Stephen Cooper**

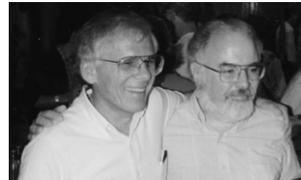

Helmstetter (left) developed the 'baby machine' for studying synchronized cell populations [70] (Section 2.1.5). He characterized the chromosome replication cycles with extremely careful measurements using his invention and autoradiography techniques. Helmstetter and Cooper (right) together developed the textbook model of the bacterial cell cycle named after them [71, 72] (Section 2.2.4).

**"Willie" Donachie**

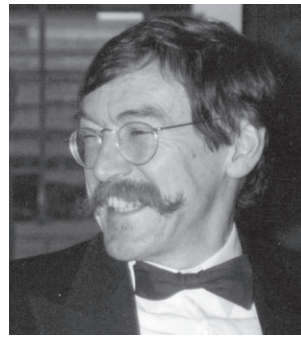

Perhaps best known for his theoretical insights on the constant 'initiation mass' during the bacterial cell cycle [73] (Section 2.2.5), Donachie contributed to a range of problems in bacterial physiology, in particular, to our understanding of DNA replication initiation [74] and cell division [75–81].

**Hans Bremer**

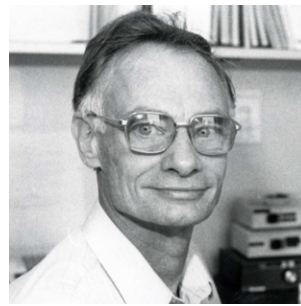

Bremer made considerable advances in our understanding of physiological regulation of ribosome synthesis [82–88]. A great legacy of his work is the review article co-authored with Pat Dennis cataloging, in a self-consistent fashion, the changes in various macromolecular components and kinetic parameters as growth rate is modulated by nutrient change [10].

*2.1.2. Chemostat: continuous cell culture.* In contrast to batch culture in flasks as in Figure 2A, a continuous cell culture device allows uninterrupted cell growth over many generations in steady-state by maintaining a constant environment through continuous dilution. The chemostat represents the most sophisticated continuous cell culture method, firstly adapted by Novick and Szilard in 1950 [89–93]. The chemostat works by replacing the culture with fresh medium at a fixed rate, called the 'dilution rate'. In this way, the steady-state growth rate of cells can be set to any prescribed rate below the maximal growth rate attainable in the medium [69]. The chemostat was widely used to investigate bacterial physiology, metabolism and biosynthesis by many researchers [34, 49, 94–103].

*2.1.3. Radioactive pulse labelling and autoradiography: quantifying macromolecular synthesis rates.* DNA is now known to be closely related to growth rate and cell size. In 1950s, however, DNA replication was still largely a mystery, though obviously of great importance given the role of DNA in the transmission of hereditary information [104–106]. Although progress had been made using nuclear staining methods [107, 108] or *in vitro* biochemical assays [109], the technological breakthrough for studying DNA replication was the use of radioactive pulse-labelling and autoradiography. During a pulse-labelling experiment, radioactive isotopic molecules involved in nucleotide biosynthesis, such as $N^{15}H_4Cl$ or $[^{14}C]$thymine, are added to the growth medium and briefly incorporated into newly synthesized DNA. After a few minutes, the labelling is stopped by transferring cells to a radioactive-free medium. The proportion of labelled-to-unlabelled DNA in the cell culture will change over time, and the DNA synthesis rate (or period of DNA synthesis) can be quantified by measuring the radioactivity incorporation rate via autoradiography or other quantitative assays [110–113].

The famous Meselson-Stahl experiment in 1957-1958 pulse-labelled the DNA of *E. coli* cells by $N^{15}H_4Cl$ and used density-gradient centrifugation to differentiate the parent and daughter chromosomes, concretely supporting the semi-conservative model of DNA replication raised by Watson and Crick [115, 116]. Later, John Cairns pulse-labelled the DNA of *E. coli* by $[^3H]$thymine and used autoradiography to directly show that the chromosome is a single-stranded, sequentially replicated molecule and replication of DNA starts from a single origin [117, 118]. Independently, other researchers reached the same conclusion by using either radioactive labelling [119–130], genomic marker transformed by bacteriophage [131–137] or bromouracil labeling (5-

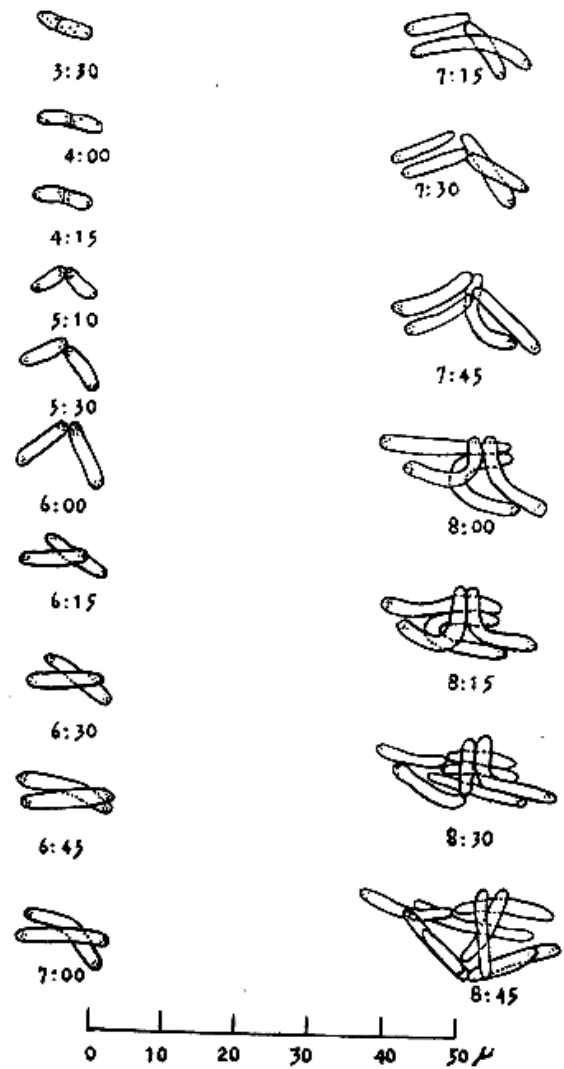

Figure 4: **Hand-drawn figures of *B. megaterium* to measure cell size, by Henrici in his 1928 book.** The microcolony of cells were observed continuously under microscope for some hours, and captured by camera lucida drawing (adapted from [114]).

bromouracil is a nucleobase analog and can be differentiated by density measurement) [138–142].

Maaløe's lab and others also employed radioactive labelling since the 1950s to quantify the DNA synthesis rate and its relation to the division cycle in bacterial cells [109, 110, 143–145]. One of the great puzzles was that DNA synthesis, while continuous through the cell cycle during rapid growth, exhibited quiescent gaps during slow growth conditions [110, 119, 146]. Furthermore, the DNA content was found to be proportional to the nutrient-imposed growth rate at steady state [2]. A solid model explaining how DNA synthesis coordinates with growth was still lacking. In 1963, while a post-doc in Maaløe's lab, Charles Helmstetter first came up with a method of studying age-synchronized cells – the 'baby machine' (see Section 2.1.5) – that would ultimately provide a model for this coordination [147, 148]. He and Steven Copper, another post-doc in the lab, used the baby machine along with radioactive pulse labelling, to uncover the fundamental relations in *E. coli* DNA replication under a variety of growth conditions [72, 111]. We will discuss their results in Section 2.2.4 followed by the baby machine in Section 2.1.5.

### 2.1.4. Microscopy, motion pictures and coulter counter: measuring size and generation time of individual cells.

One of the earliest attempts to measure the size of individual bacterial cells is seen in Henrici's 1928 book [114], where he reported the observed changes in cell size in a growth-phase-specific manner and measured cell size using bright-field microscopy observations (Figure 4). In Kelly and Rahn's 1932 work [149], they manually sketched every 5 minutes of the growth of *Enterobacter aerogenes* and *Bacillus cereus* under the microscope, and reconstructed the lineage tree up to four generations. Although their initial interest was to see if cells die, they noted significant cell-to-cell variability in the growth rate even under uniform conditions. Baynes-Jone and Adolph independently used motion pictures to conduct time-lapse imaging for *E. coli* [150].

Later, camera-based micrography and videography were more often employed to capture morphology and growth (*e.g.*, [11, 151]). Cells were typically fixed for size measurement [152] or grown in a micro-chamber supplied with fresh medium during imaging [11, 14, 16, 17]. By photographing individual cells over a whole generation from birth to division, the cell-to-cell variability can be well-quantified, and yields distributions of cell properties such as cell size [12, 13, 152, 153] and generation time [14–18, 154–156]. Electron microscopy was also employed by some researchers to precisely measure the cell size distribution [152, 157–162]. More recently, fluorescence microscopy has become a standard technique to acquire both cell size and intracellular protein dynamics, due in large part to the availability of strains expressing fluorescently labeled proteins [163–165].

In parallel with microscopy, Kubitschek in 1958 first employed the Coulter counter to measure individual bacterial cell size [166]. The Coulter counter measures the resistance of a conducting solution along a microchannel when a cell passes through. The number of cells in a solution is counted by the number of resistance pulses as all cells flow through, and the cell volumes can be inferred from the amplitude of each pulse. The Coulter counter was often used as a high-throughput method to determine cell size distribution [167–171]. The principle of Coulter counter was later

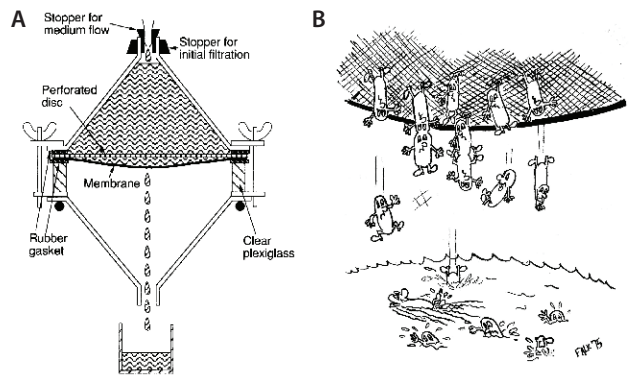

Figure 5: **Baby machine for age-synchronized sampling. A.** The schematic diagram of the membrane elution apparatus, adapted from [172]. **B.** A cartoon for baby machine, adapted from [70] with permission.

used in other techniques for single-cell measurements, *e.g.*, flow cytometer (see Section 2.1.6).

### 2.1.5. Baby machine: in search of synchrony.

The original motivation of synchronizing the division cycle of a population is to capture the behavior of the ideal 'average cell' by aligning the division cycle of all cells [173–175]. Early attempts at synchronization included temperature shock or nutritional shock [107, 176, 177]. The idea being that the heat shock or nutrient starvation would arrest the division cycle at a certain point, and that the cell cycles would be synchronized upon resumed growth. Unfortunately, shocks and shifts are not able to synchronize the population because they do not lead to a narrowing of the age distribution (Section 2.2.1) [4, 172, 178]. Furthermore, we now know that shocks induce transient stress-responses in the cell, perturbing cell physiology.

While developing a method of synchronization based on size-fractionation, Charles Helmstetter hit upon an alternative that did not narrow the age distribution, but rather sampled a narrow strip from the age distribution. His device came to be called the 'baby machine.' Briefly, cells were first filtered through a membrane with pores smaller than the size of the cells. The membrane is then flipped upside down, and most cells remain attached to the underside of a membrane. As growth media flows through the membrane, cells continue to grow and newly born cells are eluted (Figure 5). In a small volume of effluent, Helmstetter had effectively millions of identically-aged single cells [147, 148, 179–182]. This method of synchronous-age sampling became the standard for studying DNA dynamics [183–186]. In 1968, Helmstetter and Cooper used the baby machine to establish their model of multifork replication in rapidly growing *E. coli* cells (see Section 2.2.4) [72, 111, 178].

*2.1.6. Flow cytometry: measuring DNA content and cell cycle parameters.* Flow cytometry was developed in 1960s and was immediately applied to the sorting of particles, macromolecules and cells in large numbers based on physical and chemical properties. In a typical flow cytometer, a cell suspension flows along an ultra-thin channel that accommodates one cell at a time. The cell size is inferred from light scattering, and cellular components such as nucleic acid are stained fluorescently, illuminated by laser and detected in a spectrum-specific manner. Flow cytometry was first used to characterize the bacterial cell in late 1970s [187, 188]. Bailey and others measured the cellular composition, including protein and nucleic acid contents, of *B. subtilis* at high-throughput ($> 10^3$ cells per second) [187, 189, 190]. Paau and colleagues studied cell size and nucleic acid content of several bacterial species [188]. The high-throughput of flow cytometry enables the analysis of distributions of cell properties, *e.g.*, the distribution of DNA content per cells, and quantifying cell cycle parameters in an accurate and non-invasive fashion (see the definitions of cell cycle parameters in Figure 2F).

Steen, Boye and Skarstad pioneered the measurement of the bacterial cell cycle by flow cytometry [191–197]. First, the bulk DNA content of an exponentially growing population is measured. By applying the canonical age distribution (Section 2.2.1) and Helmstetter-Cooper's model for the DNA content per cell (Section 2.2.4), the cell cycle parameters can be calculated from the fit to the distribution of DNA content [194, 195]. As an independent check, Steen *et al.* measured the timing of replication initiation in antibiotic-treated cells that were separated into two populations: before initiation of DNA replication and after [197–202]. They used rifampicin, which at sub-lethal dosage halts the initiation of a new round of DNA replication but permits completion of ongoing replication. They also used cephalexin, which stops cell division. Therefore, cells treated with rifampicin will contain integer multiples of one chromosome equivalent DNA, and the populations before and after initiation can be separated by flow cytometry. The initiation timing can then be calculated from the ratio of the two populations. Subsequently, flow cytometry has been widely used to examine the bacterial cell cycle in a variety of growth conditions and genetic backgrounds [203–215].

*2.1.7. Thymine deficient mutants and antibiotics: perturbation and growth inhibition experiments.* The Copenhagen school laid the groundwork for quantitative studies of bacterial physiology. In addition to nutrient limitation and shift-up experiments reviewed in the Section 2.1.1, Maaløe's laboratory developed perturbation methods to study bacterial growth and biosynthesis. One example is the use of a thymine deficient mutant of *E. coli*. As one of the four nucleotide bases in DNA, external supply of thymine is required for the deficient mutant strain to sustain its DNA synthesis and survive. Otherwise, the DNA synthesis rate will be impeded in a thymine-dependent manner. Maaløe and others initially used this strain to study the relationship between DNA, RNA and protein synthesis [119, 144, 216].

Donachie in 1969 reported an important thymine starvation experiment: in the absence of thymine, mutant strains not only stopped their DNA synthesis but also cell division. Upon the re-addition of thymine, both DNA synthesis and cell division resumed, although with a constant time-delay in between. Thus he concluded that cell division requires completion of DNA synthesis [75]. Pritchard and Zaritsky performed a thymine limitation experiment where they titrated the thymine concentration in the growth medium. They found that the DNA synthesis rate was altered in a thymine-dependent manner. Intriguingly, the growth rate was not significantly affected, and seemingly decoupled from DNA synthesis [217]. Subsequently, Zaritsky and colleagues investigated other consequences of thymine limitation, including effects on cell size, cell shape, replication initiation and so on [38, 218–224]. Thymine limitation became an important method for unraveling the connections between DNA synthesis and other aspects of bacterial cell physiology [225–237]. Beyond thymine deficiency, a number of mutant strains were used to study the coordination between growth, cell cycle and cell size of bacteria (see, for example, [238–245]).

In parallel with genetic perturbations, antibiotics were used to perturb bacterial physiology. Rye and Wiseman performed a survey of the effects of multiple antibiotics on growth rate and cell size; this is the earliest experiment using antibiotics to investigate cell size control [246, 247]. Antibiotics were instrumental in elucidating the coordination between the inhibition of biosynthesis and other aspects of cell physiology [193, 248–250]. Sub-lethal concentrations of antibiotics remain a powerful tool for quantitative biologists to study bacterial physiology [251, 252].

*2.1.8. Computer simulation: testing models against experimental data.* Alongside technological innovations born on the lab bench, computer simulation (which became prevalent in 1960s-1970s) facilitated the quantitative study of bacterial physiology. Margolis and Cooper first ran a computer program to simulate bacterial growth and cell cycle in 1970, two years after the publication of Helmstetter-Cooper model. Computer simulations at that time were used to numerically investigate models or make qualitative predictions [253–255]. Koch, in his 1977 work, for the first time ran a computer simulation to rigorously evaluate different models by fitting to experimental data [256]. After that, more researchers employed computer simulations to clarify their understanding of bacterial systems. For example, Bremer and his colleagues performed simulations to differentiate models of biosynthesis by comparing theoretical predictions with RNA and protein synthesis rates measured in his lab [113, 235]. Skarstad and others used computer programs to simulate DNA content distributions and fit these to their flow cytometry data [195, 214]. Subsequently, computer simulations have played an essential role in testing various hypotheses underlying the regulation and coordination of bacterial physiology, including cell cycle [207, 257–260], replication initiation [261], cell-size control [9, 262] population dynamics [103, 263] and so on [264, 265].

## 2.2. Part II: Major models and conceptual advancements

Early work in microbiology was complicated by a lack of well-defined state variables and standard reference conditions. Up to the middle of the $20^{th}$ century, influenced by the work Buchanan [8] and Henrici [114, 266], the microbial 'life cycle' was thought to echo our own human development: cells inoculated to fresh media from an overgrown culture start small and sickly, but become large during vigorous exponential growth only to shrink into the frailty of old age as the culture becomes overgrown. According to Henrici [266],

> It is quite evident that similar laws govern the development of both the multicellular organism and the population of free unicellular individuals. To some extent a culture of bacteria...behaves like an individual, and we may look upon the progressive cell changes [during the microbial life cycle] as the same sort of phenomenon as the cytomorphosis occurring in a multicellular animal.

Implicitly, the bacterial culture is thought of as a multicellular aggregate of undifferentiated bacteria.

Jacques Monod's review in 1949 [6] made a clear case that, with the properly-chosen state variables, simple quantitative relations could be discerned in the complex 'phases' of bacterial growth. Perhaps the most enduring insight from Monod's early work on bacterial physiology is the hyperbolic dependence of the exponential growth rate $\lambda$ on the concentration of a growth-limiting substrate $S$,

$$\lambda = \lambda_{max}^0 \frac{S}{S + K_D}, \qquad (1)$$

where the phenomenological parameters $\lambda_{max}^0$ and $K_D$ are properties of the bacterial strain and the growth-limiting nutrient $S$. Eight years later, it was Campbell [267] who brought the exponential growth rate $\lambda$ to the forefront of physiological studies by defining the notion of 'balanced growth' (Box 2).

In balanced growth, all of the complexity of cellular regulation and adaptation operates in a concerted manner to ensure that every constituent in the cell doubles at the same rate. Furthermore, balanced growth was no longer seen as a characteristic of one of the many 'phases' of growth the bacterium must pass through – it was a steady-state that could be maintained by dilution for as long as the investigator wished. According to Elio Schaechter, "the difference between 'exponential phase' and 'balanced growth' is the difference between watching apples fall and thinking of gravity [268]." The scientific focus shifted from the phases of the growing culture to the individual bacteria in balanced growth.

With the definition of a standard reference state of growth, the study of bacterial physiology entered a golden age. The following decade (1958-1968) saw seminal advances in the understanding of cellular growth, which continue to be a source of wonder and inspiration for over 50 years. In the following section, we will briefly review the milestones in the study of bacterial growth and reproduction from that period.

In pursuit of a mechanistic explanation for the pauses in growth observed upon a change in nutrients, Monod shifted from bacterial physiology as his primary focus. Nevertheless, he left a lasting legacy on the field, advocating for, and pioneering, many of the core analytic methods used to study bacterial growth physiology. Although molecular mechanisms of gene regulation became a major research theme of the Pasteur group (leading to Jacob, Monod and Lwoff winning the Nobel prize in 1965), elucidation of the larger context of that regulation was never abandoned, and Monod returned time and again to help shape our present view of bacterial growth (see, for example, Section 2.2.3)

---

**Box 2 – Exponential vs. balanced vs. steady-state growth**

Exponential growth, balanced growth, and steady-state growth are often used as synonyms with each other. However, there are subtle but important differences as explained below.

**Exponential growth**. During the growth of a cell culture, an exponential growth phase is reached if during that period, the number of cells in the culture follows the equation,

$$N(t) = N_0 e^{\lambda(t - t_0)} \qquad (2)$$

where $\lambda$ is a constant; $t$ is the time; $t_0$ is a reference time and $N_0$ is the cell number at $t_0$ [269].

**Balanced growth**. Allan Campbell in his 1957 paper [267, 268] defined balanced growth as follows:

> *Growth is balanced over a time interval if, during that interval, every extensive property of the growing system increases by the same factor.*

A cell culture can be in balanced growth, even if individual cells do not show balanced growth. A good example is *E. coli* cells in slow growth conditions without exhibiting overlapping cell cycles, where the rate of DNA synthesis in individual cells is discontinuous although their growth is continuous.

**Steady-state growth**. In Painter and Marr's 1968 paper, they defined steady-state growth as

> *The distribution of each intensive random variable (e.g., cell age or cell protein) does not depend on the time.*

Note that the exponential growth of a cell culture does not imply steady-state, whereas steady-state always implies both exponential and balanced growth [269, 270]. One obvious example is that, when cell division is blocked by specific antibiotics without affecting growth (*e.g.*, sub-lethal dosage of cephalexin or penicillin), the population follows balanced growth while distribution of cell size is clearly variant over time [270].

*2.2.1. Age and size distributions of a growing population.* In Section 1, we defined the age of the cell (Figure. 2D-E). The age distribution is the basis of understanding the quantitative properties of any steady-state growth. For example, it was essential for the analysis of Helmstetter's baby machine experiments (Section 2.2.4). The age distribution has a long history [12, 271–277]. Koch expressed in 1976 his amusement citing a full list of independent derivations of the age distribution [278]:

> *Workers too numerous to mention have independently derived the limiting law describing the distribution of ages in an asynchronous population of exponentially growing cells where all cells have precisely the same doubling time. The earliest published paper with the seeds of this derivation appears to be that of J. G. Hoffman in 1949 (2).*
> ...
> *This list presents clear evidence that biology as a discipline is markedly different than physics,*

> *where it would be inconceivable for example that a dozen Pauli's might discover, and publish as news, their exclusion principle in different journals over a 19-year period. The basic, more general principle that leads to this distribution was stated by Euler in 1760 (see 1970, Theor. Pop. Biol. 1:307).*

In what follows, we present one form of the derivation by Powell [279]. His opening remark captures the heart of the age distribution:

> *The age distribution in a growing culture has a curious and interesting property which is not generally known; roughly speaking, the youngest organisms are present in greatest number.*

Intuitively, for a population of exponentially growing cells, newborn cells are twice as abundant as those about to divide, which should be explicitly reflected in the formulation of age distribution. Specifically, denote by $\varphi(a)da$ the probability to find a bacterium with age between $(a, a + da)$, and denote by $\rho_{\tau_d}(\tau)d\tau$ the probability that a given bacterium divides at age $a \in (\tau, \tau + d\tau)$; we call $\varphi(a)$ the *age distribution* of the population, and we call $\rho_{\tau_d}(\tau)$ the *doubling time distribution*. Throughout, we assume the culture is in a steady-state of exponential growth $N(t) = N_0 e^{\lambda t}$, and that the age distribution is likewise at steady-state, *i.e.* $\varphi(a, t) \to \varphi(a)$.

Because $\rho_{\tau_d}(\tau)$ is a normalized probability distribution, the cumulative density $F_>(\tau)$,

$$F_>(\tau) = \int_\tau^\infty \rho_{\tau_d}(\tau')\, d\tau',$$

is a measure of the proportion of bacteria in the population with doubling times greater than $\tau$. In particular, if a bacterium has attained an age of $a$, then the probability that it will attain an age $a + t$ without dividing is given by the ratio,

$$\begin{array}{l}\text{Given bacterium reaches an age } a, \\ \text{probability no division in } (a, a+t)\end{array} = \frac{F_>(a+t)}{F_>(a)}.$$

This expression can be understood as an application of Bayes rule: the *joint* probability that the doubling time $\tau$ is greater than $a+t$ and $a$, denoted by $P(\tau > a+t, \tau > a)$, is the product of the *conditional* probability that $\tau > a+t$ given that $\tau > a$, denoted $P(\tau > a+t | \tau > a)$, with the probability that $\tau > a$, denoted $P(\tau > a)$,

$$P(\tau > a + t, \tau > a) = P(\tau > a + t | \tau > a)P(\tau > a).$$

Rearranging for the conditional probability,

$$P(\tau > a + t | \tau > a) = \frac{P(\tau > a + t, \tau > a)}{P(\tau > a)}.$$

But $t > 0$, so the $\tau > a$ condition is automatically satisfied if $\tau > a + t$ and the joint probability reduces to $P(\tau > a + t, \tau > a) \equiv P(\tau > a + t)$. Finally, the singlet probabilities can be written in terms of the cumulative density $F_>$,

$$P(\tau > a + t | \tau > a) = \frac{P(\tau > a + t)}{P(\tau > a)} = \frac{F_>(a + t)}{F_>(a)}.$$

For a culture of $N$ bacteria, the *number* of bacteria with age between $(a, a + da)$ is $N\varphi(a)da$, so the *number* of bacteria still undivided at $a + t$ is given by,

Number of bacteria still undivided at age $(a + t)$ $= N\varphi(a) \, \frac{F_>(a + t)}{F_>(a)} \, da.$

But during the interval of time $t$, the total culture population has grown to $Ne^{\lambda t}$; the *fraction* of survivors is then given by

Fraction of bacteria still undivided at age $(a + t)$ $= \dfrac{N\varphi(a) \, \frac{F_>(a+t)}{F_>(a)} \, da}{Ne^{\lambda t}}$

$= \varphi(a) \, \dfrac{F_>(a + t)}{F_>(a)} \, e^{-\lambda t} da,$

with age between $(a + t, a + t + da)$. This is exactly $\varphi(a + t)da$, and so we have the difference equation,

$$\varphi(a) \, \frac{F_>(a + t)}{F_>(a)} \, e^{-\lambda t} = \varphi(a + t), \tag{3}$$

for all $t$.

It is difficult to solve this equation directly for the age distribution $\varphi(a)$; if we look instead at the small-time limit $(t \to 0)$ we can get some insight into the solution. In the limit $t \to 0$, using the leading-order Taylor polynomial approximations,

$$F_>(a + t) = F_>(a) + tF_>'(a) + \mathcal{O}(t^2),$$

$$\varphi(a + t) = \varphi(a) + t\varphi'(a) + \mathcal{O}(t^2),$$

$$e^{-\lambda t} = 1 - \lambda t + \mathcal{O}(t^2),$$

the difference equation (Eq. 3) reduces to a separable first-order differential equation for $\varphi(a)$,

$$\frac{\varphi'(a)}{\varphi(a)} = \frac{F_>'(a)}{F_>(a)} - \lambda + \mathcal{O}(t^2),$$

or,

$$\frac{d}{da} \ln \varphi(a) = \frac{d}{da} \ln \left[ F_>(a) e^{-\lambda a} \right] + \mathcal{O}(t^2).$$

Integrating both sides, (and dropping $\mathcal{O}(t^2)$ terms),

$$\varphi(a) = \varphi(0) e^{-\lambda a} F_>(a) = \varphi(0) e^{-\lambda a} \int_a^\infty \rho_{\tau_d}(\tau') d\tau', \tag{4}$$

where the integration constant $\varphi(0)$ is chosen to normalize the age distribution,

$$\int_0^\infty \varphi(a) da = 1.$$

Notice that both the exponential and the integral are positive and decrease monotonically with $a$, and so the maximum of $\varphi(a)$ is $\varphi(0)$, *i.e.* irrespective of the doubling time distribution, the population is mostly composed of younger cells. Powell [279] provides a derivation for the constant of integration $\varphi(0)$ for an arbitrary doubling-time distribution $\rho_{\tau_d}(\tau)$; for our purposes it is sufficient to take the simplified distribution $\rho_{\tau_d}(\tau') = \delta(\tau - \tau')$, *i.e.* all bacteria in the population divide when their age is exactly equal to the doubling time, $a = \tau$. In that case,

$$\varphi(a) = \varphi(0) e^{-\lambda a} \int_a^\infty \delta(\tau' - \tau) d\tau'$$

$$= \begin{cases} \varphi(0) e^{-\lambda a}, & 0 \le a < \tau \\ 0, & \text{otherwise} \end{cases}$$

Using the normalization condition to determine $\varphi(0)$, we have,

$$\varphi(a) = \frac{\lambda}{1 - e^{-\lambda \tau}} e^{-\lambda a}, \quad 0 \le a < \tau,$$

or, because $\lambda = \ln 2 / \tau$,

$$\varphi(a) = \frac{(2 \ln 2)}{\tau} 2^{-a/\tau}, \quad 0 \le a < \tau. \tag{5}$$

This is the idealized, or canonical, age distribution (see red line in Figure 2E, right). Based on this canonical age distribution, when assuming that cell elongates exponentially, the canonical distribution for cell length $\rho_l(l)$ across the population is given by,

$$\rho_l(l) = \frac{l_d}{l^2} \tag{6}$$

(see red line in Figure 2E, left).

### 2.2.2. The Copenhagen School of Bacterial Physiology.

Bacterial growth and bacterial physiology had been studied for many years, but the late 1950s marks a watershed period. Building upon their rigorous experimental methods discussed in the previous section, in 1958, Maaløe's lab published back-to-back papers that are recognized as the gold-standard of what a quantitative approach could achieve (Stephen Cooper has called them the 'fundamental experiments of bacterial physiology' [172]). This, and subsequent work (along with the scientists passing through Maaløe's lab), became known as the 'Copenhagen school' of bacterial physiology. One of the major tenants of the school was: 'Look - but don't touch!' and great pains were taken to minimize perturbations to balanced growth during experimental observation [280]. As a consequence, the data collected by the Copenhagen school was unprecedented in its accuracy and reproducibility. To the analysis, Maaløe brought to bear his prodigious mathematical skills and his extraordinary intuition for the inner-life of bacteria [4].

The first of the ground-breaking 1958 papers from the Copenhagen school focuses on the macromolecular composition of *Salmonella typhimurium* in balanced growth, with growth rate modulated by the nutrient composition of the medium and by temperature [2]. Using 20 different growth media, Schaechter *et al.*found that the macromolecular composition of Salmonella is largely dependent upon growth rate alone. Suppose, for example, you have two flasks full of different media – one with a poor carbon source and rich nitrogen source, the other with a rich carbon source and a poor nitrogen source – but designed so that the cells double every hour. Despite huge differences in how the nutrients are processed, the large scale composition, including DNA/cell, RNA/cell, protein/cell, mass/cell, are all the same. At this macroscopic level, the bacteria growing in the two flasks are indistinguishable.

Moreover, when plotted against doubling rate $\mu = \lambda/\ln 2$, the mass/cell is roughly exponential, *i.e.*, mass/cell $\propto 2^\mu$ (Figure 6A (blue)). That is, fast-growing cells are bigger. At a doubling rate of doubling/20 minutes, cells are twice as large as those growing at a rate of doubling/30 minutes, and four-times as large as those growing at a rate of one hour/doubling. Notice something about the units – the proportionality should really be written $\propto 2^{\mu/\mu_0}$ where $\mu_0 = 1$ doubling/hour. What is the significance of this timescale, $\mu_0 = 1$ doubling/hour? It would take 10 more years for the work of Cooper and Helmstetter to explain from where $\mu_0$ comes.

A second empirical relation observed by Schaechter *et al.*is that the RNA/cell increases more rapidly than mass/cell (Figure 6A (green)), *i.e.*, $e^{2.85}$. Taking the $\log_2(2.85) \approx 1.5$, this expression can be concisely written as RNA/cell $\propto 2^{1.5\mu}$ (the RNA/cell data could as well be fitted by RNA/cell $\propto (a + b\mu) \, 2^{\mu/\mu_0}$; see Section 5). Finally, the DNA/cell increased with doubling rate more gradually than mass/cell (Figure 6A (gold)), per unit increase in the doubling rate, the DNA/cell increases by $\times 1.73$. Taking the $\log_2(1.73) \approx 0.8$, this expression can be concisely written as DNA/cell $\propto 2^{0.8\mu}$. In all, they conclude that the rate of increase in per-cell abundance with growth rate is: RNA > mass > DNA.

Although changes in temperature affect the growth rate, the macromolecular composition was

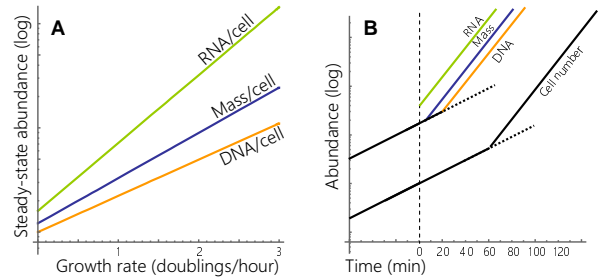

Figure 6: **Schaechter-Maaløe-Kjeldgaard experiments. A. Balanced growth.** [2] When growth rate is changed by the quality of the available nutrients, the per-cell abundance of RNA, Mass and DNA scale approximately exponentially with the doubling rate $\mu$: RNA $\propto 2^{1.5\mu}$, Mass $\propto 2^\mu$ and DNA $\propto 2^{0.8\mu}$. **B. Nutrient shift-up.** [3] At time $t = 0$, the culture is shifted from glucose minimal medium (doubling time 50 minutes) to broth (doubling time 20 minutes). The transition to the post-shift rate of accumulation is abrupt (almost discontinuous for RNA), and occurs at 5 minutes for mass, 20 minutes for DNA and 70 minutes for cell numbers. The timing of these transitions is invariant to the details of the per- and post-shift media, and determines the slopes of the Mass/cell and DNA/cell lines in panel A.

unchanged: "In all cases, the growth rate at $25\,^{\circ}\mathrm{C}$ was about half that at $37\,^{\circ}\mathrm{C}$; nevertheless, mass, RNA, DNA and number of nuclei/cell remained constant for a given medium... In fact, our data suggest that more extensive analyses of $25\,^{\circ}\mathrm{C}$ cultures would permit graphs to be constructed which would be identical with those of [Figure 6A] if the growth rate values on the abscissa were reduced to half. Thus, within the temperature range studied,

*The size and chemical composition of the cells are related to the growth rate only in so far as it depends on the medium.*

The second paper in this series studies *shifts* in growth media [3]. Here, we focus on nutrient 'up-shift', transitioning from slower to faster growth rates; transitioning from glucose minimal medium (50 minute doubling time) to rich broth (22 minute doubling time), they observed that (1) the synthetic activities respond chronologically as rate increase in RNA, mass, DNA, then cell division; (2) rate of mass accumulation transitions at about 5 minutes, DNA at 20 minutes and cell division at 70 minutes post-shift, and these time intervals are *irrespective of the details of the pre- and post-shift media*; (3) the shifts in synthesis rates, when they occur, are very abrupt so that mass, DNA and cell numbers are piece-wise exponential functions – RNA shifts more abruptly, and is practically discontinuous at the time of shift.

The observation that DNA and cell division shift at 20 minutes and 70 minutes post-shift irrespective of the details of the growth media (*i.e.*irrespective of the cell growth rate) is remarkable. On the other hand, Maaløe and his colleagues realized that because the transition to the new synthesis rate is so abrupt, the timing of the transition is implicit in the steady-state growth dependence in the per-cell abundance as shown in Figure 6A. A shift-up moves the per-cell abundance of constituent $Y$ (*e.g.*DNA, mass) from doubling rate $\mu_1$ to $\mu_2$. Suppose we follow the mass (or OD) per cell. Denote by $m(t)$ the mass-per-cell; in the rich medium at doubling rate $\mu_2$, the mass-per-cell $m_2(t)$ is given by the ratio,

$$m_2(t) = \frac{OD(t)}{N(t)},$$

where the optical density $OD(t)$ is proportional to the dry mass per mL, and $N(t)$ is the cell number per mL. Denote by $t_A$ the time post-shift that $OD(t)$ attains its new accumulation rate $\mu_2$, then post-shift

$$OD(t) = OD_1(t_A)2^{\mu_2(t-t_A)},$$

where $OD_1(t) = OD_1(0)2^{\mu_1 t}$ is the pre-shift mass density. Similarly, for the cell number $N(t)$,

$$N(t) = N_1(t_B)2^{\mu_2(t-t_B)},$$

where $N_1(t) = N_1(0)2^{\mu_1 t}$ is the pre-shift number density, and $t_B$ is the time post-shift that $N(t)$ attains its new accumulation rate $\mu_2$. After both $OD$ and $N$ have transitioned to their new accumulation rate $\mu_2$, the mass-per-cell is at steady-state $m_2(t) \to m_2$; putting these expressions together,

$$m_2(t) = \frac{OD(t)}{N(t)} = \frac{OD_1(0)2^{\mu_1 t_A}2^{\mu_2(t-t_A)}}{N_1(0)2^{\mu_1 t_B}2^{\mu_2(t-t_B)}},$$

or,

$$m_2 = \frac{OD(0)}{N(0)} 2^{(t_A-t_B)[\mu_1-\mu_2]} = m_1 2^{(t_B-t_A)[\mu_2-\mu_1]}$$

because $OD(0)/N(0) = m_1$ is the steady-state mass-per-cell prior to the shift. Taking the log of both sides,

$$\Delta \log_2 m = \Delta\mu(t_B - t_A) \implies \frac{\Delta \log_2 m}{\Delta \mu} = (t_B - t_A).$$

But this is just the slope of the log-plot of the steady-state mass-per-cell. How do the two compare? The mass (or OD) transitions about 5 minutes post-shift, the DNA about 20 minutes post-shift and the cell number about 70 minutes post-shift. That gives a slope of,

$$(t_B - t_A) \approx 70 - 5 = 65 \text{ minutes}$$

for the mass-per-cell (60 minutes from the slope of the steady-state data); whereas

$$(t_B - t_A) \approx 70 - 20 = 50 \text{ minutes}$$

for the DNA-per-cell (48 minutes from the steady-state data) – so both estimates agree very well (within 5-10%).

The two Maaløe papers from 1958 established, like nothing before, the value of studying of bacterial physiology in balanced growth, and demonstrated that even nutrient shifts could be best understood in that context. Stephen Cooper has written eloquently, and at length, about the Schaechter-Maaløe-Kjeldgaard experiments [4, 5]. He sees in them a necessary re-examination of the phases of growth described by Buchanan, Henrici and Monod [5]:

> ...the classical bacterial growth curve is really a laboratory artifact of using overgrown cultures taken from the previous day to start up a growing culture. I suggest that the results of Schaechter-Maaløe-Kjeldgaard indicate that one should teach the shift-up and shift-down results in classes, and then consider the classical growth curve as a special case of shift-ups and shift-downs.

### 2.2.3. The role of RNA and ribosomes in protein synthesis.

By the late 1950s, reseachers observed that the RNA content (RNA/cell) is larger in 'growing' cells than in 'non-growing' cells, but there was no consensus yet on what that increase in RNA meant [2, 281, 282]. At the time of their study, Neidhardt and Magasanik were able to write that "[e]mbarrassingly little is known of the role played by RNA in the growth and metabolism of bacteria... Since the ribosomal RNA constitutes the bulk of the total RNA, one hint of the role of ribosomal RNA may be found in the observation that the rate of protein synthesis and the total amount of RNA are concomitant variables in most biological systems [58]."

Neidhardt and Magasanik's paper had several important consequences. First, they made it clear that 'growing', and 'non-growing' are not well-defined terms (as did the Schaechter, Maaløe and Kjeldgaard paper that preceded it): the physiological state of 'non-growing' cells is very much dependent upon how the cells are introduced to the stationary state, and the RNA/cell, or more usefully the RNA/protein, varied several fold depending upon the exponential growth rate. In fact, the correlation between the RNA/protein ratio and growth rate was approximately linear for doubling rates above 0.6 doublings/hour (Figure 7).

In addition, they measured the fraction of total RNA that is ribosomal RNA, and found that the fraction (86%) is growth rate independent. Taken together, the growth rate is positive-linearly correlated with the mass fraction of ribosomes; that is enough for them to conclude that the ribosome plays a catalytic role in protein synthesis (see Section 5 below).

To lend further support to their conclusion, they observed the dynamics of different cell constituents during a nutrient up-shift. As in the Schaechter-Maaløe-Kjeldgaard experiments (Figure 6B), they observed a rapid and immediate increase in the synthesis of RNA, then protein and DNA synthesis began to increase – "The results of this experiment make it unlikely that high RNA content is simply a consequence of fast growth rather than a necessary antecedent. But if this conclusion is correct, then one should be able to vary the nature of the supplement without changing the results of the experiment." This is indeed what they observed.

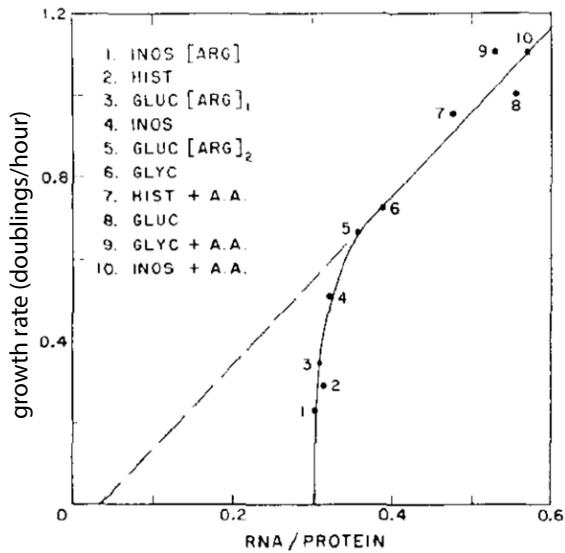

Figure 7: Above about 0.6 doublings/hour, the RNA/Protein ratio is linear. Neidhardt and Magasanik took this as evidence that ribosomes play a catalytic role in protein synthesis [58].

Neidhardt & Magasanik established the role of the ribosome in catalyzing protein synthesis. From this work emerged the hypothesis of Maaløe that ribosomes operate at maximum rate, and that growth rate is proportional to the protein mass fraction of ribosomes [55]. Later work demonstrated that the rate of protein synthesis per ribosome is not growth-rate independent [283, 284], but that this growth-dependence can be reconciled with the linearity between RNA/Protein and the growth rate observed by Neidhardt & Magasanik [285]. According to Maaløe, to grow faster the cell must increase its protein mass fraction of active ribosomes [55, 285, 286], but we now appreciate that this increase must come at the expense of synthesizing other proteins. The implicit constraints on protein synthesis imposed by the demand for ribosomes produce global growth-rate dependence in the expression of most proteins. Consequences of this indirect regulation is one of the newly-emergent themes in growth physiology (see Section 5).

The idea that the ribosome played a catalytic role in protein synthesis was very much in the air at that time. Although the dominant view in 1960 was that 'each gene controls the synthesis of one specialized ribosome, which in turn directs the synthesis of the corresponding protein - a scheme which could be epitomized as the one gene-one ribosome-one protein hypothesis [287],' contemporary data, particularly from the Monod group, had challenged that view.

In 1957, Pardee, Jacob and Monod [288] demonstrated the existence of a protein regulator that inhibited transcription (what is now called a 'transcription factor' or more specifically, a 'repressor'). But the prevailing view at the time was that if a gene were turned on, a gene-specific ribosome would need to be made before protein synthesis could begin. Pardee, Jacob and Monod observed no such delay.

In the spring of 1960, along with Sydney Brenner, Fracis Crick, Leslie Orgel and Ole Maaløe, Jacob discussed this problem and came shortly to realize that the ribosome plays a catalytic role in protein synthesis, with genetic information carried by an unstable RNA intermediate (now called 'messenger RNA' or 'mRNA') [289, 290]. That summer, Jacob and Sydney Brenner, in the lab of Mathew Meselson at Caltech, established the existence of mRNA directly, and crushed the 'one gene-one ribosome-one protein' hypothesis forever [288]. In the words of Francis Crick, '[o]nce it was realized that the ribosome was basically a reading head the world never looked the same again [291].'

### 2.2.4. The Helmstetter-Cooper E. coli cell cycle model.
Before reviewing the cell cycle model in *E. coli*, it is worth noting that many evolutionary divergent bacterial organisms such as the Gram-negative *E. coli* and Gram-positive *B. subtilis* exhibit multifork replication. This is in stark contrast to the cell cycle of eukaryotes and some other bacterial species, where cell cycle 'check points' ensure replication cycles do not overlap. *C. crescentus* is a model bacterial organism that shares similar cell cycle features, and interested readers may wish to read review articles from Shapiro group [292–294].

As discussed in Section 2.1.3 and Section 2.1.5, in 1968, Helmstetter and Cooper used the baby machine and radioactive labelling to elucidate the mode of DNA synthesis in *E. coli*. Their back-to-back papers, similar to Schaechter, Maaløe and Kjeldgaard's dual works in 1958, became another cornerstone in the history of bacterial physiology. In the baby machine experiment, cells in steady state have an approximately-exponential age distribution prior to their attachment to the membrane (Eq. 5, Figure 8B). As older mothers divide, their daughters are washed away by the flow of media. As a result, the cell concentration in the effluent is inverted in time with respect to the age distribution of the mothers (Figure 8C). Prior to immobilization, the mother cells are 'pulse labeled' with radioactive thymidine (*i.e.*, exposed to a saturating amount of the radioactive nucleotide for several minutes before being fixed to the membrane). If there is a change in the rate of DNA synthesis during the cell cycle at age $a_i$ after division (measured as a fraction of the generation time) (Figure 8D), then the daughter cells in the effluent should exhibit a step-increase in the radiolabel but with the time axis inverted relative to the mothers (Figure 8E).

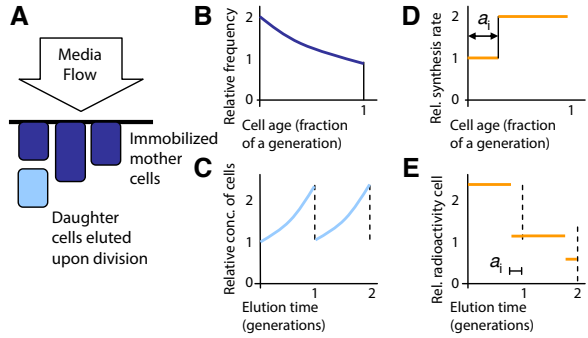

Figure 8: **DNA synthesis in age-synchronized cultures. A. Baby machine.** Mother cells (dark blue) are immobilized to the underside of a membrane through which media flows. Newborn cells (pale blue) are shed into the effluent. **B. Age distribution of mothers.** For exponentially-growing cells, newborn cells are twice as likely as those about to divide. **C.** In the effluent, the age distribution is inverted in time – first daughters from old mothers, then daughters from young mothers. **D. DNA synthesis rate in the mothers.** A step increase in DNA synthesis rate, corresponding to initiation of a round of DNA replication, occurs at an age $a_i$. **E.** As in panel B, the step-increase in DNA synthesis rate (measured using radioactive nucleotides) is inverted in the daughters and occurs a time $a_i$ *before* the division event. Panels B-E redrawn from [148].

During rapid growth (*i.e.*, doubling times below 60 minutes), Helmstetter & Cooper observed regular, periodic changes in the DNA synthesis rate during the growth cycle; however, the age of the bacterium when the initiation event occurs exhibits puzzling discontinuities – jumping from 0 to 1 at doubling times of about 30 minutes (as well as at doubling times of about 60 minutes using Helmstetter's data from more slowly growing bacteria [148]) (Figure 9B). Their resolution of this puzzle was to propose that the bacterium is initiating multiple simultaneous rounds of DNA replication (Figure 9A). In the follow-up paper [72], Cooper & Helmstetter proposed a two-timer model (Figure 9),

- **Cell division time** $\tau_d$

    (*i.e.*, the time between birth and division)

- **Cell cycle time** $\tau_{cyc} = \mathbf{C} + \mathbf{D}$

    **C** – the time to replicate the chromosome
    **D** – the time to segregate the chromosomes and divide

They found that under conditions of rapid growth imposed by the quality of the nutrient environment, the **C** period is about 40 minutes and the **D** period is about 20 minutes, irrespective of growth rate. As a consequence, cells growing at a doubling time longer than the cell cycle time ($\tau_{cyc} \approx 60$ minutes) had a latency period of ($\tau_d - \tau_{cyc}$) before DNA synthesis began. With this simple phenomenological model, they were able to fully characterize the patterns in DNA synthesis observed in the baby machine cells.

The essential idea is to set cell division as a reference point, and trace backward in time the onset of the replication cycle. This leads to 'tiling' a continuous series of cell-replication events, each of inter-division time $\tau_d$, with strips of length $\tau_{cyc}$. Consider a cell during its growth cycle under conditions of balanced exponential growth; how many generations ago was DNA replication initiated in order to ensure timely completion before division? If the inter-division time is long ($\tau_d \gg \tau_{cyc}$), then DNA replication can be initiated at some point during the present division cycle. If the inter-division time is short ($\tau_d < \tau_{cyc}$), then DNA replication initiation must begin in a previous generation. Exactly how far back depends upon the ratio between the two timers,

$$\begin{array}{l}\text{Generations } \textit{before} \text{ division that} \\ \text{DNA replication is initiated}\end{array} = \frac{\tau_{cyc}}{\tau_d} - 1.$$

Here, this number between 0 and 1 means that initiation started in the mother generation, and between 1 and 2 means in the grandmother generation, and so on (see Figure 9A). This same reasoning applies even if the inter-division time is greater than the cell cycle – in that case, a negative generation corresponds to DNA replication initiation occurring in the same cell cycle that it is destined to terminate. For example, if the inter-division time is 100 minutes and the cell cycle time is $\tau_{cyc} = 60$ minutes, then the generations before division that DNA replication is initiated is $-0.4$, or 0.4 of a generation (*i.e.*, $0.4 \times 100$ minutes = 40 minutes) *after* the cell is born in the current generation.

The simultaneous running of parallel replication forks makes the original data difficult to interpret. What is observed is changes in the DNA replication rate. When the age of the cell at the moment of DNA replication initiation is plotted as a function of the inter-division time, there is a discontinuity at around 30 minutes (Figure 9B). The discontinuity at 30 minutes occurs because (for doubling times less than 30 minutes) the newly initiated round of DNA replication is destined to conclude in the grand-daughter, rather than the daughter (as it was for inter-division time between 30-60 minutes). For inter-division times greater than $\tau_{cyc} = 60$ minutes, there are of course no further discontinuities because the cell self-initiates DNA replication destined to terminate in the same cell cycle. The apparent discontinuities in the age-of-initiation was the major conceptual challenge that Helmstetter and Cooper overcame with their simple two-timer model.

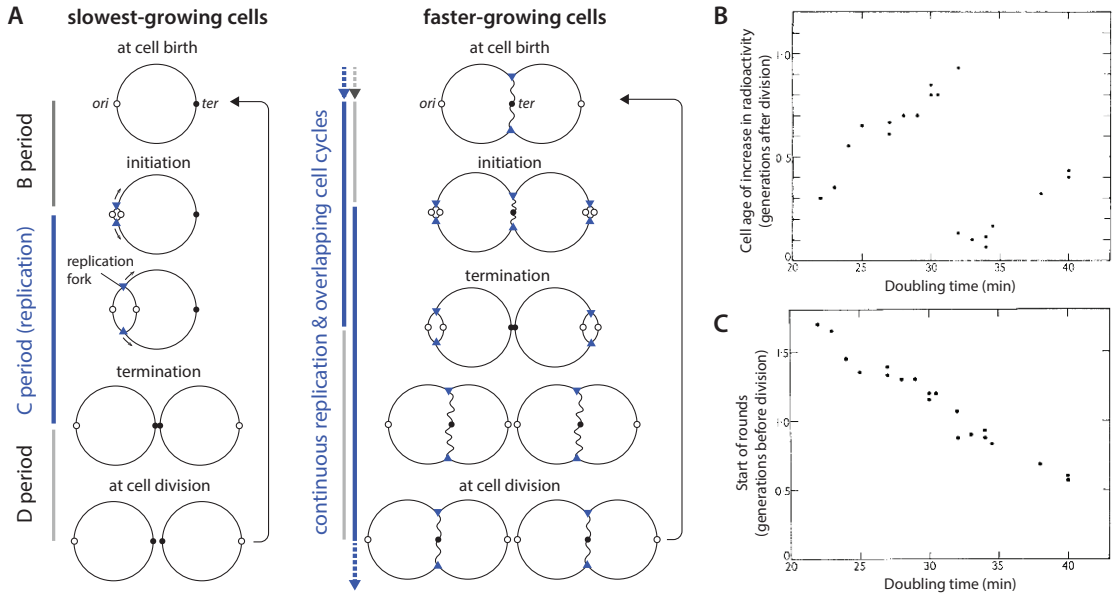

Figure 9: **Multiple rounds of DNA replication**. **A.** During slow growth (doubling time greater than 60 minutes, *upper*) there is only one round of DNA replication proceeding during the cell cycle. DNA replication is initiated at a point on the chromosome called the **origin** (filled circle), and replication proceeds simultaneously in both directions along each half of the chromosome. The site of new DNA synthesis is called the **replication fork** (grey triangle). DNA replication is terminated when the forks reach the **terminus** (octagon). During moderately rapid growth (doubling time 30-60 minutes, *lower*), there are two overlapping rounds of DNA replication (the lagging forks are initiated to terminate in the daughter). Notice that the number of origins is $2^{n_i}$, where $n_i$ is the number of overlapping rounds of replication ($2^0$ if the DNA is not being replicated); the number of forks is always twice the difference between the number of origins and the number of termini. **B.** Helmstetter & Cooper [111] observed abrupt changes in the DNA synthesis rate through the cell cycle, interpreted as initiation of new rounds of DNA replication. **C.** Given that full replication of the chromosome takes about 40 minutes under Helmstetter & Cooper's growth conditions [111], they could infer the number of generations prior to division that the newly initiated round was destined to conclude.

We can derive an explicit form for the discontinuous 'age -at-initiation' plot by removing the generation markers (mother, grandmother, great-grandmother, etc.) from an expression for the generations *after* division. Taking the negative of the previous expression,

$$\text{Generations \textit{after} division that DNA replication is initiated} = 1 - \frac{\tau_{cyc}}{\tau_d}.$$

Here, a negative generation corresponds to initiation in a previous generation. The empirical observations only record DNA replication rate changes in the present generation. To bring the age-of-initiation back into the range $(0, 1)$ we must remove the integer generations;

| If | Then |
|----|------|
| | DNA replication initiation is |
| $0 < \tau_{cyc}/\tau_d \le 1$ | self-initiated |
| $1 < \tau_{cyc}/\tau_d \le 2$ | initiated in the mother |
| $2 < \tau_{cyc}/\tau_d \le 3$ | initiated in the grandmother |

The upper-limit of the inequalities can be interpreted as the number of overlapping rounds of DNA replication at birth. For the age-of-initiation, we are only interested in the fractional part of $\tau_{cyc}/\tau_d$, and so we write compactly in several equivalent forms,

$$a_i = \begin{array}{l} \text{Cell age at which DNA} \\ \text{replication is initiated} \end{array} \qquad (7a)$$

$$= \begin{array}{l} \text{Number of overlapping} \\ \text{rounds of DNA replication} \end{array} - \frac{\tau_{cyc}}{\tau_d} \qquad (7b)$$

$$= \lceil \frac{\tau_{cyc}}{\tau_d} \rceil - \frac{\tau_{cyc}}{\tau_d} = 1 - \text{Frac} \left[ \frac{\tau_{cyc}}{\tau_d} \right] \qquad (7c)$$

where $\lceil \ \rceil$ is the ceiling function, and Frac [] takes only the fractional part of the argument. Given the canonical age distribution and the Helmstetter-Cooper model for DNA replication, it is straightforward to derive the average origins-per-cell, replication forks-per-cell, DNA-per-cell, and other associated quantities [70].

Bremer & Churchward [295] visualized how the average DNA replication constituents (origins, termini, and genome-equivalents) increase in time in a steady-state population. This intuitive and elegant method is reminiscent of that of Donachie [73], who showed that the nutrient growth law and the Helmstetter-Cooper model are consistent with the independence of 'initiation mass' for DNA replication on the growth rate (Section 2.2.5).

In balanced exponential growth, all components of the cell accumulate at the same rate. Looking at an aliquot of culture medium, the number of origins ($O$), the number of termini ($T$) and the number of cells ($N$) all increase exponentially at the same rate,

$$O(t) = O_0 2^{t/\tau_d}, \quad T(t) = T_0 2^{t/\tau_d}, \quad N(t) = N_0 2^{t/\tau_d},$$

where $\tau_d$ is the doubling time. The key insight of

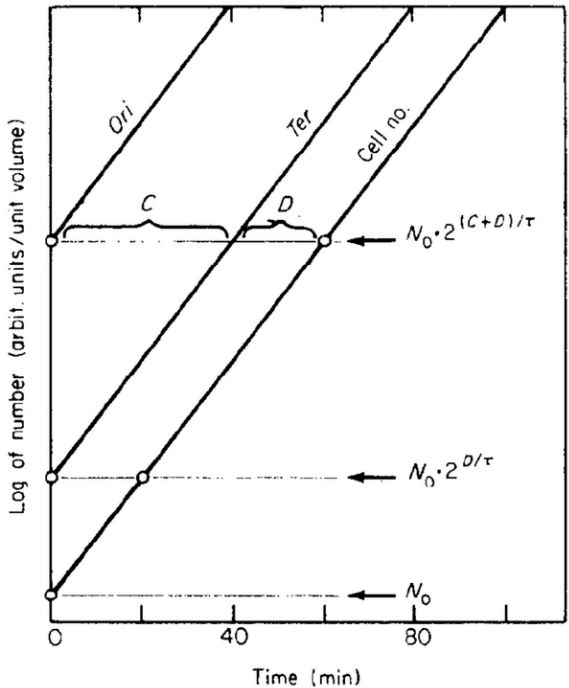

Figure 10: **Dependence of DNA replication on doubling time and cell cycle parameters**. Number of origins, termini, and cells in an aliquot of exponentially-growing cell culture. In balanced growth, the rate of accumulation of all three is given by the doubling rate $\mu = 1/\tau_d$. As a result, when drawn on a $\log_2$-linear plot against time, they appear as parallel lines. The spacing between the lines corresponds to the time it takes to convert from origin to terminus (**C**-period), and convert from terminus to cell division (**D**-period). Redrawn from [295].

Bremer and Churchward is to graph all three on a log-linear plot (Figure 10). From the causal ordering (origin turns into terminus turns into cell) and the definition of the **C**- and **D**-periods as the time for each conversion to occur, they arrive at an expression for the ratios ($ori$/cell; $\bar{O}$) and ($ter$/cell, $\bar{T}$). From the plot,

$$O_0 = N_0 2^{(\mathbf{C}+\mathbf{D})/\tau_d} \quad \text{and} \quad T_0 = N_0 2^{\mathbf{D}/\tau_d},$$

leading to,

$$\frac{ori}{\text{cell}} = \bar{O} = 2^{(\mathbf{C}+\mathbf{D})/\tau_d} \quad \text{and} \quad \frac{ter}{\text{cell}} = \bar{T} = 2^{\mathbf{D}/\tau_d}. \quad (8)$$

To relate these to the DNA content per cell, they first note that the number of forks per cell, $\bar{F}$, is twice the difference between the origins and the termini (see Figure 9B),

$$\bar{F} = 2(\bar{O} - \bar{T}).$$

The rate (per cell) of DNA synthesis in genome equivalents is then,

$$\frac{d\bar{G}}{dt} = \bar{F} \frac{1}{2\mathbf{C}} = \frac{1}{\mathbf{C}} \left[ 2^{(\mathbf{C}+\mathbf{D})/\tau_d} - 2^{\mathbf{D}/\tau_d} \right],$$

*i.e.* half-a-genome per **C**-minutes multiplied by the number of forks. In exponential growth at rate $\lambda$, the rate of DNA synthesis will likewise be exponential. When normalized to total cell number, the DNA per cell, $\bar{G}$, is given by,

$$\frac{\text{genomes}}{\text{cell}} = \bar{G} = \frac{1}{\lambda} \frac{d\bar{G}}{dt} \quad (9)$$

$$= \frac{1}{\mathbf{C}\lambda} \left[ 2^{(\mathbf{C}+\mathbf{D})/\tau_d} - 2^{\mathbf{D}/\tau_d} \right] = \frac{1}{\mathbf{C}\lambda} \left[ e^{(\mathbf{C}+\mathbf{D})\lambda} - e^{\mathbf{D}\lambda} \right],$$

which coincides with the population-averaged result of Cooper & Helmstetter (see, also, Eq. 21 in the next section).

### 2.2.5. *Donachie's insight on constant initiation mass.*

Shortly after Helmstetter & Cooper elucidated the timing of DNA replication, Donachie (and later Pritchard) noted that the steady-state mass-per-cell data of Schaechter, Maaløe and Kjeldgaard (Figure 6A, *blue*) implied that the mass-per-origin of DNA replication is constant [73, 296]. Donachie's argument below follows from noting that the average mass-per-cell and the average origins-per-cell (Eq. 8), both increase with doubling time $\tau_d$ (in minutes) as $2^{60/\tau_d} \approx 2^{(\mathbf{C}+\mathbf{D})/\tau_d}$ when $\tau_{\text{cyc}} = \mathbf{C} + \mathbf{D} \approx 60$ minutes.

From the age-of-initiation $a_i$, Eq. 7b, we can convert to absolute time-of-initiation by multiplying with the doubling time $\tau_d$,

$$\tau_i = \tau_d \times a_i = \tau_d \times n_i - (\mathbf{C} + \mathbf{D}),$$

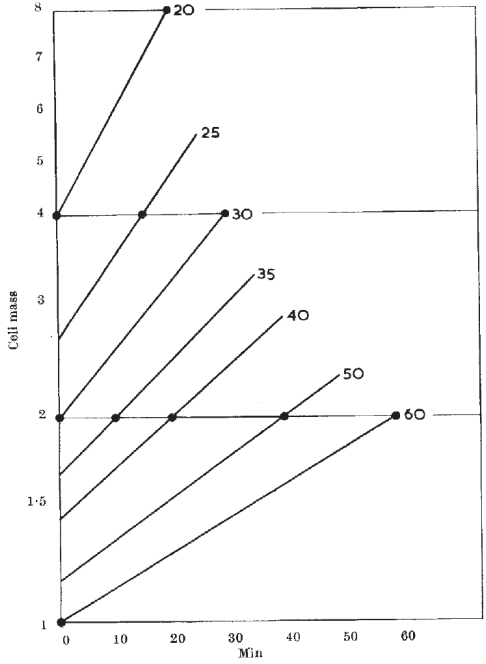

Figure 11: **Original graphics showing constant initiation mass by Donachie.** Increase in mass of individual cells with different rates of growth. The initial mass at time 0 is taken to be proportional to the average mass of cells growing at the same rate (taken from the data of Schaechter, Maaløe and Kjeldgaard). Given a constant time between DNA replication initiation and cell division ($\mathbf{C}+\mathbf{D} \approx 60$ minutes according to the data of Helmstetter & Cooper), it is possible to calculate the time when initiation occurs. These times are marked as solid circles. The masses at which initiations take place are the same or multiples of the same cell mass for cells growing at all growth rates. [73]

where $\tau_i$ is the time after division that a new round of DNA replication is initiated, and $n_i$ is the number of overlapping rounds of DNA replication. Donachie notes that the "average size of a randomly grown population of cells is proportional to the average size of the cells at the time of division." In particular, the birth-mass $M_b = \bar{M}/(2 \ln 2)$ where $\bar{M}$ is the average mass-per-cell in the exponentially-growing culture, and this relation is derived from the canonical age distribution. The initiation mass will be,

$$M_i = M_b 2^{\tau_i/\tau_d} = \frac{\bar{M}}{2 \ln 2} 2^{n_i} 2^{-(\mathbf{C}+\mathbf{D})/\tau_d}.$$

From Schaechter *et al.*(Figure 6A, *blue*), the average mass-per-cell is $\bar{M} = M_0 2^{60/\tau_d}$, so the initiation mass can be written

$$M_i = \frac{M_0}{2 \ln 2} 2^{n_i} 2^{[60-(\mathbf{C}+\mathbf{D})]/\tau_d}.$$

But $2^{n_i}$ is the number of origins at division (see

Figure 9A), which doubles upon initiation, so that

$$\frac{M_i}{\#\text{origins}} = \frac{M_0}{2 \ln 2} 2^{[60-(\mathbf{C}+\mathbf{D})]/\tau_d}. \tag{10}$$

Empirically $\mathbf{C} + \mathbf{D} \approx 60$ minutes in the experiments of Helmstetter & Cooper, and so follows Donachie's observation that mass-per-origin is approximately constant for rapidly-growing wildtype *E. coli* cells when growth rate is modulated by nutrient change. In his 1968 paper, Donachie devised a graphical method to combine the data from the papers by Schaechter *et al.*and Helmstetter & Cooper, showing the constancy of initiation mass (Figure 11). Apparently, a few months later Pritchard also reached the same conclusion [296, 297].

At first glance, it seems like an extraordinary coincidence that the growth-dependence in the mass-per-cell is precisely the same as the growth-dependence in the origins-per-cell. In fact it can be rationalized straightforwardly by the Schaechter *et al.*data for the steady-state and nutrient up-shift (Figure 6B): Upon shift-up, the cell (almost) immediately initiates a new round of replication, but the result of the up-shift is not seen in cell doubling time until $\tau_{cyc} = \mathbf{C} + \mathbf{D}$ minutes have elapsed (and the first post-shift round of replication & septation has terminated) [71]. The lag time between the post-shift increase in the rate of mass accumulation and the rate of cell doubling (*i.e.*, the time for a newly-initiated DNA fork to terminate and segregate) provides the steady-state growth-rate dependence in the mass-per-cell; the growth-dependence in the origins-per-cell follows directly from the definition of $\mathbf{C}$ and $\mathbf{D}$ (see Eq. (7*b*)).

The idea of constant initiation mass by Donachie, Pritchard and colleagues [73, 296] have been repeatedly challenged for almost half a century [185, 197, 200, 201, 298, 299]. In Section 6, we shall explain how this issue has been finally and conclusively settled in 2017 by extensive experimental data. We will also review another derivation of the constant initiation mass that is much simpler than that of Donachie and Pritchard.

### 2.2.6. Derivation of cell cycle parameters for exponentially growing population. In this sub-section, we present several useful results of cell cycle parameters.

#### (i) **Age distribution and genomic loci copy numbers in an exponentially growing population.** As mentioned in the Section 2.2.1, the age distribution of an exponentially growing population at steady-state is

$$\rho(a) = \frac{2 \ln 2}{\tau_d} \cdot 2^{-a/\tau_d} \tag{11}$$

Using the age distribution, it is straightforward to compute the average copy number of an arbitrary genomic locus $X_g$ per cell, and the result is

$$\langle X_g \rangle = 2^{\frac{\tau_{cyc} - g\mathbf{C}}{\tau_d}},\qquad(12)$$

where $g$ denotes the fractional distance from *ori* between $g_{ori} = 0$ and $g_{ter} = 1$ (here and elsewhere, we always assume that the replication fork moves at a constant speed on both chromosome arms). The average number of *ori* is given by $g = 0$,

$$\#ori = 2^{\tau_{cyc}/\tau_d}.\qquad(13)$$

(ii) **Replication period for a steady-state population.** Based on Eq. 12, the *ori* to *ter* ratio is given by

$$\frac{\langle O \rangle}{\langle T \rangle} = 2^{\frac{\mathbf{C}}{\tau_d}}\qquad(14)$$

This ratio can be directly measured experimentally using quantitative PCR (qPCR; [300]), flow cytometry (Section 2.1.6), or image cytometry (see Box 3; [301]). More generally, the copy-number ratio between two genomic loci is given by

$$\frac{\langle \text{locus 1} \rangle}{\langle \text{locus 2} \rangle} = 2^{\frac{\Delta g \cdot \mathbf{C}}{\tau_d}},\qquad(15)$$

with the duration of the $\mathbf{C}$ period is

$$\mathbf{C} = \frac{\tau_d}{\Delta g} \log_2 \frac{\langle \text{loci 1} \rangle}{\langle \text{loci 2} \rangle}.\qquad(16)$$

---

**Box 3 – Methods in measuring cell cycle parameters of a population**

**qPCR**. The $\mathbf{C}$ period of the population can be estimated by marker frequency analysis using qPCR (quantitative polymerase chain reaction). qPCR uses non-specific fluorescent dyes to intercalate with the genomic DNA extracted from a cell sample. The ratio of relative copy numbers of two loci gives the ratio of $\mathbf{C}$ period over generation time $\mathbf{C}/\tau_d$ as $\langle ori \rangle / \langle ter \rangle = 2\mathbf{C}/\tau_d$, where $\tau_d$ is the doubling time of the population. **Image cytometry.** Similar to flow cytometry, image cytometry is used to acquire both DNA content and cell morphological information via microscopy. Standard cells with known copy number of DNA are used to calibrate the DNA content in samples cells. The standard cells are grown under slow-growth conditions (non-overlapping cell cycle) and run out by using rifampicin and cephalexin. Standard cells are stained with a specific color and mixed with sample cells, and both populations are then stained

---

by DNA dye. Phase contrast and two-color fluorescent images are captured. By using $\mathbf{C}$ period obtained through qPCR, $\tau_{cyc}$ can subsequently be calculated (see Section 2.2.6).

---

(iii) **Cell cycle duration, $\tau_{cyc}$, for a steady-state population.** From the DNA content measurement, we obtained the average genome equivalent per cell. The genome content per cell can be computed by integrating the copy number of each locus over the entire chromosome:

$$\bar{L} = 2 \int_0^L n(x)\,\mathrm{d}x,\qquad(17)$$

where $n(x) = 2^{\frac{\tau_{cyc} - g\mathbf{C}}{\tau_d}}$ is the average number of locus $x$ with a chromosomal coordinate $g = x/L$ away from the *ori*, and $L$ is the total length of one arm of the chromosome (half of the chromosomal length). We thus have

$$\bar{L} = 2 \int_0^L 2^{\frac{\tau_{cyc} - g\mathbf{C}}{\tau_d}}\,\mathrm{d}x\qquad(18)$$

$$= 2 \int_0^L 2^{\frac{\tau_{cyc} - \frac{x}{L}\mathbf{C}}{\tau_d}}\,\mathrm{d}x\qquad(19)$$

$$= 2L \frac{\tau_d}{\mathbf{C}\log 2} \left( 2^{\frac{\tau_{cyc}}{\tau_d}} - 2^{\frac{\tau_{cyc} - \mathbf{C}}{\tau_d}} \right).\qquad(20)$$

Therefore, the genome content per cell is given by $\bar{L}/2L$, or

$$\bar{G} = \frac{\tau_d}{\mathbf{C}\log 2} \left( 2^{\frac{\tau_{cyc}}{\tau_d}} - 2^{\frac{\tau_{cyc} - \mathbf{C}}{\tau_d}} \right),\qquad(21)$$

which is identical to Eq. 9 (also see [72]).

Eq. 21 is the basis of experimental measurement of the $\mathbf{D}$ period. By substituting the $\mathbf{C}$ period measured from qPCR into Eq. 21, the $\mathbf{D}$ period (= $\tau_{cyc}$ - $\mathbf{C}$; time elapsed between replication termination and cell division) is given by

$$\mathbf{D} = \tau_d \log_2 \left[ \frac{\mathbf{C} \cdot \bar{G} \cdot \ln 2}{\tau_d \left( 2^{\frac{\mathbf{C}}{\tau_d} - 1} \right)} \right]\qquad(22)$$

(iv) **Cell cycle period calculated from the 'run-out' population.** Alternatively, we can calculate the cell cycle time $\tau_{cyc}$ by performing a population run-out experiment (see Section 2.1.6). After run-out, the cell population separates into two groups: cells with ages before and after the initiation time $t^*$. Thus the latter group will eventually have two times the genome equivalent of the former group. By measuring the probability distribution of cells belonging to the two groups $C_1$ and $C_2$, respectively, we can calculate the initiation time $t^*$. The time fraction of pre-initiation corresponds to the population fraction of $C_1$. Therefore, through the age probability density function $\rho(a)$ in Eq. 11, we have

$$\int_0^{t^*} \rho(a)\, da = \frac{2\ln 2}{\tau_d} \int_0^{t^*} 2^{-\frac{a}{\tau_d}}\, da \qquad (23)$$

$$= \frac{C_1}{C_1 + C_2} \qquad (24)$$

Thus,

$$t^* = -\tau_d \log_2\left[\frac{-C_1}{2(C_1 + C_2)} + 1\right] \qquad (25)$$

We can then calculate the cell cycle duration as

$$\tau_{cyc} = (\tau_d - t^*) + (n_{oc} - 1)\tau_d \qquad (26)$$

$$= n_{oc}\tau_d - t^*, \qquad (27)$$

where

$$n_{oc} = \left\lceil \frac{\tau_{cyc}}{\tau_d} \right\rceil \qquad (28)$$

and $\lceil\ \rceil$ denotes the ceiling function. Again, by substituting the **C** period from qPCR measurement, we can calculate the **D** period as

$$\mathbf{D} = n_{oc}\tau_d - t^* - \mathbf{C} \qquad (29)$$

**(v) Other cell cycle parameters.** We already calculated the copy number of any genomic locus in a steady-state population (Eq. 12). For example, we obtained $\langle O \rangle = 2^{\tau_{cyc}/\tau_d}$ and $\langle T \rangle = 2^{\mathbf{D}/\tau_d}$. We can also calculate the replication termination time as:

$$t_{ter} = \tau_d - \mathbf{D} \qquad (30)$$

From the initiation time and termination time $t_{ter}$, it is straightforward to derive the number of replication forks [63, 302]

$$\frac{\langle \text{forks} \rangle}{2} = 2^{\frac{\tau_{cyc}}{\tau_d}} - 2^{\frac{\mathbf{D}}{\tau_d}}. \qquad (31)$$

### 2.2.7. *Replication initiation control in* E. coli.

**Molecular basis of replication initiation.** With the emergence of molecular biology in the 1970s, extensive studies have identified key proteins involved in the DNA replication initiation process (*e.g.*, DnaA) and characterized their biochemical properties (*e.g.*, ATP hydrolysis of DnaA-ATP) (Figure 12). DnaA is the major player in initiation control in bacteria, and it is a widely conserved protein across species [303–308]. Upon initiation, DnaA is believed to polymerize at the origin region *ori*, change the topology of the local DNA structure, and unwind the double-strand of DNA [309–312]. It has been known that the ATP form of DnaA has higher affinity for *ori* compared to that of its ADP form [313–319]. The process of hydrolysis from DnaA-ATP to DnaA-ADP is facilitated by regulatory proteins such as Hda via RIDA (RIDA: regulatory inactivation of DnaA; the reverse process from ADP from to ATP is promoted by a specific DNA sequence DARS) [316, 320–325]. The DnaA binding sites on DNA, called 'DnaA boxes', are distributed not only at *ori* but all over the chromosome including high-affinity *datA* sites [326–329]. Both positive (*e.g.*, DiaA) and negative regulators (*e.g.*, SeqA) collaborate on the homeostasis of initiation [330–343]. The details of the molecular components involved in replication initiation can be found in several excellent reviews including [344–346].

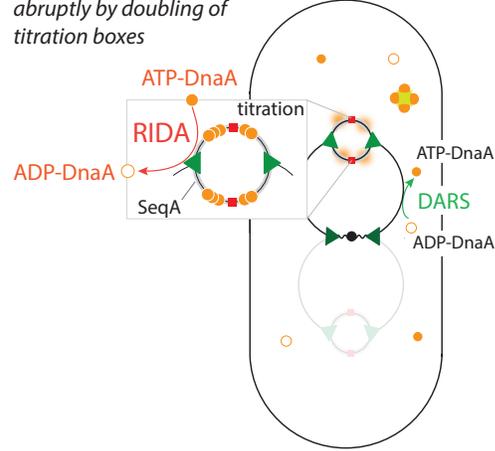

**Figure 12: A simplified illustration of the molecular mechanisms of replication initiation in *E. coli*.** This cartoon shows a cell with two overlapping cell cycles, where the triangles represent the replication forks, and the red squares represent *ori*'s on chromosome. The sites of DnaA titration boxes are not drawn.

**Modeling the constancy of initiation mass.** In 1963, Jacob, Brenner, and Cuzin proposed the influential 'replicon model' for replication initiation. The model assumes two components, a cis-acting 'replicator' and an 'initiator' that binds to the replicator (DNA sequence) [347]. Soon after the replicon model, an 'initiator threshold' model was proposed and elaborated in the 1960s and 70s, becoming the standard model in chromosome replication.

One of the basic assumptions underlying the initiator threshold model is that initiator proteins are constantly synthesized at the same rate as cellular growth, so that they occupy a constant fraction of the total proteins throughout the division cycle. Another key assumption is that soon after replication initiation, the potential for new initiation drops sharply. Perhaps the most important assumption is that replication initiates when a fixed number of initiators accumulate per origin. These assumptions are consistent with those in the historial 'structural model' proposed for eukaryotes [348–350].

Helmstetter and Cooper incorporated the initiator threshold assumption into their cell cycle model and successfully explained the timing of replication cycle upon nutrient shift-up (see Section 2.1.1) [33]. Similar ideas were also embraced by Donachie and Pritchard in the late 1960s soon after the Helmstetter-Cooper model [73, 296].

Models proposed afterwards are still based on the initiator threshold idea [56, 296, 351–353]. However, the biological assumptions underlying these models differ in subtle but important ways, leading to different molecular mechanisms and the degree of robustness as explained below (Figure 13).

#### (i) Inhibitor titration model (Pritchard, 1969) [296]

This earlier model assumes 'inhibitor' proteins that inhibit initiation when they are present above a threshold level. They hypothesized that inhibitors are expressed as a transient burst, leading to repression of initiation or re-initiation. The inhibitor concentration decreases solely via dilution through growth, and at some point during growth crosses the threshold level and replication initiates. The initiator concentration is constant as mentioned above. A major caveat of this model is that it requires fine-tuning of inhibitor expression, because the amount of inhibitors expressed in the burst will determine the initiation timing in the next generation, which does not agree with data obtained afterwards. For example, Margalit and Grover combined evidence from experiments and simulations to refute this model. They found that this model cannot reproduce the generation time distributions obtained from experimental data [354–356].

#### (ii) Autorepressor model (Sompayac and Maaløe, 1973) [56]

The main motivation of Sompayrac and Maaløe was to develop a model that can lead to constancy of initiator proteins independent of growth rate. They proposed a simple model based on auto-repression. Specifically, autorepressors and initiators are under the control of a same promoter, so that their ratio is constant during growth. Because the concentration of initiator is held constant by auto-repression (thus the total number increases in proportion to the cell size), initiators accumulate at each origin at the same rate as the growth rate in the absence of inhibitor proteins. Initiation is triggered when a fixed number of initiators have accumulated per origin. Sompayrac and Maaløe explicitly state that a consequence of the model is that a con-stant cell volume is added per origin since the previous initiation, corresponding to the number of initiators accumulated in each division cycle between two consecutive initiations. This prediction is compatible with the adder principle that we will discuss in Section 4.

#### (iii) Initiator titration model (Hansen, 1991) [351].

In contrast to the previous models, where the regulatory effector molecules were hypothetical, this model was inspired by the identification of key molecular players in DNA replication initiation. The *datA* boxes are located near *ori* and other loci in the chromosome. Once replication initiates and the *datA* boxes near the origin are duplicated, the level of DnaA at the origin drops by the additional *datA* boxes, preventing re-initiation [328, 351]. During steady-state growth, the frequency of accumulation and titration of DnaA at the origin is periodic at the same frequency as initiation.

### 2.2.8. Limitations of the initiation control models until the 1990s.

As more detailed genetic, molecular, and biochemical information about DnaA became available in the 1980s and 1990s [205, 334, 358–372], limitations of the initiation control models proposed earlier also became apparent [373–379]. Donachie and Blakely critically evaluated the extant data and noticed the limitations of the previous models described above. First, *de novo* protein synthesis and growth is required for initiation. This would mean that either initiators are made between each initiation and then inactivated after initiation [73], or an inhibitor is made after each initiation and diluted by growth (inhibitor titration model). Second, even when the intracellular concentration of DnaA is increased by as much as fivefold, initiation of chromosome replication occurs only slightly earlier [358, 380]. By contrast, a reduction in DnaA concentration causes an increase in the initiation mass in proportion [252, 328]. Third, the ratio between the number of DnaA molecules and the number of copies of *ori* is not itself a determinant of the time of initiation, because initiation takes place synchronously even when extra copies are present on plasmids or chromosome [381–383].

Donachie and Blakely searched for a proxy that changes with respect to initiation [352]. They specifically considered DnaA-ATP and DnaA-ADP and their interconversions, and how their copy numbers per cell change during the cell cycle based on the extant knowledge about DnaA. They found the ratio of the number of DnaA-ATP to DnaA-ADP per cell changes non-monotonically and peaks between birth and division. This observation led to a new hypothesis that replication initiates at the peak of the ratio (Figure 15). Here, the key underlying idea is that both DnaA-ATP and DnaA-ADP compete for the origin, but only DnaA-ATP is competent for initiation. In other words, DnaA-ADP is effectively an initiation inhibitor, thus the note by Donachie and Blakely, "[i]t is salutary to note that the model incorporates both the accumulation of an activator (DnaA-ATP) and the production and dilution of a competitive inhibitor (DnaA-ADP) during cell growth, the main features of which were once seen as opposing models."

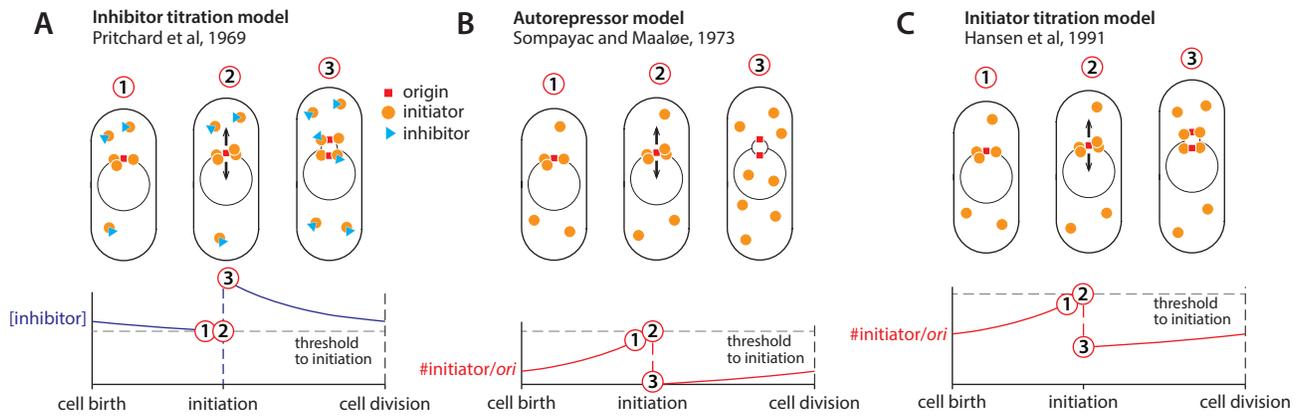

Figure 13: **Models of replication initiation control.** Graphs qualitatively show the ideas of (A) inhibitor titration model, (B) autorepressor model and (C) initiator titration model. Note that in all three models, the concentration of initiator is assumed to be constant throughout division cycle.

Although the model of Donachie and Blakely has many compelling features, it remains hypothetical; to the best of our knowledge, no one has attempted to confirm or refute the model directly.

### 2.3. 1970s - 1990s: the age of molecular and cell biology

The golden era of bacterial physiology between 1940s and 1970s coincided with a golden era of molecular and cell biology, but from the 1970s onward, fundamental research on bacterial physiology fell out of the mainstream and entered a 'dark age.' On the other hand, molecular and cell biology was in its 'exponential growth phase.' Part of the reason was certainly the shift of focus toward molecular mechanisms of gene regulation, and the maturation of tools including PCR, gel electrophoresis, blotting methods and molecular cloning. As a result, thousands of genes, proteins and individual pathways were discovered. It was this information explosion of molecular mechanisms that allowed the resurgence of the study of bacterial physiology in the last decade. In this subsection, we will briefly go through those advancements in molecular and cell biology during 1970s - 1990s, which are relevant to the bacterial physiology and cell-size control.

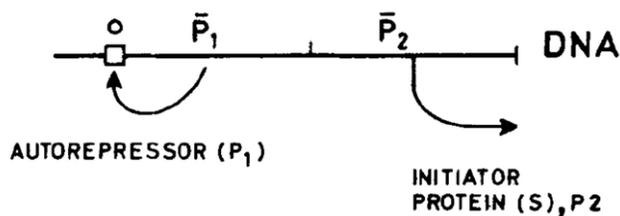

Figure 14: Model operon of the autorepressor model by Sompayrac and Maaløe [56]. Both the autorepressor ($P_1$) and the initiator ($P_2$) are under the control of the same promoter, so that the copy-number homeostasis of the initiator proteins is ensured by autorepression by $P_1$. Autorepression of an initiator protein DnaA has been shown experimentally by Andrew Wright's group [357].

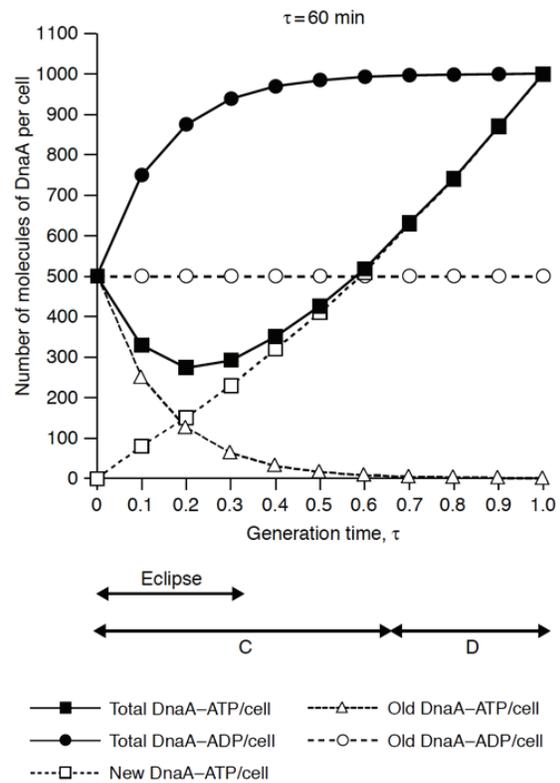

Figure 15: **Models of DnaA-ATP to DnaA-ADP ratio controlling initiation.** This figure is adapted from [352]

### 2.3.1. Does E. coli elongate exponentially?

A classic question in bacterial cell biology is whether *E. coli* elongates linearly or exponentially. This debate is considered settled by modern single-cell data [9, 384–386], and there is little doubt that an exponential-fit is a good description for elongation of *E. coli*. Nevertheless, it is useful to revisit this question for its historical importance.

In the literature, researchers often attribute the difficulty of settling linear vs. exponential growth to their indistinguishability. Their maximum difference is about 5%, which was indeed considered too small in the early days. This technical problem has largely been resolved by modern experimental methods including video microscopy, microfluidics, and high-throughput image analysis. Even measurement of buoyant mass of individual live cells is currently possible using a micro-cantilever [387].

In the 1960s, Kubitschek performed important pioneering experiments to understand the kinetics of global biosynthesis. First, he measured the growth rate of a synchronized cell culture of *E. coli* using three strains and found that the growth rate (measured in terms of cell counts) is linear with respect to time [388]. Next, he conducted a series of measurements of the uptake rate of nutrient radio-labeled compounds. He found that

> [u]ptake rates were constant during more than the first two-thirds of the cycle, or reasonably so, for all of these compounds: glycine, leucine, glucose, acetate, phosphate, sulfate, and thymidine. [389]

These two papers were published in 1968 in *Biophysical Journal* and formed the basis of Kubitschek's proposal that individual *E. coli* cells elongate linearly.

Since the beginning of the 1970s, he fought hard to defend his linear model. [His debate with Cooper has been well-documented, and interested readers may wish to read Cooper's book and references therein [172, 390].] His experiments consistently led him to conclude that the number of nutrient uptake sites in *E. coli* remains constant in each generation and doubles near cell division [226, 391, 392]. Since he hypothesized that cellular elongation rate is proportional to the number of uptake sites, a constant number of uptake sites would mean constant elongation rate. Over the years, he refined his linear elongation model by adopting Donachie's notion and idea of a 'growth zone' (and 'unit cell'; see below), which fit his view [393]. The work by Sargent also fell into this category [393–395].

Throughout the 1980s, Kubitschek maintained his view. Sometimes he labeled linear growth model with 'bilinear' growth, but it was essentially the same in that the slope of the 2nd linear regime should reflect the increase in the uptake rate [396, 397].

It is still unclear why his radio-labeling experiments always produced data that was consistent with his linear model, other than that the measurements relied on indirect methods. The story of Kubitschek is a misfortune, especially because he as a physicist was one of the pioneers who used the Coulter counter to measure the cell size distribution. His other contributions to bacterial physiology are also notable [166, 169, 171, 393, 398–404].

### 2.3.2. 'Unit cell' by Donachie.

An interesting development in this period, inspired by Kubitschek's linear growth model was the 'unit cell' model by Donachie and colleagues [78, 405, 406]. They proposed that *E. coli* grows at one of the cell poles before reaching a 'unit cell' size, and at both poles when the cell is larger than the unit cell (Figure 16).

> *A Unit Cell Model of Bacterial Growth:* We may summarize our observations on the growth of cells of E. coli as follows. (1) The growth of cells is always unidirectional if they are less than a certain critical length (about 3.4 μm) and always bidirectional if they are more than

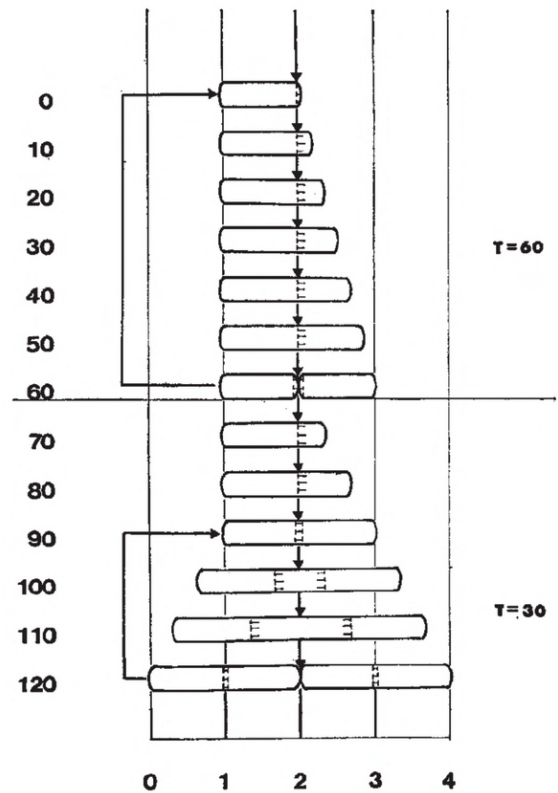

**Figure 16: Donachie and Begg's model of growth and 'unit cells.'** The shaded area is the growth zone, defining one unit cell. Adapted from [405] with permission.

*this length. (2) When growth is unidirectional, extension always takes place from the pole of the cell that was formed in the proceeding cell division... To provide a conceptual framework for this set of observations, we have developed a simple model of cell growth. According to this model, cells of* E. coli *have a minimum length of about 1.7 µm. We shall refer to a cell of this length as a 'unit cell.'* ...

While Donachie and colleague were motivated by erroneous experimental data, the concept of a fundamental unit of cell size is relevant in understanding cell size control. We will come back to this issue in Section 6.

*2.3.3. Cell division and FtsZ.* Cell division, as the concluding event in the cell cycle, received almost as much attention as DNA replication initiation [74, 77, 407–409]. FtsZ is a tubulin analog which plays a central role in the division control which was first discovered by Joe Lutkenhaus in the Donachie lab [410]. It was found to form a ring-like structure at the division site, called the 'FtsZ-ring' or 'Z-ring' [411]. It is a highly conserved proteins across species of bacteria, and as an essential protein, FtsZ mutants are unable to initiate cell division [412]. The FtsZ monomer binds and hydrolyzes GTP to conduct conformational change and self-assembly into protofilaments in cell (see good reviews in [413–415]). FtsZ tethers to the inner membrane at the division site together with more than ten other proteins, and is known to be negatively regulated by MinC, SulA and others in *E. coli*. Moreover, Z-ring is thought to control cell division via crosstalks to DNA replication, chromosome segregation, cell size and growth rate (see the Section 2.3.4 below) [413, 416–423]. Recently, FtsZ was also found to be crucial in coordinating the cell wall synthesis at the division site in both *E. coli* and *B. subtilis* [424, 425].

*2.3.4. Cell division control: Nucleoid occlusion, Min system, metabolic sensors and more.* Unlike the eukaryotic cell cycle whose regulation is highly reliant upon checkpoints - regulatory logic gates with restrictive conditions that must be satisfied in order for the cell cycle to pass through - the cell cycle in many bacterial species like *E. coli* does not have well-defined checkpoints. How cell cycle events are coordinated with one another continues to be a fascinating and largely open question [79, 426–433]. In the previous Section 2.2.7, we reviewed how replication initiation is regulated by direct molecular mechanisms and coordinated with growth rate (global biosynthesis rate). As mentioned in Section 2.3.3, cell division was thought to be determined by the preceding steps in cell cycle, which may function as 'checkpoints.' One prevailing idea is that cell division can take place only if the sister chromosomes have segregated and are no longer occluding the middle region along the cell's long axis (or midcell). This is known as the 'nucleoid occlusion model' [434, 435]. Although this mechanism is more about physical effects, some molecular evidence has been identified which associates cell division with DNA replication. For example, SlmA is a nucleoid-associated protein which disrupts Z-ring assembly if the chromosomes are still near the midcell division site [436].

Another well-studied regulatory system is MinCDE system in *E. coli* [416, 437]. In this three-component system, MinC is a FtsZ assembly inhibitor which associates with MinD on the inner membrane, and the MinC-MinD complex is dissociated by a cytoplasmic protein MinE. Altogether, the MinC-MinD complex performs a pole-to-pole oscillation with a defined wavelength and period, which helps place the Z-ring at the midcell site and ensures the precision in septum positioning [438, 439]. The pattern formation of Min proteins has been extensively studied *in vivo* and *in vitro*, and the mathematical modeling of Min oscillation is another brilliant example in quantitative bacterial physiology. Here we list several more papers on Min system for readers who are particularly interested in this topic [440–447]. In addition to nucleoid occlusion and the oscillatory Min system, SulA, an inhibitor to FtsZ polymerization when DNA damage occurs (SOS reponse) in *E. coli*, serves as a coordinator of cell division with DNA replication [438]. Recently, 'metabolic sensors' (OpgH in *E. coli* and UgtP in *B. subtilis*) were reported to link cell division control to cellular metabolism [448–451]. These metabolic sensors are nutrient-sensitive and they negatively regulate Z-ring assembly, further inhibiting cell division [448, 452–455]. Not limited to these, more regulatory systems have been discovered recently, and it seems more complicated than thought to reach a simple conclusion on how cell division is coordinated [456–458].

*2.3.5. Cell envelope: Cell-wall and Fatty acid synthesis.* The gram-negative bacterium, such as *E. coli*, is enclosed by a cell envelope which is composed of inner and outer cell membranes separated by the periplasmic space and a layer of cell wall in between. How exactly the synthesis of cell envelope is coordinated with growth and cell cycle still remains elusive [459–467]. The *E. coli* cell wall is one stress-resistant layer of peptidoglycan (or murein). The peptidoglycan layer is a covalently bonded network of long, rigid glycan strands cross-linked by relatively short and flexible peptide bridges. It is a strong but elastic network that provides mechanical strength to counteract internal turgor pressure and prevent cell lysis [69, 172]. With the cooperation of cell wall synthesis machinery (composed by many proteins), *E. coli* grows by inserting peptidyglycan into distributed sites along the lateral cell wall [468], which is in line with the exponential elongation of the rod shape (see Section 2.3.1). During cell division, the cell wall synthesis machinery also localizes at the midcell site and further closes up the septum by invagination [469–472]. Recent findings indicate that this septal growth of cell wall is directed by FtsZ [424, 425]. A model based on surface-to-volume ratio suggests that excess cell wall precursors may serve as a thresholding factor for division to take place [473].

The inner and outer membranes in *E. coli* consist of phospholipids and membrane proteins, and they are direct products of fatty acids metabolism and the synthesis of transport proteins [474, 475]. These membranes are critical in transport and osmotic homeostasis, but their roles in cell growth, cell cycle and cell size control are still poorly understood even though various models have been proposed [315, 474, 476–482]. It has been demonstrated that fatty acid synthesis is closely associated with amino-acid starvation (stringent response), and further regulates cell size [483]. Also recent results demonstrate that fatty acid flux sets cell envelope capacity, which in turn dictates cell size [484].

*2.3.6. Cell shape control and MreB.* Cell shape control, or morphogenesis, in rod-like bacteria species such as *E. coli* and *B. subtilis* has been a long-standing question in bacterial physiology, along with cell-size control [1, 51, 52, 485–495]. In contrast to cell-size control where total cell volume varies in different nutrient conditions, the question of cell-shape control can be simplified to how the aspect ratio (cell length versus cell diameter) is maintained or varied under different growth conditions [496]. The maintenance of rod-like shape has been found to be closely related to cell wall synthesis. For example, MrdB and PBP2 are proteins involved in the peptidyglycan synthesis, and their mutants were known to have spherical cell shape [497]. The shape of thymine-deficient cells were shown, though not convincingly, to be rounder or more irregular when thymine supply became limited [221, 223, 231, 498]. This is possibly due to the role of thymine in metabolism related to peptidyglycan synthesis. MreB is another important cytoskeleton protein, whose deletion leads to transformation of cell shape to spherical [499, 500]. MreB, beneath inner membrane, co-localizes with MrdB, PBP2 and other cell wall synthesis machinery parts are recently shown to direct the insertion of newly synthesis peptidyglycan while undergoing circumferential movements [501–503]. Interestingly, recent evidence has demonstrated that MreB-regulated shape is actually decoupled from growth rate and cell size control [504].

*2.3.7. Stringent response and cell-size control.* In nutrient-poor conditions, the exponential growth rate is reduced, with a commensurable decrease in ribosome abundance. The coordination among nutrient status, ribosome biogenesis and the growth rate is carried out by the stringent response [505–508]. A major player in the stringent response is (p)ppGpp (guanosine pentaphosphate or tetraphosphate), a phosphorylated derivative of GTP. The cytoplasmic concentration of (p)ppGpp increases when the cell is starved for amino acids or energy. The regulatory molecule (p)ppGpp acts on a suite of metabolic proteins, though its most direct function is to inhibit RNA synthesis (primarily ribosome biogenesis by inhibiting rRNA transcription), which in turn slows down growth [509–521]. In addition, (p)ppGpp was found to regulate multiple cell cycle events including replication and cell division [522, 523]. For replication initiation in *E. coli*, the (p)ppGpp concentration correlates negatively with the transcription of *dnaA* though the mechanism remains unclear [366, 511]. For cell division, in addition to the metabolic sensors discussed in Section 2.3.4, (p)ppGpp appears to serve as an additional mediator between chromosome segregation and the cell division machinery (*e.g.*, FtsZ) [524, 525]. Furthermore, the stringent response is thought to directly regulate cell size via fatty acid synthesis [484], as discussed in Section 2.3.5. Despite these many interactions, a complete picture of how the stringent response is connected to cell-size control under general growth conditions is still lacking.

*2.3.8. Further reading.* Due to space limitations, this review focuses on quantitative approaches to bacterial physiology. For those interested in more molecular aspect of the related issues, we refer the reader to excellent review articles written by important contributors. A very partial list is as follows.

- Cell size control: Petra A. Levin [449, 526]
- Replication initiation: James Berger, Tsutomu Katayama, Anders Løbner-Olesen, Kirsten Skarstad [344–346, 527, 528]
- Cell wall synthesis and cell morphology: Kerwyn Casey Huang, Waldemar Vollmer and Kevin Young [529–533]
- Cell division machinery: Piet de Boer, Jeff Errington, Harold Erickson, Elizabeth Harry, Joe Lutkenhaus and William Margolin [413–415, 435, 534–538]
- Stringent response and rRNA transcription control: Richard Gourse [516, 519]

- Chromosome organization: Suckjoon Jun & Andrew Wright, Nancy Kleckner and David Rudner [539–541]

- Bacterial cytoskeleton: Ethan Garner and Zemer Gitai [542–544]

- Stress response: Carol Gross [545, 546]

- Single-molecule approach to protein synthesis: Jonathan Weissman [547]

- Evolutionary aspect: Richard Lenski [548, 549]

- Circadian clock in bacteria: Susan Golden [550, 551]

- Bacterial membrane: Natividad Ruiz [552]

- Biophysics of cellular dynamics: Joshua Shaevitz, Julie Theriot and Martin Howard [553–557]

- Novel techniques for bacteria study: Grant Jensen, Cees Dekker and Paul Wiggins [558–561]

- Cell cycle control and cellular dynamics in *C. crescentus*: Christine Jacobs-Wagner and Lucy Shapiro [292–294, 562, 563]

### 2.4. 2000s - present: Back to the origin

*2.4.1. Issues in gene expression.* With the development of fluorescent protein reporters by Roger Tsien and co-workers in the 1990's, the early 2000's saw a resurgence of interest in phenotypic variability among members of an isogenic population. In contrast to the earlier work focused on variability in cell morphology (accessible through bright-field microscopy), the new-wave of variability studies could monitor fluctuations in protein expression both among cells and through time [24–30, 564].

One of the key ideas to come out of that period was the recognition that fluctuations in protein expression had a much wider variance than simple Poisson statistics, attributable to the amplification of small fluctuations in transcription by dozens of ribosomes working in series to translate mRNA into protein [26, 565]. Furthermore, by adjusting the transcription and translation rate, it is possible to independently tune the mean and variance of a given protein. There continues to be considerable interest in determining mechanisms by which cells suppress or exploit intrinsic fluctuations in gene expression.

An attractive hypothesis is that constraints imposed by cell physiology and growth lay down the deterministic landscape of available phenotypes in a given environment and fluctuations in gene expression then facilitate transitions among these possibilities. How physiology constrains gene expression fluctuations, however, remains a largely unexplored question.

*2.4.2. Resurgence and reassessment of the growth law and the cell cycle model.* Microbial physiology fell out of fashion as the new era of molecular biology started in the 1970s (Figure 3). The upside of the tremendous advances in molecular biology was that mechanism of biological processes finally came with an explosion of new molecular insights occurring across the entire field of biology. The downside was that biology became overspecialized and its style of research drifted far and fast away from the physical sciences.

The tide has turned back again in the past few years, and we are currently witnessing strong resurgence of interest in quantitative microbial physiology. The scope of research in microbial physiology these days spans from single-cell to populations, and from molecules to the whole cell, with a completely new set of available technologies coming from both biology and physics. At the single-cell level, we now have a fairly satisfactory phenomenological description of how individual cells maintain size homeostasis (Section 4). At the population level, a coarse-grained proteome picture has good predictive power for allocation of cellular resources under different growth conditions (Section 5). Furthermore, we now understand that the nutrient growth law by Schaechter, Maaløe, and Kjeldgaard and the Helmstetter-Cooper model are in fact a special case of a more general coordination principle between replication initiation, replication-division cycle, and the global biosynthesis rate (Section 6).

Before getting there, we will first start by introducing in the next Section the old concept of 'variability' that disappeared during the 1970s but returned in the 1980s. The concept of variability and single-cell physiology will naturally lead us to modern microbial physiology.

## 3. Variability and single-cell physiology

While the study of individual cells is often regarded as an invention of modern systems biology inspired by stochastic gene expression and cell-to-cell variability [24–30] understanding single-cell behavior has been the ultimate goal of bacterial physiology since the birth of the field [57, 63, 69, 172].

As mentioned earlier in Section 1, cell-to-cell variability was apparent to early bacterial physiologists because the measured steady-state age and cell size distributions deviated from the predictions of the deterministic model more significantly than experimental uncertainties (Figure 2E). For these reasons, the deterministic version was specifically called 'canonical' such as 'canonical age distribution' or 'canonical size distribution.'

Physiological control, particularly control of growth and division, can be formally described using stochastic variables and the language of non-equilibrium statistical mechanics as pioneered in the 1950s and 1960s [14, 95, 168, 566–573]. In this approach, different size-control models can be implemented as constraints imposed upon the general formalism. For example, cell size control by a "sizer" exclusively depends on the absolute size of the cell, whereas a "timer" on the age of the cell. The "adder," which we will discuss in Section 4, formally measures the amount of growth from a specific point during the cell cycle (*e.g.*, cell division or replication initiation) to divide. This section provides an overview of the non-equilibrium statistical mechanics formalism applied to size control.

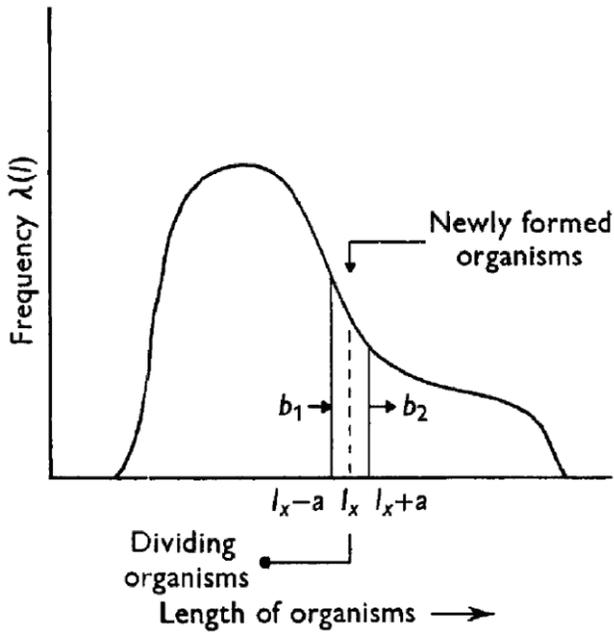

Figure 17: **A hypothetical steady-state size distribution (Figure from Collins and Richmond [574])**. In this figure, $l$ is the cell length and $l_x \pm a$ ($l_x \pm dl$ in the main text) denotes a small range of cell length around an arbitrary cell length $l_x$. The total number of cells between $l_x - a$ and $l_x + a$ is given in terms of the difference between incoming flux $b_1$ and outgoing flux $b_2$ of cell populations (see main text). In our text, we use $\rho(l)$ instead of $\lambda(l)$ for the probability density function of cell length $l$, and $\lambda$ for the growth rate of an exponentially multiplying population in steady state growth.

### 3.1. Non-equilibrium statistical mechanics formalism for size control

The first complete formalism was presented in the Appendix of Collins and Richmond [574]. The flux argument to derive the growth (elongation) rate $V_x$ is particularly intuitive and elegant, and is worth the reading. Consider an arbitrary general steady-state distribution of cell length $\rho(l)$ (Figure 17). The total cell number $N$ in the culture increases exponentially as $N = N_0 e^{\lambda t}$, where $\lambda$ is the growth rate of the population. If $N_{dl}$ organisms lie between $l_x$ and $(l_x + dl)$ in length, and if $N_{dt}$ organisms grow to length $l_x$ in a small time $dt$, then the group of organisms of length between $l_x$ and $l_x + dl$ is replaced $N_{dt}/N_{dl}$ times within $dt$. For this to occur, organisms of length $l_x$ must be able to increase in length by the amount $dl N_{dt}/N_{dl}$ times during $dt$. That is

$$V_x = \frac{N_{dt}}{N_{dl}} \frac{dl}{dt}. \tag{32}$$

**Growth rate** $\lambda$ is the exponent in $N(t) = N_0 e^{\lambda(t-t_0)}$ for a population of cells in exponential growth (see Box 2).

**Elongation rate** $\alpha$ is defined as the rate of increase in length normalized by the cell length,

$$\alpha = \frac{1}{l}\frac{dl}{dt}, \tag{33}$$

namely the "instantaneous elongation rate." For exponentially elongating cells, $l(t) = l_0 e^{\alpha t}$, the exponent is therefore the elongation rate. In steady-state, the average elongation rate $\langle \alpha \rangle$ is equivalent to the growth rate $\lambda$ of the population, because $\langle \alpha \rangle = \left\langle \frac{1}{l}\frac{dl}{dt} \right\rangle = \langle 1/\tau_d \rangle \ln 2 = \lambda$.

**Elongation velocity** $V$ is defined as the increase rate in cell length, which is defined as, $V = \frac{dl}{dt} = l \times \alpha$.

From the definition of $N_{dt}$, we have $N_{dl} = N\rho(l_x)dl$. The number of organisms that reach $l_x$ during $dt$, $N_{dt}$, equals the number of organisms formed with $l < l_x$ during $dt$ by birth, minus the number of organisms disappear from $l < l_x$ during $dt$ by division, minus the number of organisms disappear from $l < l_x$ during $dt$ by growth or "drift."

Therefore, $N_{dt} = N\rho(l_x)dl$ from the definition of $N_{dl}$ and $N_{dt}$ is given by

$$N_{dt} = \lambda N dt \left[ 2\int_0^{l_x} \Psi(l)dl - \int_0^{l_x} \rho_l(l)dl - \int_0^{l_x} \rho(l)dl \right], \tag{34}$$

where $\Psi$ is the birth size PDF, $\rho_l$ is the division size PDF, and $\rho$ is the size PDF of a steady-state population. The factor of 2 in the first term on the *rhs* is to account for the birth two daughter cells from a dividing mother cell.

Based on this argument, Collins and Richmond obtained the following,

$$V_x = \tag{35}$$

$$\lambda \left[ \underbrace{2 \int_0^{l_x} \Psi(l) dl}_{\text{birth (thus 2×)}} - \underbrace{\int_0^{l_x} \rho_l(l) dl}_{\text{division}} - \underbrace{\int_0^{l_x} \rho(l) dl}_{\text{drift}} \right] \Big/ \rho(l_x),$$

which is essentially a flux equation to ensure the conservation of the cells of size $l$.

Tyson and Diekmann [575, 576] revived the formalism by Collins and Richmond almost three decades later. They rewrote Eq. 35 in a differential form as

$$\frac{d}{dt}[v(l)\eta(l)] = 4\gamma(2l)\eta(2l) - \gamma(l)\eta(l) - \lambda\eta(l), \tag{36}$$

where $v(l) = dl/dt$. The new division rate function $\gamma(l)$ is the PDF for a cell of size $l$ at time $t$ to divide in the time interval $(t, t + dt)$. The additional factor 2 in the first term on the *rhs* accounts for the creation of two identical daughter cells by the birth process. From practical point of view, the differential form of Eq. 36 by Tyson and Diekmann is more convenient to work with than the integral form by Collins and Richmond.

The formalism represented by Eqs. 35 and 36 has been utilized in a series of papers and books that include [575–583]. Additional features, such as fluctuations in the rate of elongation $V_x$ or in the sizes of the offsprings due to the positioning of the septum can be included in the formalism (see, *e.g.*, the Supplementary Material in [581]). In general, this is not needed for the study of *E. coli* (and likely other rod-shaped bacteria) because the aforementioned fluctuations are significantly smaller than that of the division size $l_d$ or the generation time $\tau_d$.

As mentioned previously, different size-control scenarios can be implemented via the elongation rate $v(l)$ and the division rate $\gamma(l)$. Typically, $v(l)$ is assumed to be proportional to $l$ for exponential elongation, whereas $\gamma(l)$ encodes the assumptions specific to the cell-size control model. For example, Tyson and Diekmann tested the "sloppy size control" model originally suggested by Koch and Shaechter [570] and Powell [571]. This notion refers to the hypothesis that there is a critical size for division ("sizer"; see below Section 3.2), but due to intrinsic stochasticity the division length is variable around its mean (the latter being strictly controlled). Here, again, it is important to notice that stochasticity at the cellular level was already appreciated and theories developed more than a half century ago are still relevant.

The reader should be aware of a subtle but important assumption in Tyson and Diekmann's approach [576], which enforces "causality" by strictly preventing an overlap between the newborn size and division size distributions.

### 3.2. Size control models: sizer and timer

The formalism in Eq. 36 can be used analyze idealized size control models such as "sizer", "timer", or "adder." The version presented below is by Diekmann *et al.*[575] and Bradde and Vergassola [9].

**Sizer.** The main assumption of the "sizer" model is that cells divide after reaching a threshold size drawn from a probability distribution that depends on growth conditions.

Because the cell width of a typical model organism such as *E. coli* or *B. subtilis* remains nearly constant during elongation, we shall use cell length $l$ as a proxy of cell size. Lengths of the cells at birth and division are denoted by $l_b$ and $l_d$, respectively. The number of individuals in a population of bacteria having length $l$ at time $t$ is denoted by $\eta(l, t)$. The sizer mechanism posits that the rate of division of the cells depends on their size (length) $l$ only, and the equation that governs the evolution in time of $\eta(l, t)$ reads:

$$\partial_t \eta(l, t) + \partial_l(v(l)\,\eta(l, t)) = \tag{37}$$

$$4v(2l)\gamma(2l)\eta(2l, t) - v(l)\,\gamma(l)\eta(l, t),$$

where $v(l) = dl/dt$. The *rhs* in (37) is the total time-derivative, with the $l$-derivative drift term accounting for the elongation of the cells. The *rhs* of (37) arises from the division of cells: $\gamma(l)$ is the division rate or the "division rate function" [see, *e.g.*, [575]]. The function is defined as the local Poisson rate of cell division, *i.e.*, $\gamma(l)\,dl$ is the probability that a cell of length $l$ divides while growing from $l$ to $l + dl$. Therefore, the probability that a cell with initial size $l_b$ has not divided and has reached size $l_d$ is $\exp\left[-\int_{l_b}^{l_d} \gamma(l')\,dl'\right]$. Finally, the conversion of the division rate from unit size to unit time involves the factor $|dl/dt| = v(l)$, which explains its presence in the second term (of loss) on the *rhs* of (37).

The first term in the *rhs* of (37) describes the gain in the number of cells in the length range $(l, l+dl)$. The gain originates from those bacteria in the size range $(2l, 2l + 2dl)$ which divide and thus halve their size (fluctuations in the size of the offsprings are neglected here for simplicity but will be discussed later). The structure of the first term is thus analogous to the second term, with $l$ replaced by $2l$ in the arguments of the functions. The factor 4 arises from the product of two factors of 2: the first is due to the $2dl$ range of dividing cells and the second is due to division producing two offspring.

The solution to the equation (37) depends on the specific form of the division rate $\gamma(l)$ and the elongation law $v(l)$. It is possible to extract a few general relations as follows.

First, integrating (37) over $l$, we obtain the equality

$$\partial_t \int dl\, \eta(l,t) = \int dl\, v(l)\, \gamma(l)\eta(l,t)\,, \qquad (38)$$

where the *rhs* is the total number of individuals which divide in the time interval $[t, t+dt]$. For the rate of growth of the number of individuals in the population, we obtain then

$$\partial_t \ln \int dl\, \eta(l,t) = \int dl\, v(l)\, \gamma(l) \frac{\eta(l,t)}{\int dl\, \eta(l,t)} \equiv \langle v(l)\gamma(l)\rangle\,. \qquad (39)$$

At long times, we require that $\eta(l,t)/\int dl\, \eta(l,t)$ will reach a steady-state distribution that we denote by $\rho(l)$. The corresponding steady-state rate of growth of the population is then denoted $\lambda = \int dl\, v(l)\gamma(l)\rho(l)$.

Second, multiplying (37) by $l$ and integrating, we obtain for the average size

$$\partial_t \langle l \rangle = \langle v(l)\rangle - \langle v(l)\gamma(l)\rangle\, \langle l \rangle\,. \qquad (40)$$

In steady-state, the two terms on the *rhs* of (40) will balance. In other words, for exponential elongation $l(t) = l_b e^{\kappa t}$, the consistency condition $\langle l\,\gamma(l)\rangle = 1$ is satisfied as $v(l) = \kappa l$ from (40).

Finally, multiplying (37) by $l^q$ and integrating, we obtain a division series of relations for higher-order moments. In the steady-state and for an exponential elongation $v(l) = \kappa l$, the relations read $(q-1)\langle l^q \rangle = (1 - 2^{1-q})\langle l^{q+1}\gamma(l)\rangle$, where the integer $q > 1$ and we have made use of (40).

**Timer.** The "timer" model posits that cell division is controlled by the age of the cell, *i.e.*, the time elapsed since its birth. The state of cells is therefore described by their size $l$ and age $\xi$. The corresponding number of cells at time $t$ is denoted by $\eta(l,\xi,t)$, and the equation for the evolution of $\eta$ reads:

$$\partial_t \eta(l,\xi,t) + \partial_l\left[v(l)\eta(l,\xi,t)\right] + \partial_\xi \eta(l,\xi,t) = -\gamma(\xi)\eta(l,\xi,t)\,; \quad \eta(l,0,t) = 4\int \gamma(\xi')\eta(2l,\xi',t)\,d\xi'\,. \qquad (41)$$

The *lhs* is the total time derivative: the drift in $l$ is due to the elongation of cells $dl/dt = v(l)$ while the drift in $\xi$ is due to the aging of cells $d\xi/dt = 1$ (the discontinuity in $\xi$ occurring at division will be addressed momentarily). The *rhs* of the first equation in (41) is the loss term due to the division of cells. The Poisson division rate function (division rate) $\gamma$ depends now on the age of the cell, $\xi$. Furthermore, since $d\xi/dt = 1$ there is no additional factor coming from the conversion of the rate of division from unit age to unit time [see (37)].

From the definition of the division rate it follows that

$$\rho_{\tau_d}(\tau) = \gamma(\tau)\, e^{-\int_0^\tau \gamma(\xi')\,d\xi'}$$
$$\Rightarrow \gamma(\tau) = \frac{\rho_{\tau_d}(\tau)}{1 - \int_0^\tau d\xi'\rho_{\tau_d}(\xi')}\,, \qquad (42)$$

where $\rho_{\tau_d}(\tau)$ is the probability density for the generation time $\tau_d$ of a given cell and the derivation proceeds as for (42) Finally, the last equation in (41) is the boundary condition that accounts for newborn cells having all the same age $\xi = 0$, irrespective of their size $2l$ that gets halved. The integral $\int d\xi\, \gamma(\xi)\eta(2l,\xi,t)$ represents the total number of cells that divide in the unit time; as in (37), the factor 4 is the product of the factor 2 resulting from the $2\,dl$ width of the range of dividing cells and the factor 2 due to division producing two newborn daughters cells.

The dynamics of the age of the cells is independent of their size, as can be easily seen by integrating (41) over $l$. The resulting equations for the marginal distribution $\eta(\xi,t) = \int \eta(l,\xi,t)\,dl$ reads

$$\partial_t \eta(\xi,t) + \partial_\xi \eta(\xi,t) = -\gamma(\xi)\eta(\xi,t) \qquad (43)$$

$$\eta(0,t) = 2\int d\xi\, \gamma(\xi)\, \eta(\xi,t)\,. \qquad (44)$$

Furthermore, integrating (41) over $l$ and $\xi$, we obtain for the population growth rate $\lambda$,

$$\partial_t \int \eta(l,\xi,t)\,dl\,d\xi = \int \gamma(\xi)\eta(l,\xi,t)\,dl\,d\xi$$
$$= \int \gamma(\xi)\eta(\xi,t)\,d\xi\,, \qquad (45)$$

which depends on the marginal distribution $\eta(\xi,t)$ only. The well-known solution [279] to (43) is $\eta(\xi,t) = e^{\lambda t}\tilde{n}(\xi)$, where

$$\tilde{n}(\xi) = A e^{-\int_0^\xi \gamma(\xi')\,d\xi' - \lambda\xi} \qquad (46)$$

$$2\lambda \int_0^\infty e^{-\lambda\xi} e^{-\int_0^\xi \gamma(\xi')\,d\xi'}\,d\xi = 1\,. \qquad (47)$$

The first equation is simply obtained by integrating the differential equation in (43) and the constant $A$ depends on the initial size of the population. The second relation is obtained from the boundary condition in (43) by an integration-by-parts. A similar integration-by-parts provides a check that the second equation in (47) is equivalent to the growth rate relation $\lambda \int \tilde{n}(\xi)\,d\xi = \int \tilde{n}(\xi)\gamma(\xi)\,d\xi$ derived in (45).

As for the dynamics of the size of cells, we can obtain the equation for the mean size multiplying (41) by $l$ and integrating over $l$ and $\xi$. The resulting expression reads

$$\partial_t \langle l \rangle = \langle v(l) \rangle - \lambda \langle l \rangle\,, \qquad (48)$$

where $\langle l \rangle \equiv \int l\, \eta(l,\xi,t)\, dl\, d\xi / \int \eta(l,\xi,t)\, dl\, d\xi$ and similar definitions apply for other averages. For a stationary state, the *lhs* is required to vanish, *i.e.*, elongation should balance the growth. In the case of linear elongation $v(l) = \mathrm{const.}$, a steady-state distribution is reached and $\langle l \rangle = \mathrm{const.}/\langle \gamma(\xi) \rangle$, where we used $\lambda = \langle \gamma(\xi) \rangle$. In contrast, for exponential elongation $v(l) = \kappa l$, a steady-state distribution exists only for the special choice $\kappa = \lambda$. If the equality is not satisfied, then the average size grows to infinity or decay to zero exponentially fast (see below).

The lack of control of the cell size by the timer mechanism was remarked in [584, 585] and can be understood intuitively by considering the sizes at birth $l_b^{(n)}$ and $l_b^{(n+1)}$ over two consecutive generations $n$ and $n+1$. For an exponential elongation, $\ln l_b^{(n+1)} - \ln l_b^{(n)} \simeq \kappa \tau_d - \ln 2$, where $\tau_d$ is again the generation time and $\kappa$ the elongation rate in the $n$-th generation. Exponential elongation of the size of the cells requires that the average value of $\kappa \tau_d$ be precisely-tuned in order to avoid a systematic drift of $\ln l_b$. Notice that even in the absence of drift, the long term behavior of $\ln l_b$ will be analogous to a random walk (assuming that the values of $\kappa \tau_d$ fluctuate and decorrelate over the generations). Therefore, the variance of the size of the cells will grow across the generations and no effective control of the size of the cells is achieved by the timer mechanism. Diluting bacteria by washing them out, *e.g.*, a term $-D\eta(l,\xi,t)$ is added to the equation (41) [575], will not modify the previous conclusion unless dilution is coupled to the size, *i.e.*, the dilution rate depends on $l$.

In practice, the elongation rate deviates from a linear behavior at very small and at very large sizes so that the logarithm of the size will not go to zero or diverge to infinity. However, its behavior will depend very sensitively on the details of the elongation law at very small and large sizes [575, 581] and the resulting size distributions are generally significantly wider than the actual data (not shown). Most importantly, the timer mechanism disagrees with the experimental data insofar as it predicts the conditional distribution $P(\tau_d|l_b)$ of the generation time $\tau_d$ *vs* the initial size of cells $l_b$ should be independent of $l_b$ [9]. Conversely, data shown in [9] (Figure 18) indicate a clear dependency on $l_b$.

**Mixed models.** Diverse combinations of sizer and timer mechanisms are conceivable. A well-known instance is the bilinear model proposed in [78]. The proposed dynamics is summarized as follows: cells grow linearly in time $v(l) = u$ until they reach a threshold length $l_\Lambda$. After reaching $l_\Lambda$, cells keep growing for a fixed amount of time ($\sim$ 20 mins) at a velocity $2u$ (see [172, 388, 396] for a discussion of bilinear vs exponential elongation). Ref. [78] clearly poses relevant questions and deserves all the influence it had in the field; however, the specific mechanism which was proposed is not supported by modern experimental data [9, 384].

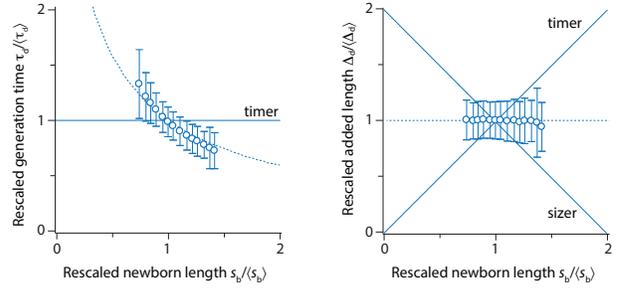

Figure 18: **Correlation of added size and newborn size, and correlation of generation time and newborn size.** Data is from [9].

Before the "adder" model was (re)discovered [9, 586, 587], other combinations of age and size control had been proposed. The intuitive motivation comes from plots like (Figure 2D) in the main text, where a behavior intermediate between a timer and a sizer is observed (see Figure 18). The most recent proposition is [580], where a control mechanism operating "concertedly" (and not sequentially as in [78]) on the size and the age of the cell is discussed. No major inconsistency with experimental data (for a single growth condition) is observed by introducing a model where the division rate function $\gamma(l,\xi)$ depends jointly on the size and on the age of the cells and by best fitting the function to the data. However, a simpler adder model that we discuss below explains these observations and beyond.

### 3.3. The genius of Arthur Koch: how mathematical insight led to biological insight, and vice versa

Among the giants in microbial physiology, Arthur Koch's depth, breadth, and especially his originality (and generosity) is unparalleled [69, 256, 278, 570, 572, 588–603]. One of the main questions that occupied him for a long time was, "How does a cell know how big it is?" [256] His first serious work on the subject was with Elio Schaechter [570, 589] based on the analysis of the time-lapse experiment of the growth of individual bacteria. Schaechter's group adopted the imaging technique developed by Mason and Powelson [604] to visualize the *E. coli* and *B. subtilis* nucleoids *in vivo* by growing them in a medium with a refractive index near that of the cellular cytoplasm. They studied typically between 100-200 cells, comparable to the standard of microbiology in the 1990s and 2000s before the use of microfluidics that allowed tracking of tens of thousands of cells [384, 605].

The key results from Schaechter *et al.*'s experiments is that the coefficient-of-variation (CV) of the generation time is about 20%, whereas is is about 10% for the division size. They also measured the correlation of various measurables between mothers and daughters, and also between daughters [589].

In his review article *'Does the Initiation of Chromosome Replication Regulate Cell Division?'* in 1977 [256], Koch presented a series of brilliant arguments that laid the constraints on models of cell size and cell cycle control based upon measurements of the coefficient-of-variation. One particular example is worth mentioning here.

> *Had the cell a division mechanism inherently consisting of a series of stages in the cell cycle, each to be timed from the completion of the last stage independently of the cell size at that time, then one would have expected the cell size at division to have a larger (or equal) coefficient of variation than that for the age distribution, since random sources of uncorrelated variation only add. Consequently any mechanism which does not lead to negative correlations between parts of a cell cycle or between cycles of related individuals is excluded. A chance fluctuation in timing of one phase must sooner or later lead to a fluctuation of opposite sign at some later phase. This logic eliminates several previously considered pure branching processes such as those proposed by Rahn (1931) [606] and Kendall (1948) [567], and demands statistical models containing deterministic elements.*

Based on the observation that the CV of division size is smaller than the CV of generation time, Koch further correctly predicted a negative correlation of the generation time between mothers and daughters, and a positive correlation between sisters. The basic idea is to consider two consecutive generations as a single generation of one hypothetical cell such that the division of the real mother corresponds to approximately at half-way the growth of the hypothetical cell. We can then apply Koch's argument for negative correlations of the duration of two parts of the same cell cycle, *i.e.*, the generation times of the mother and daughter (the first half and the second half of the "cell cycle" of the hypothetical cell). We illustrate this graphically In Figure. 19.

Later, Koch's interest went deeper into major processes underlying cellular reproduction such as cell division, chromosome replication initiation and termination, and a hierarchy in the triggering of these processes. His approach was, again, based on the variability of individual processes and their comparisons and correlations. He was fond of, and trusted, the autoradiographic and size data from steady-state populations by Chai and Lark, as well as by Forro [125, 143, 152]. To test various extant models, Koch employed numerical simulations and compared them against the data [256]. To our knowledge, this was one of the first works (if not *the* first) in biology that a researcher used computer simulations to test models against the data, and clearly Koch understood the real value of simulations was in excluding incorrect models.

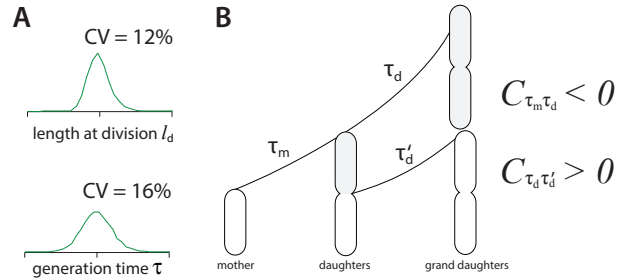

**Figure 19:** **Koch's predictions on the negative correlation between mother and daughter generation times.** (A) The coefficient-of-variation (= $\sqrt{\text{mean/variance}}$, CV) of generation time is larger than the CV of division length. (B) Based on (A) and his explanation why the duration of parts of the same cell cycle must be negatively correlated, Koch deduced that the generation time correlation between mother and daughter should be negative, whereas that of two daughters must be positive.

The reader may have noticed that Koch was in favor of a sizer model, and indeed he and Schaechter introduced the notion of 'critical size' for cell division. This was a result of their ranking of biological processes in their order of CV, and their hypothesis that the process with the smallest CV corresponded to key control point. This reasoning led Koch to propose that a sizer mechanism provides a better characterization of size control than a timer mechanism. Furthermore, he concluded based on his analysis of the autoradiographic data and computer simulations that

> *On the basis of these several lines of evidence I feel fairly confident, therefore, in concluding that the initiation of rounds of DNA synthesis is neither well controlled with respect to the cell size, nor to cell age....*

In our view, this conclusion was premature, and he would have benefited from single-cell data that became available in the past few years. For example, we now know that *E. coli* and many other bacteria do not employ a sizer to control their size. Instead, they are "adders" as we explain in the next section. Nevertheless, his intuition that there may be multiple triggers, instead of a single trigger, underlying cell size and cell cycle have not been fully resolved, and should be considered seriously by future researchers.

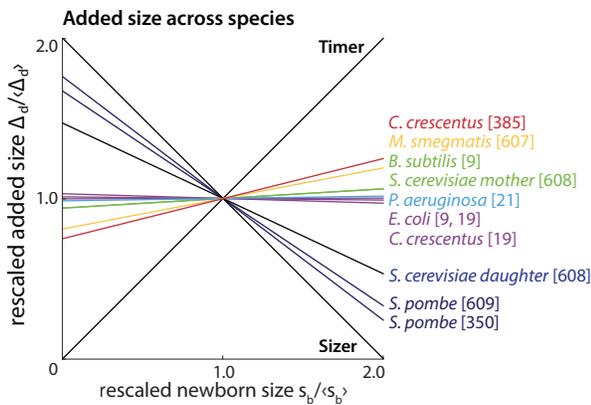

Figure 20: **The adder principle appears in distinct organisms [22].** This graph is summarized from references [9, 19, 21, 350, 385, 607–609].

## 4. Adder as a new phenomenological paradigm of cell size homeostasis

The field of cell-size control has been rapidly transforming in the past few years due to the (re)discovery of the "adder" principle [9, 19–23]. This was made possible because of the new single-cell growth and division data (*e.g.*, [384]) that were not available previously.

The adder model posits that cells add constant size $\Delta_d$ between birth and division, irrespective of the birth size. Many evolutionary divergent organisms from bacteria to eukaryotes have been shown to be adders (Figure 20). While the adder principle is difficult to understand from a molecular point of view, it provides a very intuitive explanation for how cells maintain size homeostasis. This is illustrated in Figure 21.

This section explains the history, the experiments, and the modeling relevant for the adder principle of size homeostasis.

### 4.1. History of adder

In 1993, Koppes and colleagues published a paper entitled "Mathematics of cell division in *Escherichia coli*," where they (despite their passive Dutch voice) explicitly proposed and tested using extant data the current form of the adder (they originally called the model "incremental size model" before the model was termed "adder" [9, 20]).

> *There is another way of viewing the control of cell division, the incremental-size model. This model states that a growing cell divides after having increased its size by a critical amount; here it is*

## Size convergence by adder

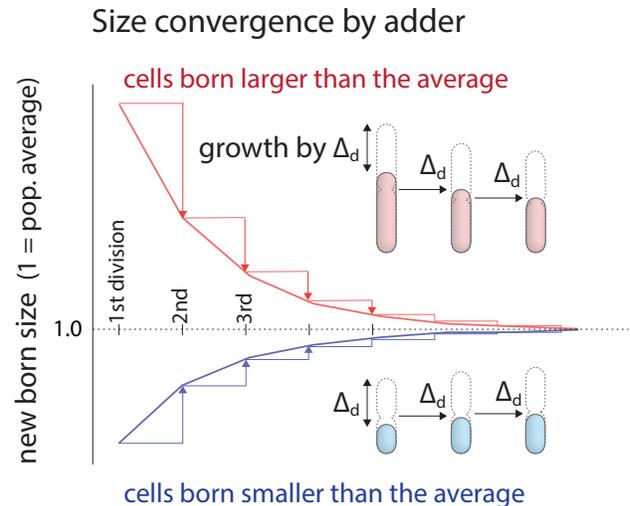

Figure 21: **The convergence of cell size by the adder principle.** A cell born larger than the population average, adds a fixed size $\Delta_d$ and divides in the middle. The daughter cell is smaller than the mother. The daughter cell also grows by $\Delta_d$ and divides in the middle, and becomes even smaller. This continues until the daughter's newborn cell size becomes the same as $\Delta_d$ itself. The same convergence principle works the same way for cells born smaller than the population average.

> *the increment that is considered to be the same, on the average, for all the cells in the culture regardless of their size at birth...*

It is unclear how Koppes came up with the model. As mentioned previously, the adder is highly unintuitive from the biological point of view. In fact, Koppes and Voorn themselves expressed the same sentiment in their article in 1997 [587]:

> *How a bacterial cell manages to grow such a fixed amount of mass or surface is, in our opinion, equally hard to imagine as it is trying to grow to a predetermined size.*

One possibility is that Koppes and colleagues in the Dutch school in the 1980s had attempted to apply the sloppy size control model (which is essentially a sizer with noise) to the extant data, and realized that the model is inconsistent with the data and thus explored different models. A possible inspiration during the exploration may have been the work by Sompayrac and Maaløe (1973) [56] discussed in Section 2.2.7. Koppes and colleagues were clearly aware of the work:

> *Initiation would thus occur when cellular volume had increased by a fixed amount per origin since the previous initiation.*

Ideally, comprehensive single-cell growth data would have provided unambiguous support for a specific division model. Unfortunately, such data as shown in Figure 23 (data from Jun lab [9]) was unavailable in the early 1990s. Instead, Koppes and colleagues developed a mathematical scheme (especially with Grover) to calculate the correlation coefficient $r$ between size at birth $l_b$ and size at division $l_d$ based on the coefficient-of-variation (CV) of $l_b$, $l_d$, and the generation time $\tau_d$, which was available in [586, 610–614]. They obtained $r = $ "0.52 and 0.56 for interdivision times of 21 min and 125 min, respectively." [586]. This is close to the prediction of the perfect adder, $r = 1/2$ [9, 615], but significantly larger than the prediction of the sizer model, $r = 0$ (by the very definition of a sizer). This led Koppes and colleagues to the following conclusion:

> *The thirty year old so-called sloppy-size model could be rejected, whereas the newly-developed incremental-size model was accepted (by lack of alternative).*

That *E. coli* is an adder became evident from the large amounts of single-cell data in [384], which was significantly extended to other nutrient conditions [9, 19] as well as to evolutionary distant Gram-positive *B. subtilis* [9].

### 4.2. Mother machine, single-cell growth experiments, data, and choice of control parameters

Technological development in high-throughput microscopy, imaging, image analysis, and microfluidics were critical in the development of modern single-cell bacterial physiology [386, 387, 616–618]. There are many excellent reviews on the technology side with biological applications, and we refer the reader to them [605, 619–621]. In this subsection, we will use only one example known as the "mother machine" (Figure 22). The mother machine device was introduced by the Jun lab in 2010 [384, 605], and has been extended to other organisms (Figure 22) [622–627].

The mother machine allows tracking of thousands of mother cells for hundreds of consecutive generations (Figure 22). A typical timelapse sequence of growth and division is shown in Figure 2D. Six parameters can be immediately defined and deduced from the sequence. The birth size $l_b$, division size $l_d$, the size added between birth and division $\Delta_d = l_d - l_b$, the relative septum position $l_{1/2}$, the generation time $\tau_d$, and the instantaneous elongation rate $\alpha = \frac{1}{l}\frac{dl}{dt}$. Obviously, not all six parameters are independent from one another. From a biological and a modeling point of view, it is important to determine the minimal set of parameters controlled by the cell. A useful approach is to plot correlations among all pairs of the potential control parameters as shown in Figure 23.

In this example, the instantaneous elongation rate $\alpha$ is nearly uncorrelated with other parameters. It is therefore reasonable to assume that the cell must control $\alpha$ independent of other parameters. The other extreme example is the generation time $\tau_d$, which is correlated with all other parameters except the septum position $l_{1/2}$. Therefore, $\tau_d$ is a less-than-ideal control parameter from a modeling perspective, and biologically, as the timer mechanism is not a good strategy to maintain cell-size homeostasis in exponentially elongating bacteria (Section 3.2).

For the adder strategy, we choose $\Delta_d$ as the other control parameter and describe the growth and division dynamics using the "kinetic" parameter $\alpha$ and the "spatial" parameter $\Delta_d$. The results and predictions of the modeling are tested against the data in the following sections.

### 4.3. Modeling the adder

The general formalism for the adder is the same as the one used to characterize the sizer and the timer summarized in Sections 3. Because the adder model posits that the mechanism of control involves a single variable, the added mass at division $\Delta_d$, the dynamics of size control can then be described using two variables the instantaneous elongation rate $\alpha$ and the added mass $\Delta_d$. Note that the binary division is assumed to be perfect given that the distribution of septum position is the narrowest among all measurable physiological variables (Figure 23).

If $l(t)$ is the length of a cell at the current time $t$, its added length is denoted $\Delta(t) = l(t) - l_b$, where $l_b$ is the length at birth. The density of cells $\eta(l, \Delta, t)$ having length $l$ and added size $\Delta$ obeys the following continuity equation

$$\partial_t \eta(l, \Delta, t) + \partial_l\left[v(l)\eta(l, \Delta, t)\right] + \partial_\Delta\left[v(l)\eta(l, \Delta, t)\right] = -\gamma(\Delta)v(l)\eta(l, \Delta, t)\,; \tag{49}$$

$$v(l)\eta(l, 0, t) = 4\,v(2l)\int_0^\infty \gamma(x)\eta(2l, x, t)\,dx\,. \tag{50}$$

As before, the *lhs* in (49) is the total time-derivative and the two drift terms are due to the elongation of the cells, *i.e.* $dl/dt = v(l)$ and $d\Delta/dt = v(l)$. The *rhs* accounts for the division of the cells. The Poissonian division rate function $\gamma(\Delta)$ depends now on the added size $\Delta$. Proceeding as for (42), we obtain the relation

$$\rho_{\Delta_d}(\Delta) = \gamma(\Delta)\,e^{-\int_0^\Delta \gamma(x)\,dx}$$
$$\Rightarrow \gamma(\Delta) = \frac{\rho_{\Delta_d}(\Delta)}{1 - \int_0^\Delta dx\,\rho_{\Delta_d}(x)}\,, \tag{51}$$

between $\gamma(\Delta)$ and the distribution $\rho_{\Delta_d}(\Delta)$ for the size added at division ($\Delta_d = l_d - l_b$) of individual cells. By individual cells we mean that cells should be weighted equally, tracking them individually and avoiding known bias effects related to the speed of reproduction [279]. As in the equation (37), the conversion of the rate of division to unit time involves the Jacobian $|d\Delta/dt| = v(l)$, that appears then in the right hand side of (49). Finally, (50) is the boundary condition that accounts for cells having all $\Delta = 0$ at birth, irrespective of their size $2l$ that gets halved. The integral $\int \gamma(x)\eta(2l, x, t) dx$ represents the total number of cells that divide in the unit time; as in (37), the factor 4 is the product of the factor 2 resulting from the $2 dl$ width of the range of dividing cells and the factor 2 due to division producing two newborn daughters cells.

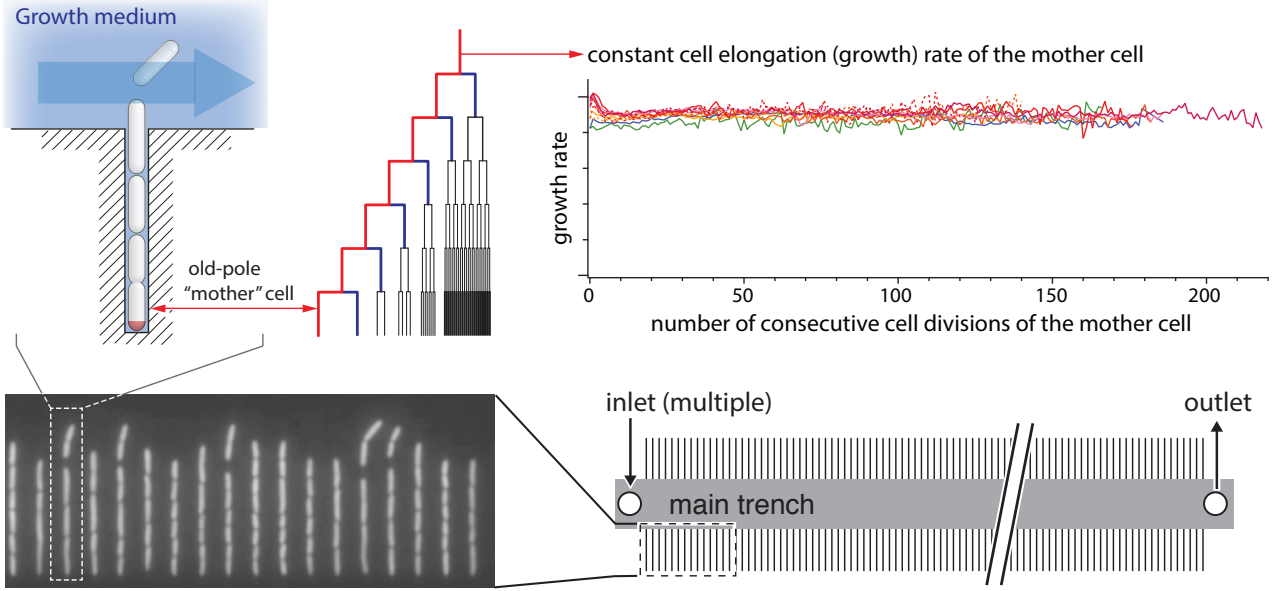

Figure 22: **The microfluidic mother machine.** Each mother machine device consists of thousands of long, narrow growth channels. The physical dimensions of the growth channels are such that *E. coli* cells fit snuggly. The cell at the deadend of the growth channel inherits the same cell pole from previous generation upon division, thus the "mother" cell. *E. coli* cells growing in the mother machine do not show any sign of aging in terms of their instantaneous elongation for hundreds of generations.

A series of relations analogous to those obtained for the sizer model can be derived from (49) for a general division rate $\gamma$. Integrating (49) over $l$ and $\Delta$ and using the boundary condition (50), we obtain

$$\partial_t \ln \int \eta(l, \Delta, t) dl d\Delta$$
$$= \int v(l)\gamma(\Delta) \frac{\eta(l, \Delta, t)}{\int \eta(l, \Delta, t) dl \, d\Delta} dl \, d\Delta$$
$$\equiv \langle v\gamma \rangle. \tag{52}$$

Multiplying (49) by $l$ and integrating over $l$ and $\Delta$, we derive for the average size

$$\partial_t \langle l \rangle = \langle v(l) \rangle - \alpha \langle l \rangle, \tag{53}$$

where the rate of growth of the population $\alpha = \langle v(l)\gamma(\Delta) \rangle$ from (52). For a linear elongation rate, $v(l) = u$, (53) gives for the size at equilibrium $\langle l \rangle = 1/\langle \gamma(\Delta) \rangle$. For an exponential elongation rate, $v(l) = \alpha l$, (53) yields $\alpha = \alpha$, *i.e.*, the value $\langle l\gamma(\Delta) \rangle = 1$ for the correlation between the size $l$ and the division rate function $\gamma(\Delta)$.

Finally, we can obtain the following series of relations for higher order moments at the steady state:

$$(1 - m)\langle l^m \Delta^q \rangle - q\langle l^{m+1} \Delta^{q-1} \rangle$$
$$= 2^{1-m}\delta_{q,0}\langle l^{m+1}\gamma(\Delta) \rangle - \langle l^{m+1}\Delta^q\gamma(\Delta) \rangle \tag{54}$$
$$m \geq 0, q \geq 0$$

where we specified relations for the case of exponential elongation $v(l) = \alpha l$ and we have used $\alpha = \alpha$ and $\langle l\gamma \rangle = 1$ derived previously.

*4.3.1. Comparison with experiments.* The comparison with experimental data for the adder model proceeds as for the sizer mechanism. The elongation rate $v(l) = \frac{dl}{dt}$ and the division function $\gamma(\Delta)$ are extracted from experimental data, namely from the distribution of the sizes at division $l_d$ and the distribution of the instantaneous elongation rates $\alpha$. These are then used to simulate the cell size control process at the level of individual cells. Finally, we compare statistical observables alternative to those used for calibration, in order to assess the validity of the model.

The calibration of the model proceeds as follows. The instantaneous elongation rate $\alpha$ for individual bacteria is obtained by exponential fits of the experimental curves of size versus time. The probability densities $\rho_\alpha^{ex}(\alpha)$ of the resulting elongation rates in the various growth conditions are shown in

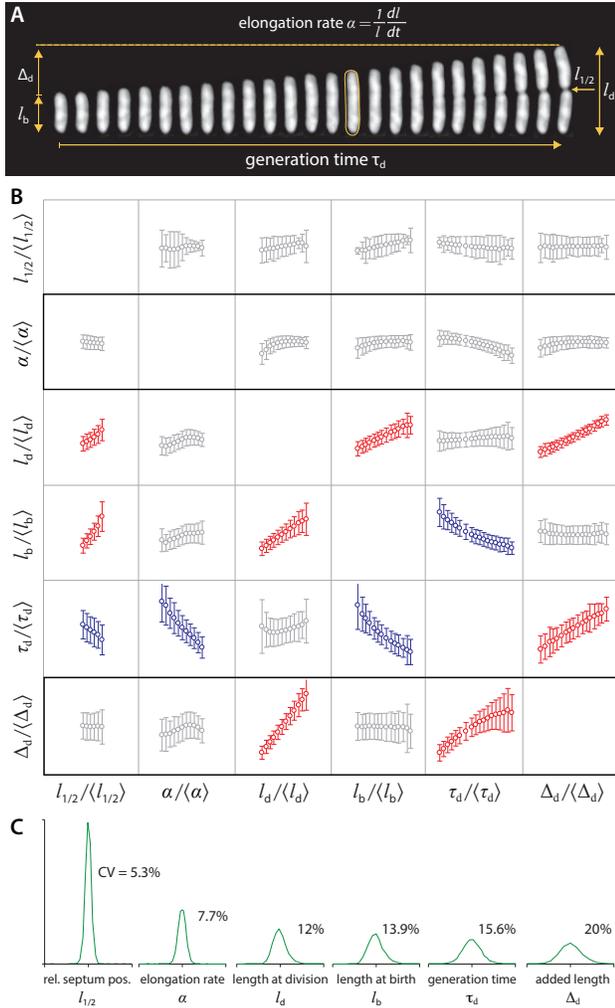

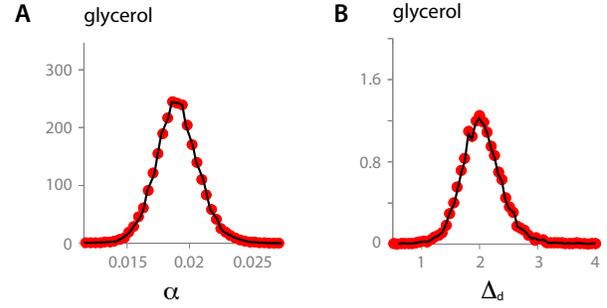

Figure 24: **Calibration of the adder model for the control of the cell size.** The instantaneous elongation rate of the cells is an independent, identically-distributed (iid) random variable drawn from the *E. coli* experimental distribution, $\rho_\alpha^{ex}(\alpha)$ (red dots) in the plots of panel (A) for the seven different growth conditions presented in Figure 28. Correlations among elongation rates of mother and siblings are weak and thus not taken into account. The black curves are the results of the numerical simulations. The division rate $\gamma(\Delta)$ is computed as detailed in the text (see Eq. 51) from the distribution $\rho_{\Delta_d}^{ex}(\Delta)$ of the increments at division $\Delta_d = l_d - l_b$. In panel (B) we show the experimental distributions for the added size at division $\Delta_d$ (red dots) for the same growth condition as in panel (A). The curves in black are the results of numerical simulations of the model detailed in the text. Their agreement with the experimental curves confirms that the parameters of the model are appropriately calibrated. Similar curves are obtained for *B. subtilis*.

Figure 23: **Single-cell growth data obtained from a mother machine experiment. A.** The graphical definitions of six physiological parameters for single-cell growth. **B.** All correlations between six normalized parameters, $l_{1/2}/\langle l_{1/2}\rangle$, $\alpha/\langle\alpha\rangle$, $l_d/\langle l_d\rangle$, $l_b/\langle l_b\rangle$, $\tau/\langle\tau\rangle$ and $\Delta_d/\langle\Delta_d\rangle$, are shown as a $6\times6$ matrix of subplots. The growth condition is MOPS with 0.2% glucose. In the matrix, the positive correlations are color-coded to red, negative to blue and nearly-uncorrelated to grey. **C.** The distributions of all six paramters in the ascending order of their relative widths.

Figure 24. In the numerical simulations one draws values of $\alpha$ randomly from $\rho_\alpha^{ex}(\alpha)$, neglecting (as for the sizer case) weak correlations between the instantaneous elongation rates of mother and daughter cells. The division rate $\gamma(\Delta)$ is computed using (51) with the probability of the size increments at division $\Delta_d \equiv l_d - l_b$ read directly from the experimental data $\rho_{\Delta_d}^{ex}(\Delta)$ (see Figure 24).

The distribution of the instantaneous elongation rates $\alpha$ and the division function $\gamma(\Delta)$ extracted from the experimental data are used to simulate the dynamics of a bacterial colony. Each cycle of elongation of a cell proceeds at the constant (random) rate $\alpha$ and division occurs with the Poissonian rate $\gamma(\Delta)$, which depends on the size increment $\Delta$ only. After an initial transient, distributions for the various observables reach a stationary form and the resulting numerical distributions for the added size at division $\Delta_d = l_d - l_b$ and the instantaneous elongation rate $\alpha$ (used to calibrate the model) compare to the experimental distributions as shown in Figure 24 over a range of growth conditions.

**Distributions of size and age at division.** A first test for the adder model is provided by the curves in Figure 25 showing the agreement of the distributions for the final size $l_d$ and for the generation time $\tau_d = \ln(l_d/l_b)/\alpha$ (where the values of the various quantities are those of individual bacteria). The corresponding results for *B. subtilis* are shown in Figure 26. Most importantly, the model by its very definition agrees with the $l_b$-independent curves of the conditional probability $P(\Delta_d|l_b)$ shown in the reference [9].

**Correlations: size across generations.** An additional test for the model comes from the correlations of the size among genealogically related cells. Some of the correlations below have been calculated in Ref. [615] by a slightly different procedure. Correlations between the added length of the mother $\Delta_{\mathrm{d}}^M$ and the daughter, $\Delta_{\mathrm{d}}^D$ are experimentally found to be small so we shall neglect them hereafter.

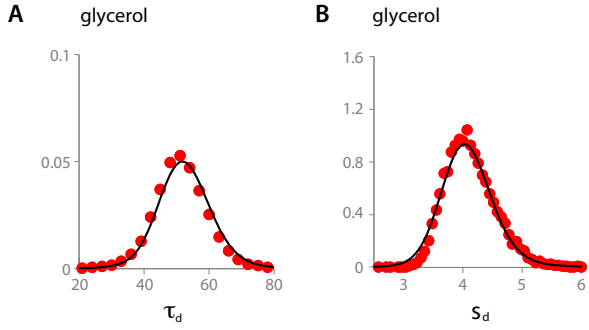

Figure 25: **Test of the adder model for the control of the cell size.** The model calibrated as in Figure 24 is simulated numerically and the *E. coli* distributions of the generation time $\tau_{\mathrm{d}} = \log_2(l_{\mathrm{d}}/l_{\mathrm{b}})/\alpha$ and the size $l_{\mathrm{d}}$ at division of the cells are reported in panels (A) and (B), respectively, for one representative growth condition. Red dots refer to experiments while black curves are the numerics. The agreement of theoretical predictions with experimental data substantiates the validity of the adder mechanism for the control of the cell size.

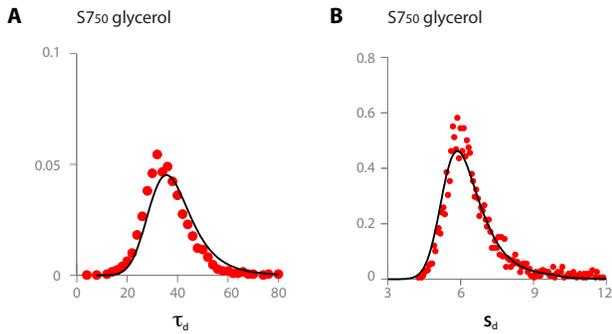

Figure 26: **Test of the adder model in *B. subtilis*.** As in Figure 25, the distributions of the generation time $\tau_{\mathrm{d}} = \log_2(l_{\mathrm{d}}/l_{\mathrm{b}})/\alpha$ and the size $l_{\mathrm{d}}$ at division of the cells are reported in panels (A) and (B), respectively, for one representative growth condition. Red dots refer to experiments while black curves are the numerical predictions.

We begin with the correlation $\mathcal{C}_{dd}^{(p)} \equiv \langle l_{\mathrm{d}}^M l_{\mathrm{d}}^{D(p)} \rangle$ between the size at division of the mother and its $p$-th generation descendant. For example, $p = 1$ gives the correlation between the sizes at division of mother and daughter cells. The decay of the correlation function

is computed by the formula

$$
\begin{aligned}
\mathcal{C}_{dd}^{(p+1)} &= \left\langle l_{\mathrm{d}}^M l_{\mathrm{d}}^{D(p+1)} \right\rangle \\
&= \left\langle l_{\mathrm{d}}^M \left( \frac{l_{\mathrm{d}}^{D(p)}}{2} + \Delta_{\mathrm{d}}^{(p+1)} \right) \right\rangle \\
&= \frac{1}{2} \left[ \mathcal{C}_{dd}^{(p)} + \langle l_{\mathrm{d}} \rangle^2 \right] .
\end{aligned}
\tag{55}
$$

where we have used that $\langle \Delta_{\mathrm{d}} \rangle = \langle l_{\mathrm{b}} \rangle = \langle l_{\mathrm{d}} \rangle/2$, as can be easily derived from $l_{\mathrm{d}} = l_{\mathrm{b}} + \Delta_{\mathrm{d}}$ and $\langle l_{\mathrm{b}} \rangle = \langle l_{\mathrm{d}} \rangle/2$. For the connected part of the correlation function $C_{dd}^{(p)} \equiv \mathcal{C}_{dd}^{(p)} - \langle l_{\mathrm{d}} \rangle^2$, it follows that

$$
\begin{aligned}
C_{dd}^{(p)} &\equiv \langle l_{\mathrm{d}}^M l_{\mathrm{d}}^{D(p)} \rangle - \langle l_{\mathrm{d}}^M \rangle \langle l_{\mathrm{d}}^{D(p)} \rangle \\
&= \frac{\sigma_{l_{\mathrm{d}}}^2}{2^p},
\end{aligned}
\tag{56}
$$

where $\sigma_{l_{\mathrm{d}}}^2 = \langle l_{\mathrm{d}}^2 \rangle - \langle l_{\mathrm{d}} \rangle^2$ is the variance in the division size $l_{\mathrm{d}}$. The comparison with the numerical simulations is shown in Figure 27. By similar arguments we can show that the correlation function for the size at birth $C_{bb}^{(p)} = \sigma_{l_{\mathrm{b}}}^2/2^p$ where the variance of the size at birth $\sigma_{l_{\mathrm{b}}}^2 = \langle l_{\mathrm{b}}^2 \rangle - \langle l_{\mathrm{b}} \rangle^2 \simeq \sigma_{l_{\mathrm{d}}}^2/4$ because the size is (roughly) halved at division. The scaling is confirmed in Figure 27. Finally, the mixed correlation $C_{bd}^{(p)} = C_{bb}^{(p)}$ because the added size is statistically independent of the initial size.

The positive correlation between the sizes across generations has an intuitive explanation. An ancestor cell bigger than the average will generate progeny that statistically relaxes to the average size as illustrated in Figure 21. The fact that the added length is independent of the initial size means that progeny will inherit only part of the ancestral size, which is successively halved as generations proceed. That accounts for the positive correlations and its $1/2$ rate of decay across generations.

Notice that the previous results give for the Pearson correlation coefficient between the size at birth and at division:

$$
\frac{\langle l_{\mathrm{b}}\, l_{\mathrm{d}} \rangle - \langle l_{\mathrm{b}} \rangle \langle l_{\mathrm{d}} \rangle}{\sigma_{l_{\mathrm{b}}} \sigma_{l_{\mathrm{d}}}} = \frac{C_{bd}^{(0)}}{\sigma_{l_{\mathrm{b}}} \sigma_{l_{\mathrm{d}}}} = \frac{\sigma_{l_{\mathrm{b}}}}{\sigma_{l_{\mathrm{d}}}} \simeq \frac{1}{2} ,
\tag{57}
$$

which accounts for the behavior of experimental data observed in the reference [9]. It also follows from $l_{\mathrm{d}} = l_{\mathrm{b}} + \Delta_{\mathrm{d}}$ and $l_{\mathrm{b}} \simeq l_{\mathrm{d}}/2$ that for the adder model

$$
\sigma_{l_{\mathrm{d}}}^2 = 4\sigma_{l_{\mathrm{b}}}^2 = \frac{4}{3}\sigma_{\Delta_{\mathrm{d}}}^2 ; \quad \langle l_{\mathrm{d}} \rangle = 2\langle l_{\mathrm{b}} \rangle = 2\langle \Delta_{\mathrm{d}} \rangle .
\tag{58}
$$

The coefficients of variation ($CV$) of the three quantities are

$$
CV_{l_{\mathrm{d}}} \equiv \frac{\sigma_{l_{\mathrm{d}}}}{\langle l_{\mathrm{d}} \rangle} \simeq CV_{l_{\mathrm{b}}} \simeq \frac{1}{\sqrt{3}} CV_{\Delta_{\mathrm{d}}} .
\tag{59}
$$

Predictions are in excellent agreement with the experimental data presented in the reference [9]. The coefficient of variation of the birth size $l_b$ is actually slightly larger than the coefficient of variation of the division size $CV_{l_d}$ because the birth size $l_b$ is also affected by the noise in the positioning of the septum. We have neglected septum-positioning noise because it is small; when included, it leads to $CV_{l_b}$ being slightly larger than $CV_{l_d}$, as shown in the reference [9].

**Correlations: generation times.** Generation times $\tau_d = 1/\alpha \ln (l_d/l_b)$ involve a logarithm, which appears to complicate the derivation of the resulting correlation functions. This problem can be circumvented by using the observation that the coefficient of variation in the birth size $l_b$ and the added mass at division $\Delta_d$ are small to develop the logarithm as a power-series. We can then derive approximate expressions for the correlation functions. Neglecting for simplicity the small noise in the instantaneous elongation rate $\alpha$, *i.e.*, $\alpha = \langle \alpha \rangle$, expansion of the doubling time $\tau_d$ and the doubling rate $1/\tau_d$ is as follows:

$$\tau_d = \frac{1}{\alpha} \ln \left( 1 + \frac{\Delta_d}{l_b} \right)$$

$$\simeq \frac{1}{\alpha} \left[ 1 + \frac{1}{2} \frac{\delta \Delta_d}{\langle \Delta_d \rangle} - \frac{1}{2} \frac{\delta l_b}{\langle l_b \rangle} \right] \quad (60)$$

$$\frac{1}{\tau_d} \simeq \alpha \left[ 1 - \frac{1}{2} \frac{\delta \Delta_d}{\langle \Delta_d \rangle} + \frac{1}{2} \frac{\delta l_b}{\langle l_b \rangle} \right], \quad (61)$$

where $\delta \Delta_d$ and $\delta l_b$ denote fluctuations with respect to their mean values and we used $\langle l_b \rangle = \langle \Delta_d \rangle$ (see (58)). Second-order terms will not be needed as they cancel out from the correlations computed below.

From Eq (60), the mean, the variance and the coefficient of variation of the doubling time $\tau_d$ are :

$$\langle \tau_d \rangle \simeq \frac{1}{\alpha} \quad (62)$$

$$\langle \tau_d^2 \rangle - \langle \tau_d \rangle^2 \simeq \frac{1}{4\alpha^2} \left[ \frac{\sigma_{\Delta_d}^2}{\langle \Delta_d \rangle^2} + \frac{\sigma_{l_b}^2}{\langle l_b \rangle^2} \right] = \frac{1}{\alpha^2} \frac{\sigma_{l_b}^2}{\langle l_b \rangle^2} \quad (63)$$

$$\Rightarrow \; CV_{\tau_d} = \frac{CV_{l_b}}{\ln 2} = \frac{CV_{\Delta_d}}{\sqrt{3} \ln 2}, \quad (64)$$

where Eq. 59 was used in the second equation. The ratios of the coefficients of variation are in excellent agreement with the experimental data presented in the reference [9].

Similarly, the mean value and the variance for the doubling rate $1/\tau_d$ read :

$$\left\langle \frac{1}{\tau_d} \right\rangle \simeq \alpha \; ; \quad \left\langle \frac{1}{\tau_d^2} \right\rangle - \left\langle \frac{1}{\tau_d} \right\rangle^2 \simeq \alpha^2 \frac{\sigma_{l_b}^2}{\langle l_b \rangle^2}, \quad (65)$$

and derive then

$$\frac{\left\langle \frac{l_b}{\tau_d} \right\rangle - \langle l_b \rangle \left\langle \frac{1}{\tau_d} \right\rangle}{\sigma_{1/\tau_d}^2} \simeq \frac{\langle l_b \rangle}{2\alpha}. \quad (66)$$

The expression (66) was used in Figure 1C of the reference [9] to fit the dependency of the initial size versus the number of divisions per hour, in given growth conditions. The fitting curve was written as $l_b = c_1 \frac{1}{\tau_d} + c_2$ and it is verified that the constant $c_1$ is then equal to the correlation (66).

We can also derive the correlation between the birth mass $l_b$ and the generation time $\tau_d$ as

$$\langle l_b \tau_d \rangle - \langle l_b \rangle \langle \tau_d \rangle \simeq -\frac{1}{2\alpha} \frac{\sigma_{l_b}^2}{\langle l_b \rangle} \quad (67)$$

$$\Rightarrow \; \frac{\langle l_b \tau_d \rangle - \langle l_b \rangle \langle \tau_d \rangle}{\sigma_{l_b} \sigma_{\tau_d}} \simeq -\frac{1}{2}, \quad (68)$$

where Eq 62 was used. The reason for the negative sign is intuitive: the elongation at a length $l$ proceeds at the rate $\alpha l$, *i.e.*, the longer the cell, the faster it grows. Therefore, if $l_b$ is larger/smaller than the mean it will take less/more time to complete the addition of the size $\Delta_d$ (independent of $l_b$).

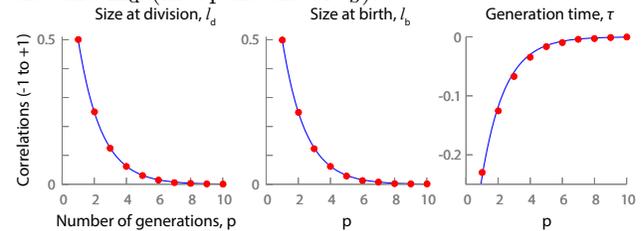

Figure 27: **Correlations in the adder model.** We simulate the process under the adder model and assume no correlation between the $\Delta_d$ of the mother and its siblings. The connected correlation function $\langle l_d^M l_d^{D(p)} \rangle - \langle l_d^M \rangle \langle l_d^{D(p)} \rangle$, divided by its value $\sigma_{l_d}^2$ for $p = 0$, is plotted as a function of the generation $p$ in panel (A). The line is the prediction derived in the text $2^{-p}$ while the dots are numerical values. The corresponding correlation for the newborn size $l_b$ is shown in panel (B). Finally, the connected correlation of the generation time defined in Eq. 69 as a function of the generations $p$ is shown in panel (c). The best fit for the decay is the exponential behavior $-0.43 \times 2^{-p}$, confirming the behavior derived in the text.

Finally, we can explicitly compute the decay of the correlations among the division times across generations. We indicate by $\tau_d^M$ the generation time of the mother and by $\tau_d^{D(p)}$ the generation time of a $p$-th generation descendent. For instance, daughters correspond to $p = 1$. We are interested in the behavior of the correlation

$$C_{\tau_d \tau_d}^{(p)} \equiv \frac{\langle \tau_d^M \tau^{D(p)} \rangle - \langle \tau_d^M \rangle \langle \tau_d^{D(p)} \rangle}{\sigma_{\tau_d}^2}, \quad (69)$$

Using Eq. (60), we can approximate the correlation by

$$C_{\tau_d \tau_d}^{(p)} \simeq -\frac{1}{[CV_{\Delta_d}^2 + CV_{l_b}^2]} \times \tag{70}$$

$$\left\langle \left( \frac{\delta \Delta_d^M}{\langle \Delta_d^M \rangle} - \frac{\delta l_b^M}{\langle l_b^M \rangle} \right) \frac{\delta l_b^{D(p)}}{\langle l_b^{D(p)} \rangle} \right\rangle, \tag{71}$$

where we have used that $\delta \Delta_d^{D(p)}$ is independent of all other fluctuations in the adder model and the expression (62) of the variance $\sigma_{\tau_d}^2$ in terms of the coefficients of variation. Note that the mean $\langle l_b^{D(p)} \rangle$ is the usual value of the mean (unaffected by the mother's fluctuations) as corrections would yield higher-order corrections.

The size at birth of a $p$-th generation descendant is $l_b^D = \frac{(l_b^M + \Delta_d^M)}{2^p} + \sum_{k=1}^{p-1} \frac{\Delta_d^{D(k)}}{2^{p-k}}$. Since the increments $\Delta_d^{D(k)}$ are independent of the mother's fluctuations in the adder model, we conclude that

$$C_{\tau_d \tau_d}^{(p)} \simeq -\frac{CV_{\Delta_d}^2 - CV_{l_b}^2}{CV_{\Delta_d}^2 + CV_{l_b}^2} \times \frac{1}{2^p}. \tag{72}$$

Using Eq. (59), we obtain for the correlation between mother and daughters $C_{\tau_d \tau_d}^{(1)} \sim -\frac{1}{4}$. The behavior (72) is consistent with the experimental data in the reference [9] and with the results of numerical simulations in Figure 27 (small corrections to the constant prefactor are due to the noise in the positioning of the septum and fluctuations in the instantaneous elongation rate $\alpha$; the constant agrees indeed with (72)).

Anticorrelations between the generation times of the mother and its descendants are intuitive. Consider a given initial size for the mother. As its generation time becomes longer, the mother will divide bigger and will then tend to have descendants with a bigger newborn size. The size at birth and the generation time of a cell tend to be anticorrelated. This follows intuitively from the definition $2^{\alpha \tau_d} = l_d/l_b = 1 + \Delta_d/l_d$ and the fact that the added size is independent of the initial size, *i.e.*, it takes less time to add the fixed amount $\Delta_d$ if the cell elongates faster [see Ref. 68 and 73 for a more formal proof]. Combining the two statements above leads to a conclusion that the division times of the mother and its descendants are anticorrelated.

**Correlations involving exponentials of the generation time.** We conclude this Section by computing some correlations involving exponentials of the generation time. The reason is that since $e^{\alpha \tau_d} = l_d/l_b$, correlations do not involve any logarithm and we can demonstrate anticorrelations without any hypothesis on the strength of the fluctuations. We first

show that the initial size and $e^{\alpha \tau_d}$ are anticorrelated in the adder model:

$$\langle l_b e^{\alpha \tau_d} \rangle - \langle l_b \rangle \langle e^{\alpha \tau_d} \rangle = \langle l_d \rangle - \langle l_b \rangle \left( 1 + \langle \Delta_d \rangle \left\langle \frac{1}{l_b} \right\rangle \right)$$

$$= \langle l_b \rangle \left( 1 - \langle l_b \rangle \left\langle \frac{1}{l_b} \right\rangle \right) \tag{73}$$

$$\leq 0. \tag{74}$$

Here, we used $\langle \Delta_d \rangle = \langle l_b \rangle = \langle l_d \rangle / 2$, and the inequality $\langle x \rangle \langle 1/x \rangle \geq 1$ holding for any positive-definite random variable $x$ (as it can be proved for example by the Cauchy-Schwarz inequality $1 = \left\langle \sqrt{x} \sqrt{1/x} \right\rangle \leq \langle x \rangle^{1/2} \langle 1/x \rangle^{1/2}$).

The exponentials of the division times of the mother and its daughters are also anticorrelated,

$$\left\langle 2^{\alpha^M \tau_d^M} 2^{\alpha^D \tau^D} \right\rangle = \left\langle \frac{l_d^M}{l_b^M} \frac{l_d^D}{l_b^D} \right\rangle$$

$$= 2 \left\langle \frac{l_d^D}{l_b^M} \right\rangle$$

$$= 2 \left\langle \frac{l_d^M/2 + \Delta_d^D}{l_b^M} \right\rangle$$

$$= 1 + 3 \langle \Delta_d \rangle \left\langle \frac{1}{l_b} \right\rangle, \tag{75}$$

where we have used again that in the adder model, the added size $\Delta_d$ is statistically independent of the birth size $l_b$. Subtracting then the disconnected contribution

$$\left\langle e^{\alpha^M \tau_d^M} \right\rangle \left\langle e^{\alpha^D \tau_d^D} \right\rangle = \left\langle \frac{l_d}{l_b} \right\rangle^2$$

$$= \left( 1 + \langle \Delta_d \rangle \left\langle \frac{1}{l_b} \right\rangle \right)^2, \tag{76}$$

and using again $\langle \Delta_d \rangle = \langle l_b \rangle$), we finally obtain

$$\left\langle e^{\alpha^M \tau_d^M} e^{\alpha^D \tau_d^D} \right\rangle - \left\langle e^{\alpha^M \tau_d^M} \right\rangle \left\langle e^{\alpha^D \tau_d^D} \right\rangle$$

$$= \langle l_b \rangle \left\langle \frac{1}{l_b} \right\rangle \left( 1 - \langle l_b \rangle \left\langle \frac{1}{l_b} \right\rangle \right)$$

$$\leq 0. \tag{77}$$

### 4.4. Collapse of the probability distributions and scaling forms

Experimental data show that the distributions of the added mass at division $\Delta_d$ from different nutrient conditions collapse onto each other when rescaled by their respective means (Figure 28) [9]. It has recently been observed that distributions of body sizes have a universal form across many species [579]. The universal form seems to be uniquely determined by the mean of the distribution, *i.e.*, when the various distributions are rescaled by their mean, they tend to collapse onto a unique curve similar to the *E. coli* data. This recent remark generalizes analogous, classical observations made for different bacterial populations and growth conditions [628, 629] to the extensive single-cell data for *E. coli* and *B. subtilis*. We review the theory that explains the property of scale-invariance that is common to all size distributions, viz. $l_b$, $l_d$ and $\Delta_d$. In other words, if one of the three spatial distributions is scale-invariant, the others inherit that property under the adder scenario for cell-size homeostasis.

The distribution $\rho_Z(z)$ of a (generic) random variable $Z$ is scale invariant if it has the form:

$$\rho_Z(z) = \frac{1}{\langle Z \rangle} \phi \left( \frac{z}{\langle Z \rangle} \right), \qquad (78)$$

where $\phi$ is an arbitrary non-negative normalized function. The statistics of $Z$ is supposed to change with the growth conditions, *e.g.*, quality if the nutrient environment, inhibition by antibiotics, etc. The non-trivial content of the scaling form (78) is that when conditions are changed, the distribution will be modified, yet its shape remains invariant when properly rescaled by the new mean value of $Z$. The form (78) also implies that, when conditions are varied, the normalized moments $\langle Z^p \rangle / \langle Z \rangle^p$ will remain constant. Finally, the scaling form (78) is equivalent to the statement that the Laplace transform $\mathcal{L}_Z(u)$ of the distribution $\rho_Z$ has the form $\mathcal{L}_Z(u) = \psi(u \langle Z \rangle)$, where $u$ is the Laplace transform variable and $\psi$ is an arbitrary function (respecting the general constraints for the Laplace transform of a probability distribution).

If the scaling form (78) holds for the distribution of the cell size either at division $l_d$ or at birth $l_b$, then it holds also for the other quantity and for the distribution of the added size at division $\Delta_d$. Indeed, if noise in the halving of the sizes at division is neglected, $l_d = 2l_b$. The distributions for $l_d$ and $l_b$ are then related as $\rho_b(l_b) = 2\rho_d(2l_b)$ and the scaling form of either one of the distributions clearly implies scale-invariance for the other. Moreover, since $l_d = l_b + \Delta_d$ and $\Delta_d$ is independent of $l_b$, we have

$$\mathcal{L}_{l_d}(u) = \mathcal{L}_{l_b}(u) \times \mathcal{L}_{\Delta_d}(u), \qquad (79)$$

where $\mathcal{L}$ indicates the Laplace transform of the respective probability distributions. Using $l_d = 2l_b$ we have $\mathcal{L}_{l_b}(2u)/\mathcal{L}_{l_b}(u) = \mathcal{L}_{\Delta_d}(u)$ and therefore the distribution of the added size inherits the scale invariance of $l_b$ (if the latter has it).

If noise in the halving in division of the size of daughters is included, the argument is slightly more involved. It is useful to use $l_d = \epsilon l_b$, where $\epsilon$ is a random variable centered around 2, and assume that the distribution of $\epsilon$ does not change as the means

$\langle l_b \rangle$ and $\langle l_d \rangle$ vary with growth conditions. Taking the logarithm of $l_d = \epsilon l_b$, we have again a sum and the Laplace transforms of the logarithms of the three variables are therefore related as in (79). The scale-invariant form (78) implies for the Laplace transform of the distribution of $\ln Z$ that $\mathcal{L}_{\ln Z} = \langle Z \rangle^{-u} \psi(u)$, where $\psi$ is arbitrary yet it does not contain $\langle Z \rangle$. Using that $\langle l_d \rangle = 2\langle l_b \rangle = \langle \alpha \rangle \langle l_b \rangle$ (which is valid in any growth condition), one can verify that if either $l_b$ or $l_d$ is scale-invariant, the other variable will inherit that property.

Finally, scale invariance (if present) is also inherited by the size distribution $\eta(l, \Delta, t)$. The equation for its dynamics is (49), which reduces to

$$\alpha \eta(l, \Delta) + \partial_t (v(l) \eta(l, \Delta)) + \partial_\Delta (v(l) \eta(l, \Delta)) \\ = -\gamma(\Delta) v(l) \eta(l, \Delta), \qquad (80)$$

in the steady-state with the growth rate $\alpha$ defined by (52). Taking $v(l) = \alpha l$ and using $\langle l \rangle \propto \langle \Delta \rangle$, one can verify that a scaling form for the steady-state size distribution $\eta(l, \Delta)$ is indeed compatible with (80).

A more explicit way to relate $\eta(l, \Delta)$ to the distributions of $\Delta$ and $l_b$ involves the integration of (80) along the characteristics and the tracking of cells from the current time $t$ back to their last division. For that purpose, it is convenient to introduce the age $\xi$ of a cell, as in Section 3.2, so that

$$\frac{dl}{d\xi} = v(l), \qquad \frac{d\Delta}{d\xi} = v(l), \qquad (81)$$

during the elongation of the cell. The initial size $l_b = l(0)$ and $\Delta(0) = 0$ are the initial conditions. The equation (80) is rewritten as

$$\frac{d\eta(l(\xi), \Delta(\xi))}{d\xi} = -F(l(\xi), \Delta(\xi)) \eta(l(\xi), \Delta(\xi)), \qquad (82)$$

where $F(l, \Delta) = \alpha + \partial_l v(l) + \gamma(\Delta) v(l)$. We can then track each cell back to its birth:

$$\eta(l, \Delta) = \eta(l - \Delta, 0) e^{-\int_0^\xi d\xi' F(l(\xi'), \Delta(\xi'))} \\ = \eta(l - \Delta, 0) e^{-\int_0^\Delta d\Delta' \frac{1}{v(l(\Delta'))} F(l(\Delta'), \Delta')} \\ = \eta(l - \Delta, 0) e^{-\int_0^\Delta d\Delta' \gamma(\Delta')} e^{-\int_0^\Delta d\Delta' \frac{\alpha + \partial_l g(l(\Delta'))}{g(l(\Delta'))}}. \qquad (83)$$

For the exponential elongation rate $v(l) = \alpha l$, equation (53) gives $\alpha = \alpha$ and we have that

$$\eta(l, \Delta) \propto \rho_b(l - \Delta) \left( 1 - \int_0^\Delta dx \rho_{\Delta_d}(x) \right) e^{-\int_0^\Delta dx \frac{2}{l(0) + x}} \\ = \rho_b(l - \Delta) \left( 1 - \int_0^\Delta dx \rho_{\Delta_d}(x) \right) \left( 1 - \frac{\Delta}{l} \right)^2, \qquad (84)$$

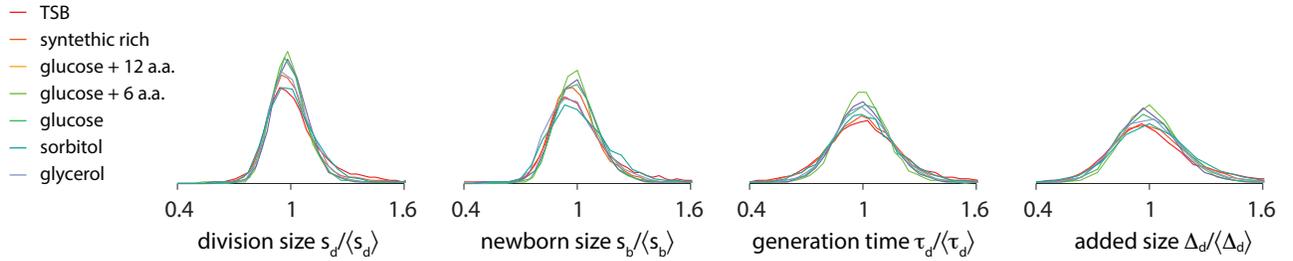

**Figure 28: Collapsed distributions of physiological paramters.** Distributions of $l_d$, $l_b$, $\tau$ and $\Delta_d$ from different growth conditions show scale invariance, *i.e.*, collapse when rescaled by theri respective means [9].

where we have used (51) to express $e^{-\int_0^{\Delta} d\Delta' \gamma(\Delta')}$ in terms of $\rho_{\Delta_d}$. It is immediately verifiable that if $\rho_b$ and $\rho_{\Delta_d}$ are scale-invariant, so, too, will the steady-state size distribution $\eta(l, \Delta)$ be. Integrating (84) over $\Delta$, we obtain for the marginal $\eta(l)$ at the steady state:

$$\eta(l) \propto \int_0^l dx \, \rho_b(l-x) \left(1 - \int_0^x dy \rho_{\Delta_d}(y)\right) \left(1 - \frac{x}{l}\right)^2 \, . \tag{85}$$

### 4.5. Other models proposed for the origin of the adder and the consideration of chromosome replication

Several models have been proposed to explain the molecular origin of the adder principle. So far all these models are a variation of a threshold concept that we discussed earlier for the onset of chromosome replication. For example, Harris and Theriot proposed a threshold hypothesis that a fixed amount of excess surface materials (*e.g.*, precursors for peptidoglycan synthesis) must accumulate to trigger cell division [473]. The Alfridge lab attempted to link the adder between cell divisions vs. between consecutive replication initiations in *Mycobacteria* [630]. The work from the Elf lab provided valuable insights by experimentally demonstrating a specific example that breaks the adder [631]. They results are based on the following three key observations:

(i) Initiation mass is constant at the single-cell level and uncorrelated with the birth size.

(ii) The cell-cycle duration $\tau_{cyc}$ is constant at the single-cell level, uncorrelated with the birth size and the initiation mass.

(iii) In slow growth conditions where cell cycles do not overlap, $\tau_{cyc}$ is linearly proportional to the generation time $\tau$. This together with the invariant initiation mass leads to deviation from the adder, because size at division is given by $S_d = S_i 2^{\tau_{cyc}/\tau_d} = S_i \times const.$ In other words, slow growing cells divide when they reach a constant size, similar to the sizer.

However, the deviation from the adder in slow growing *E. coli* cells does not necessarily mean that they switch size control mechanism based on the growth condition. Rather, this sizer-like behavior is likely an accidental consequence of global biosynthesis being limiting in slow growing cells, whereby both global biosynthesis and cell cycle progression slow proportionally leading to $\tau_{cyc}/\tau = const.$ A similar relationship between $\tau_{cyc}$ and $\tau$ has already been known for slow growing populations since the 1960s by Helmstetter and colleagues [252, 632].

### 4.6. Control of variability ("noise") and hierarchy of physiological controls

In Section 2.2.7, we discussed various models of initiation control. It is important to remember that initiation control is subject to biological fluctuations, thus note the degree of variability observed in both the single-cell and population level measurements. Possible biological consequences of variability in initiation control, and the tightness of coupling with division, has already been pointed out by Koch [256]:

> Our results, particularly the occurrence of the 1n and 3n cells, do suggest, however that cell division is not a necessary or sufficient condition for the initiation of a round of chromosomal replication. Neither must cell division await the start of a new round of DNA synthesis in both daughter chromosomes.

(Here, 1n and 3n cells mean that cells that show an odd number of replication origins, instead of the multiples of 2).

For the control of coefficient of variability, insights from plasmid copy number control are invaluable, especially the "Kinetic proof reading" type view of initiation control by Paulsson and Ehrenberg [633]. The basic thesis of this work is that cells must use multiple steps to trigger replication initiation to reduce the standard deviations of plasmid copy number distributions. $\sigma \to \sigma/\sqrt{n}$, where $n$ is the number of steps. R1 is a low-copy number plasmid, whereas ColE1 is a high-copy number. R1 and ColE1 both employ multi-step processes for initiations but they are different in important details so that the copy number control of the two plasmids exhibits different strategies.

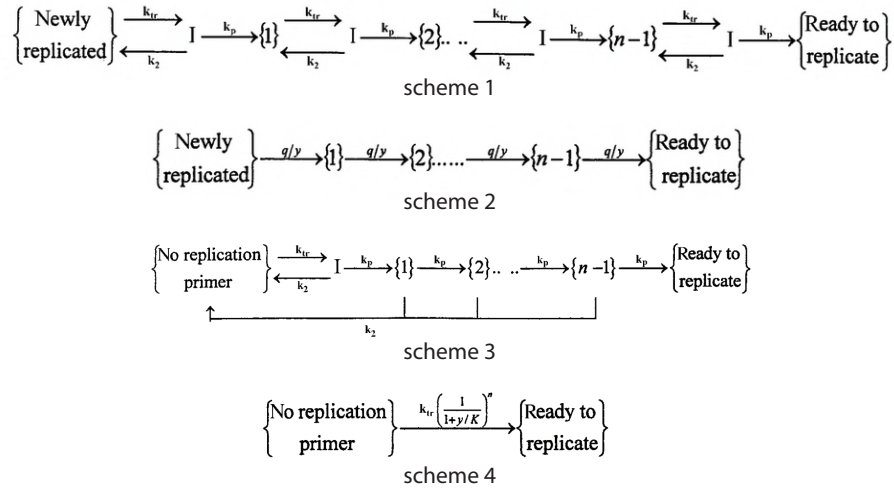

Figure 29: **Schemes of initiation control in plasmid.** The figure is adapted from [633].

**R1.** R1 initiation is controlled by two genes. RepA isSection the initiator. A large number of RepA must accumulate at the origin site oriR1. This causes conformational changes at oriR1 and allows initiation. CopA is an antisense RNA of RepA, and expressed constitutively and it has a short half-life. Thus, CopA concentration is roughly proportional to the plasmid concentration. This makes RepA synthesis rate inversely proportional to the plasmid concentration. Figure 29 Scheme 1 describes the control. $k_{tr}$ is the transcription rate for RepA and $k_p$ is the translation rate for RepA, and $k_2$ is the CopA anti-sensing kinetic parameters. Scheme 2 is a simplified version of Scheme 1, where $p = k_p/(k_p + k_2)$ and $y = [\text{CopA}] = [\text{plasmids}]$. Thus, only two parameters determine the model: $y$ the copy number concentration ($q$ is a biochemical parameter) and the total number of "hyperbolic" steps. The main advantage of this control scheme is that $CV = 1/\sqrt{n}$, where $n$ is the total number of steps to go from *newly replicated* to *ready to replicate*. Here, $n$ has to be determined by fitting the actual data, and should be considered as the number of effective steps. As a result, CV decreases as $n$ increases and the plasmid copy number decreases. The latter is particularly important for low-copy number plasmids where stability of plasmid segregation is important (as in the case for R1).

**ColE1.** ColE1 is a high-copy number plasmid. Replication initiation is controlled by a single copy of the initiator molecule RNA II. Antisense RNA (RNA I) inhibits the nascent RNA II at many subsequent transcription steps (experimental work by Tomizawa [634]). The ColE1 copy-number control scheme is illustrated in Figure 29 Scheme 3. Here, $k_{tr}$ is the rate constant for initiation of transcription. $k_p$ is the intermediate transcription rate, whereas $k_2$ is the inhibition rate by antisense RNA I. Thus, $k_2$ is proportional to RNA I concentration, and therefore to plasmid concentration, $y$. K is a compound inhibition constant when simplified from Scheme 3.

Differences in biological consequences between Scheme 2 (*e.g.*R1) and Scheme 4 (*e.g.*ColE1) is shown in Figure 29. Analytical solutions of these schemes are difficult to obtain, but simulations are fairly straightforward and offer general insight. Both schemes reduce CV as the number of intermediate steps $n$ increases. Scheme 2 (R1) has additional feature of avoiding no replications for plasmid copy number = 1 (which would be disastrous).

### 4.7. Spatial regulation: when absolute size matters

In this section, we have mainly focused on the adder principle and related issues. This may give the wrong impression that absolute size is unimportant. On the contrary, *E. coli* employs several apparatuses that have an intrinsic length scale. We have already discussed the constancy of initiation mass (Section 4.5), and we will go deeper into the subject in Section 6 that the initiation mass in fact remains invariant under extensive growth inhibition.

From cell biological point of view, cell division involves various mechanisms for spatial regulations. For example, nucleoid occlusion states that cells cannot divide the volume occupied by the chromosome (or the nucleoid) (Section 2.3.4). Thus, the physical size of the chromosome can provide a natural length scale for the size of the cell. Another important molecular apparatus is the Min system (Section 2.3.4, Figure 30). The Min system consists of three species of proteins (MinCDE), which oscillate along the long axis of the *E. coli* cells. The Min oscillations have been beautifully described in the context of dynamic instability. The basic idea is that MinD and MinE constitute a reaction-diffusion system, such that their dynamic instability leads to oscillations with a well-defined lengthscale for the standing wave. When time-averaged, the concentration of MinD is highest at the cell poles, which is important because MinC is the inhibition of FtsZ ring formation and binds to MinD. In other words, the Min oscillations dynamically guide MinC and prevent cell division at the pole. The quantitative description of the Min system is one of the most successful examples in theoretical biological physics with close dialogue with experimentation. Pioneers in the field include Piet de Boer & Hans Meinhardt and Martin Howard [441, 442]. For the readers who want to know more about this fascinating subject, we recommend a recent treatise by Erwin Frey and Cees Dekker and references therein [635, 636]. For a more general review on cell size regulation by sizer-like mechanisms, we suggest [637].

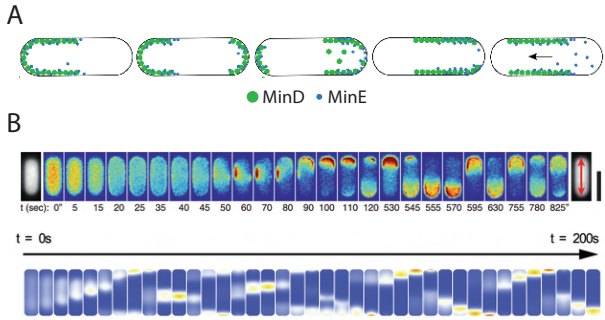

Figure 30: **Min oscillation in *E. coli*. A.** The Min oscillation has well-defined wavelength due to the reaction-diffusion of MinD and MinE. If the cell length and the wavelength of the MinD oscillation are about the same, the time average concentration of MinD will be highest at both poles. Figure is adapted from the work of the Kruse group [638]. **B.** The dynamic behavior of Min oscillation has been well manipulated and predicted in perturbed cell geometry using micro-chamber. Top: fluorescent cell image showing the MinD oscillation. Bottom: simulation. Figures are adapted from the work of the Frey group and Dekker lab [636].

# 5. Modern bacterial physiology, Part I: proteome 'sectors'

From this Section, we will turn toward contemporary views of bacterial physiology by reviewing three works recently published. Each highlights Monod's maxim that once suitable state variables are chosen, system behavior reveals itself in surprisingly-simple 'laws.' The 'law' of the present section (Section 5) is that proteome partitioning imposes strong constraints on gene expression leading to what can be called 'emergent' or 'indirect' gene regulation. The 'law' of Section 6 unites the initiation mass – unit cell, doubling rate and cell cycle time into a remarkably robust expression for the mass-per-cell over a range of physiological perturbations. The 'law' of Section 7 comes from recognizing that the temporal organization and causal relations amongall the internal de novo synthesis processes enables balanced growth and determines the growth rate. Finally, in Section 8 we speculate on how the 'laws' of proteome partition, cell-size and scheduling of cell-replication can be combined to determine the fundamental relationship between cell physiology and cell cycle control.

## 5.1. Proteome partitioning constraints on gene expression

The work of Neidhardt & Magasanik (Figure 7) was instrumental in establishing the catalytic role of the bacterial ribosome in protein synthesis. We now know that the positive linear correlation between ribosome abundance and the nutrient-mediated growth rate has deeper implications for the coupling between protein expression and growth [286, 639, 640]. Returning to the results of Neidhardt & Magasanik [58] (see Section 2.2.3), it will simplify the analysis if we convert the mass of total RNA, $M_{\text{RNA}}$, into mass of ribosomal proteins, $M_{\text{rProtein}}$,

$$M_{\text{RNA}} \ \frac{0.85\text{g rRNA}}{1\text{g RNA}} \ \frac{1\text{g rProtein}}{2\text{g rRNA}} = M_{\text{rProtein}} \quad (86)$$

The rate of protein synthesis in exponential growth is,

$$\frac{dM_{\text{P}}}{dt} = \lambda M_{\text{P}}.$$

Not all ribosomes will be actively synthesizing protein; suppose there is a number $N_{\text{Rb}}^0$ of inactive ribosomes, then

$$\frac{dM_{\text{P}}}{dt} = \lambda M_{\text{P}} = k(N_{\text{Rb}} - N_{\text{Rb}}^0),$$

where $k$ is the rate of protein synthesis per ribosome. We can further convert from ribosome *numbers* to ribosome *mass* via the mass per ribosome, $m_{\text{Rb}}$: $M_{\text{rProtein}} = m_{\text{Rb}} \times N_{\text{Rb}}$; isolating the growth rate $\lambda$ in the steady-state protein accumulation equation,

$$\frac{\lambda}{(k/m_{\text{Rb}})} = \frac{M_{\text{rProtein}}}{M_{\text{P}}} - \frac{M_{\text{rProtein}}^0}{M_{\text{P}}},$$

or,

$$\frac{\lambda}{\kappa_{\text{T}}} = \phi_{\text{R}} - \phi_{\text{R}}^{\min} \quad \Longrightarrow \quad \phi_{\text{R}} = \frac{\lambda}{\kappa_{\text{T}}} + \phi_{\text{R}}^{\min}, \quad (87)$$

where $\kappa_{\text{T}} = k/m_{\text{Rb}}$ is proportional to the average speed of a translating ribosome, and $\phi_{\text{R}}$ is the *protein mass fraction* of ribosomal proteins. Equation 87 is an empirical relationship that is observed in *E. coli* under conditions of moderate-to-fast growth (doubling times shorter than about 90 minutes, Figure 31A, *circles*) [251]; the *interpretation* that the empirical parameter $\kappa_T$ is proportional to the average speed of a translating ribosome is testable by repeating the same experiment with mutants that synthesize protein more slowly. Indeed, the linear relation between RNA/protein and the growth rate is maintained, though the slope increases with the severity of the mutation (Figure 31A, *triangles*). The slope correlates very well with the *in vitro* translation rate for ribosomes isolated from these mutants [251].

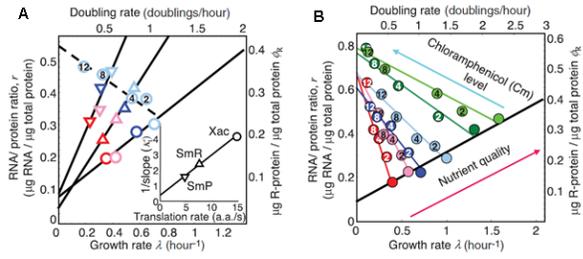

**Figure 31: Empirical growth laws in ribosome abundance. A.** When growth rate is modulated by changes in nutrient quality, the ribosome protein mass fraction exhibits a positive linear correlation with growth rate (circles). For mutants with reduced peptide elongation rate (upward triangle [moderate reduction], downward triangle [severe reduction]), the linear correlation is preserved, although the slope increases. The reciprocal of the slope correlates very well with the *in vitro* translation rates of the mutants. **B.** For a given nutrient environment (circles on solid line), when growth rate is reduced using a translation-inhibiting antibiotic (darker symbols=higher concentration), the ribosome protein mass fraction exhibits a strong negative correlation with growth rate (colored lines).

The corroboration of the interpretation of the phenomenological parameter $\kappa_T$ can be pushed further. For example, from the mutant data (Figure 31A, *eg.* blue symbols) it looks as though there is a family of lines with negative slope correlating ribosome abundance and growth rate *under translation inhibition* (Figure 31A, dashed lines). That same behaviour can be obtained from the wildtype under the action of a translation-inhibiting antibiotic (Figure 31B, *circles* – darkest symbols have the highest antibiotic concentration). In contrast to the case where growth is modulated by changes in the nutrient quality (Figure 31A, solid line), with growth modulated by translational inhibition, the ribosome abundance is *negatively correlated* with growth rate (Figure 31B; coloured lines). In fact, the correlation is so strong that the data can be adequately described by a straight line,

$$\phi_R = -\frac{\lambda}{\kappa_N} + \phi_R^{\max}, \qquad (88)$$

where here the empirical parameter $\kappa_N$ changes with the nutrient quality of the medium (*i.e.*, the drug-free growth rate), because the maximum intercept, $\phi_R^{\max}$, is only weakly growth-medium dependent over this range of growth rates. When the doubling time exceeds about 90 minutes in *E. coli* ($\tau_d > 90$ minutes, $\lambda < 0.5$/h), the RNA/protein ratio as a function of growth rate $\lambda$ begins to exhibit deviations from linearity (similar to what is evident in Neidhardt and Magasanik's data from *Aerobacter aerogenes*, Figure 7). The deviation from linearity is largely attributable to a growth-rate dependent decrease in the protein translation rate [641]. Furthermore, in the regime of slow growth, the weak growth rate dependence in the translational-inhibition intercept $\phi_R^{\max}$ is no longer negligible [642], and a more detailed partitioning of the proteome is required [643].

In moderate-to-fast growth rates (doubling times less than about 90 minutes), we have two empirical constraints on ribosome abundance and growth rate, Eq.s 87 and 88,

$$\phi_R = \frac{\lambda}{\kappa_T} + \phi_R^{\min} \quad \text{and} \quad \phi_R = -\frac{\lambda}{\kappa_N} + \phi_R^{\max},$$

that are simultaneously true under conditions where the growth is modulated by changes in nutrient quality or translational inhibition. Eliminating the ribosome abundance, $\phi_R$, we arrive at an expression for the growth rate $\lambda$ written as a parametric function of the two empirical coefficients $\kappa_T$ (characterizing the translational capacity) and $\kappa_N$ (characterizing the nutrient-processing capacity),

$$\lambda = \left(\phi_R^{\max} - \phi_R^{\min}\right) \frac{\kappa_T \kappa_N}{\kappa_T + \kappa_N} \qquad (89)$$

$$= \frac{\left(\phi_R^{\max} - \phi_R^{\min}\right)}{\frac{1}{\kappa_T} + \frac{1}{\kappa_N}} \equiv \frac{\phi_{\max}}{\frac{1}{\kappa_T} + \frac{1}{\kappa_N}}, \qquad (90)$$

where $\phi_{\max} = \phi_R^{\max} - \phi_R^{\min}$ is the dynamic range of the ribosomal proteins. The parameterization of the growth rate $\lambda$ in terms of the phenomenological constants $\kappa_T$ and $\kappa_N$ can be rearranged into a form analogous to Monod's relation, (1),

$$\lambda = \lambda_{\max} \frac{\kappa_N}{\kappa_N + \kappa_T}, \qquad (91)$$

where $\lambda_{\max} = \kappa_T \phi_{\max}$ is the maximum limiting growth rate in the perfect nutrient environment ($\kappa_N \to \infty$). When compared with Monod's relation (1), the phenomenological constant $\kappa_N$ characterizing the *quality* of the nutrient environment plays the role analogous to the concentration of a growth-limiting substrate $S$. If $\kappa_N$ is replaced by,

$$\kappa_N \mapsto \kappa_N \frac{S}{S + \tilde{K}_D}, \qquad (92)$$

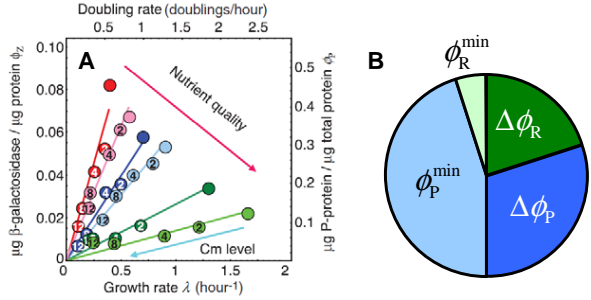

Figure 32: **Proteome partitioning constraints. A.** The protein mass fraction of an unregulated, or 'constitutive' protein exhibits near mirror symmetry with the growth dependence of the ribosomal proteins (*cf.* Figure 31B). **B.** The simplest constraint linking ribosomal and non-ribosomal proteins is to imagine the total protein mass (proteome) partitioned into two exclusive protein types: ribosome-affiliated $R$-proteins, and all other ($P$-proteins), each with a growth-dependent (dark) and growth-independent (light) component.

then Monod's relation (1) is obtained, but with an explicit dependence on the growth medium in the Michaelis constant, $K_D = \tilde{K}_D/(1 + \kappa_N/\kappa_T)$ [286].

The empirical constraints, (87) and (88), are at the level of ribosome abundance; in many applications, it is the abundance of non-ribosomal proteins (regulators, enzymes, and other genes of interest) that are the primary focus. The first step to understanding how growth-rate dependent effects are filtered through complex regulatory networks is to determine the growth-rate dependence in the expression of an *unregulated* (also called 'constitutively expressed') protein [644]. For example, we can design a strip of DNA that encodes a reporter enzyme (*i.e.*, an enzyme that is easy to measure, but not needed for growth) in such a way that the transcription and translation of this DNA does not respond to any *direct* regulation. When we measure the protein mass fraction of this enzyme, we find that the empirical constraints on ribosome mass fraction are reflected with almost perfect mirror symmetry in the mass fraction of an unregulated protein (Figure 32A). If ribosomal proteins go up, unregulated proteins go down, and *vice versa*.

The observation that identical (though anti-correlated) empirical constraints hold for unregulated protein expression (Figure 32A), suggests that the total proteome can be minimally partitioned into two exclusive protein types: ribosomal proteins with mass fraction $\phi_R$, and all other proteins, collectively referred to as 'metabolic proteins', with mass fraction $\phi_P = 1 - \phi_R$. To keep the partitioning hypothesis as general as possible, we note that the ribosomal protein fraction $\phi_R$ exhibits a growth-rate independent offset $\phi_R^{\min}$, and so we allow the same for the metabolic protein fraction

$\phi_P$. We then decompose the two protein sectors into growth-dependent and independent fractions,

$$\phi_R = \Delta\phi_R(\lambda) + \phi_R^{\min}, \phi_P = \Delta\phi_P(\lambda) + \phi_P^{\min}, \quad (93)$$

where the growth-dependent fractions are constrained,

$$\Delta\phi_R(\lambda) + \Delta\phi_P(\lambda) = 1 - \phi_R^{\min} - \phi_P^{\min} \equiv \phi_{\max}, \quad (94)$$

such that increase in one comes at the expense of a decrease in the other (Figure 32B). In terms of the proteome constraint, the empirical growth laws, 87 and 88, can be re-written as,

$$\lambda = \kappa_T \Delta\phi_R, \text{ and } \lambda = \kappa_N[\phi_{\max} - \Delta\phi_R] = \kappa_N \Delta\phi_P, \quad (95)$$

where we have identified $\phi_{\max} = \phi_R^{\max} - \phi_R^{\min}$. Growth-dependent partitioning of the proteome goes back to the seminal experimental work by Neidhardt's group in the late 1970's using 2D gel electrophoresis to separate proteins and quantify abundance [64]. A classification of protein groups based upon collective response in their growth-rate-dependent mass fraction upon perturbation of the environment has the advantage of linking gene expression directly to physiology [642, 645]. More recently, using primarily mass spectrophotometry, this type of whole-proteome profiling has been used to elucidate the regulation of the carbon-utilization hierarchy in E. coli [643], and investigate the impact of large-scale genome reduction on the physiology of B. subtilis [646].

Constraints imposed by proteome partitioning and shared cellular resources can have profound impact on networks of direct regulation, for example modifying the apparent susceptibility to antibiotics [647], or leading to complex growth-mediated feedback loops [640, 648]. How proteome partitioning constraints impact cell cycle processes, including DNA replication rate and chromosome segregation, remains one of the outstanding problems in bacterial physiology.

### 5.2. Ohmics: Electrical circuit analogies for proteome constraints

The analysis of complex systems is aided by rule-based empirical correlations: famous examples include the Woodward-Fieser rules (organic chemistry) [649, 650], Mendel's principles of heredity (genetics) [651], and Boyle's law (thermodynamics) [652]. In the exponential growth of bacteria, strong empirical correlations emerge between the macromolecular composition of the cell and the growth rate [2, 10, 251, 298, 517, 639, 653] based on reliable quantifications on biosynthesis rate under various growth conditions [82–84, 86, 229, 512, 654–663]. At the level of protein mass fraction, these correlations are mathematically identical to a voltage source connected to two series resistors; growth rate plays the role of current and the protein mass fraction plays the role of voltage drop across the resistors (Figure 33).

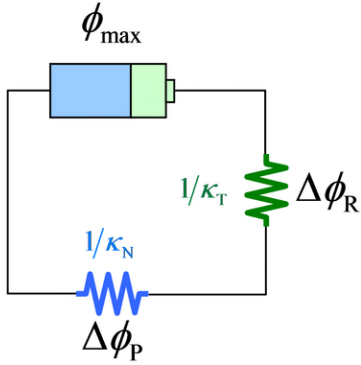

Figure 33: **Electrical circuit analogies.** The two empirical constraints on steady-state ribosome abundance (Figure 31B) and the proteome partitioning constraint (Figure 32B) are mathematically identical to the current flow through two resistors in series, $\lambda = \kappa_T \Delta\phi_R = \kappa_N \Delta\phi_P$, $\Delta\phi_R + \Delta\phi_P = \phi_{max}$, with growth rate $\lambda$ playing the role of current and $1/\kappa_T$ and $1/\kappa_N$ playing the role of resistance in each resistor. The voltage drops across the resistors are given by the mass fractions $\Delta\phi_R$ and $\Delta\phi_P = \phi_{max} - \Delta\phi_R$, respectively.

In contrast to ordinary electrical circuits, it is not possible to examine the open- and short-circuit analogues of the physiological circuit; the open-circuit corresponds to cell death, and the short-circuit would require infinitely-fast reaction rates. What is done instead is to modulate one of the conductances in the phenomenological model and extrapolate to extreme points. Over a range of growth rates, changes in nutrient quality appear to affect only the nutrient capacity $\kappa_N$, leaving the other parameters approximately unchanged (Figure 31B; colored lines with negative slope). By varying the nutrient capacity $\kappa_N$, the translational capacity $\kappa_T$ is the slope of the total ribosomal protein fraction $\phi_R$, and $\phi_R^{min}$ is the offset extrapolated in the limit $\lambda \to 0$ (Figure 31B; solid black line with positive slope). Similarly, modulating growth by translational inhibition (varying the translational capacity $\kappa_T$), allows estimation of the nutrient capacity $\kappa_N$ and the offset $\phi_P^{min}$ from the relation between ribosomal protein fraction $\phi_R$ and growth rate $\lambda$ via the proteome partition constraint $\Delta\phi_R + \Delta\phi_P = \phi_{max}$.

This analogy provides several conceptual insights into the growth constraints imposed by the demands of protein synthesis along with the overall proteome partitioning constraint. For example, the synthesis of an unnecessary protein (*e.g.*, industrial bioproducts) produces a growth-rate defect and changes in the $R$- and $P$-protein fractions that corresponds to changing the voltage source $\phi_{max} \to \phi_{max} - \phi_{OE}$, where $\phi_{OE}$ is the protein mass-fraction of the over-expressed protein (Figure 34A). Furthermore, the 'metabolic' resistor characterized by conductance $\kappa_N$ can be subdivided further, for example into a catabolic network (with efficiency characterized by $\kappa_C$), and anabolic network (with efficiency characterized by $\kappa_A$) and a remaining unassigned fraction (Figure 34B). Indirect gene regulation imposed by changes in $\kappa_C$ and $\kappa_A$, and the conceptual simplicity of the circuit framework, was instrumental in unraveling a longstanding mystery in bacterial physiology called *carbon catabolite repression* [643], referring to the apparent hierarchical utilization of different carbon sources in *E. coli*, and other organisms.

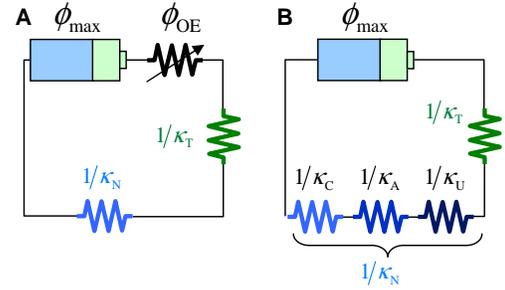

Figure 34: **Consequences of the circuit analogy. A. Over-expression.** The over-expression of an unnecessary protein produces a growth defect and re-partitioning of the coarse-grained proteome that is consistent with a voltage-sink $\phi_{OE}$ (corresponding to the protein mass fraction of the useless protein) in series with the cannonical circuit. **B. Further partitioning of the proteome.** The $P$-protein fraction can be further subdivided into functional catagories, for example catabolic proteins, anabolic proteins and the remainder (with processing efficiency characterized by $\kappa_C$, $\kappa_A$ and $\kappa_U$, respectively).

The partitioning of the 'metabolic' $\kappa_N$ resistor suggests that carbon-source co-utilization, presumably requiring a subset of non-overlapping catabolic pathways, could be represented as two resistors in parallel (Figure 35A). Mandelstam first observed a negative linear correlation between $\beta$-galactosidase expression and growth rate in different carbon sources [664]. You *et al.* [643] propose that $\beta$-galactosidase expression serves as a proxy for the catabolic protein sector (corresponding to the potential drop across the $1/\kappa_C$ resistor in the circuit analogy), and that the observed negative correlation is a consequence of proteome partitioning constraints. Extrapolating to the limit of vanishing catabolic sector (*i.e.*, $\kappa_C \to \infty$; Figure 35B) allows the contribution to growth by the non-catabolic ('notC') proteins to be expressed in terms of the maximum carbon-limited growth rate $\lambda_C$,

$$\kappa_{notC} = \frac{\kappa_T \lambda_C}{\lambda_{MAX} - \lambda_C}, \qquad (96)$$

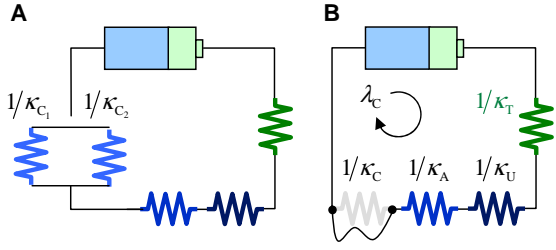

Figure 35: **Carbon co-utilization. A. Catabolic networks in parallel.** Growth on two carbon sources requiring a non-overlapping set of catabolic enzymes for processing can be represented in the circuit analogy as a pair of *parallel* resistors. **B. Characterization of the background circuit.** One of the strengths of the circuit analogy is that the background circuit can be characterized by growing in a variety of single-carbon sources (changing $\kappa_C$), and extrapolating the growth rate to the short-circuit equivalent with growth rate $\lambda_C$.

where $1/\kappa_{\text{notC}} = 1/\kappa_A + 1/\kappa_U$ and $\lambda_{\text{MAX}}$ is the maximum growth rate attained in the limit $\kappa_N \to \infty$. Denoting by $\lambda_i$ the growth rate in each carbon source individually,

$$\lambda_i = \frac{\phi_{\max}}{\frac{1}{\kappa_T} + \frac{1}{\kappa_{\text{notC}}} + \frac{1}{\kappa_{C_i}}}, \qquad (97)$$

the overall growth rate during co-utilization is,

$$\frac{\lambda}{\lambda_C} = \frac{\hat{\lambda}_1 + \hat{\lambda}_2 - 2\hat{\lambda}_1\hat{\lambda}_2}{1 - \hat{\lambda}_1\hat{\lambda}_2}, \qquad (98)$$

where $\hat{\lambda}_1 = \lambda_1/\lambda_C$ and $\hat{\lambda}_2 = \lambda_2/\lambda_C$ are the individual growth rates $\lambda_i$ normalized to the carbon-limited maximum $\lambda_C$. The circuit parameterized in this way quantitatively reproduces the coarse-grained proteome partitioning and growth rate observed during growth on two carbon sources [665].

Extending the circuit analogy, the empirical growth laws can be connected to the detailed workings of metabolism via coarse-graining akin to a Thévenin equivalent circuit. Metabolic reactions are typically catalyzed by protein enzymes. Consider, for example, a reaction converting substrate $a$ into product $b$, catalyzed by an enzyme $E$. Motivated by the growth laws which relate the growth rate to the active protein fractions $\Delta\phi_R$ and $\Delta\phi_P$, we assume that the reaction rate is proportional to a growth-rate dependent fraction of the enzyme $\Delta\phi_E$,

$$j_E = \kappa_E \Delta\phi_E,$$

where $\kappa_E$ is the effective catalytic rate constant suitably converted to units of mass fraction. The effective catalytic rate constant includes contributions from substrates, products and cofactors that depends upon enzyme properties as well as growth-condition-dependent metabolite concentrations.

The total mass fraction of the enzyme $\phi_E$ may include a growth-independent offset $\phi_E^{\min}$ that does not participate in driving the reaction flux. The total enzyme mass fraction is then written in terms of the reaction flux $j_E$, the catalytic rate constant $\kappa_E$ and the growth-independent offset $\phi_E^{\min}$ as,

$$\phi_E = \frac{j_E}{\kappa_E} + \phi_E^{\min}.$$

In the electrical circuit analogy, the potential drop across this reaction is decomposed into a potential drop $\Delta\phi_E$ across a resistor with conductance $\kappa_E$ and the voltage sink $\phi_E^{\min}$. Imagine a whole network of reactions, interconnected with one-another. Visualized as a graph, each substrate is represented by a node, and each reaction is represented by an edge. Assuming the same enzyme-catalyzed reaction rate as above, the protein cost (or potential drop) along the $l^{th}$ edge is written as,

$$\phi_{E_l} = \frac{j_{E_l}}{\kappa_{E_l}} + \phi_{E_l}^{\min},$$

where $j_{E_l}$ is the reaction flux along that edge. An important distinction between electrical elements and enzyme-mediated reactions is that Ohm's law follows directly from the physical properties of a resistor, whereas in a biological system, flux-balance is achieved through regulation.

The metabolic protein fraction $\phi_P$ is a sum of all of the individual metabolic enzymes, $\phi_P = \sum_l \phi_{E_l}$. We will use a reduction similar to a Thévenin equivalent circuit to relate the empirical parameters $\phi_{\max}$ and $\kappa_N$ to the parameters $\phi_{E_l}^{\min}$ and $\kappa_{E_l}$ that characterize the reactions in the network. The growth-rate independent offset $\phi_P^{\min}$, which in this case is simply the sum of the individual offsets $\phi_P^{\min} = \sum_l \phi_{E_l}^{\min}$, is found by taking the limit $\kappa_T \to 0$. As the conductance $\kappa_T$ vanishes, the potential drop $\Delta\phi_P$ likewise vanishes, $\Delta\phi_P \to 0$, and the potential drop across the protein synthesis machinery reaches a maximum, $\Delta\phi_R \to \phi_{\max}$. We can then infer the coarse-grained effect of the offsets $\{\phi_{E_l}^{\min}\}$ as parameterized by $\phi_P^{\min}$ via the proteome constraint $\phi_{\max} = 1 - \phi_R^{\min} - \phi_P^{\min}$ (Figure 32B). In principle, the enzyme-offsets $\{\phi_{E_l}^{\min}\}$ (and by extension the coarse-grained parameter $\phi_{\max}$) should exhibit some growth-medium dependence arising from different strategies for processing external nutrients. For *E. coli*, however, the phenomenological parameter $\phi_{\max}$ exhibits remarkably-little dependence over a large range of growth rates (see Figure 31B).

With the offsets eliminated, the system is reduced to a purely resistive network. Algebraically, the network can be formally reduced to a single resistor, and in this way the explicit correspondence between the phenomenological nutrient capacity $\kappa_N$ and the catalytic rate constants $\{\kappa_{E_l}\}$ can be determined. For a given network connectivity and distribution of effective catalytic rate constants $\kappa_{E_l}$, invoking flux-balance at every node in the network allows estimation of each reaction flux $j_{E_l}$, and consequently the active enzyme cost associated with the reaction $\Delta\phi_{E_l}$. The nutrient capacity $\kappa_N$ is by definition the proportionality constant between the growth rate $\lambda$ and the active metabolic protein fraction, $\kappa_N = \lambda / \sum_l \Delta\phi_{E_l}$, or more explicitly in terms the effective catalytic rate constants,

$$\kappa_N^{-1} = \sum_l \frac{j_{E_l}}{\lambda} \kappa_{E_l}^{-1}.$$

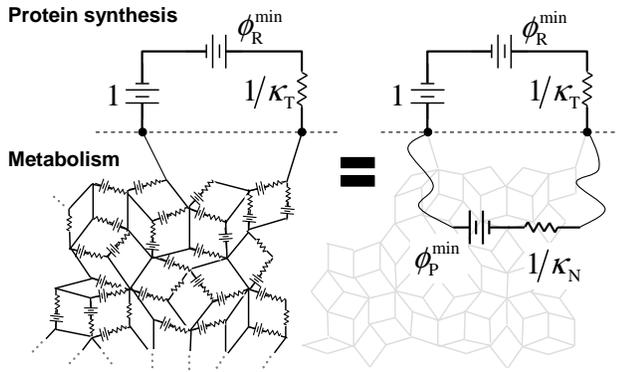

Figure 36: **Equivalent circuits**. A tangled network of resistors and batteries is indistinguishable from a single battery in series with a single resistor. For an enzyme-catalyzed reaction network, a similar equivalent representation is made possible by decomposing the mass fraction of each enzyme $\phi_{E_l}$ into growth-rate dependent and growth-rate independent parts: $\phi_{E_l}(\lambda) = \Delta\phi_{E_l}(\lambda) + \phi_{E_l}^{\min}$. The effective growth-independent offset in the metabolic protein fraction is simply a sum of the individual contributions $\phi_P^{\min} = \sum_l \phi_{E_l}^{\min}$ and the nutrient capacity $\kappa_N$ corresponds to the effective conductance of the network defined as the proportionality between the growth rate $\lambda$ and the total active enzyme cost: $\kappa_N = \lambda / \sum_l \Delta\phi_{E_l}$. In this way, enzyme-catalyzed networks of arbitrary complexity can be coarse-grained into simple equivalent circuits at the expense of introducing lumped phenomenological parameters.

The empirical nutrient capacity $\kappa_N$ is then interpreted as a flux-weighted sum of the effective catalytic constants in the metabolic network. The effective catalytic constants $\kappa_{E_l}$ include the catalytic rate of the enzyme modified by substrate and cofactor abundance; substrate abundance could depend upon the steady-state level of other enzymes in the network, breaking the linearity of the flux expression. A partial justification for treating the effective catalytic rates $\kappa_{E_l}$ as constants independent of enzyme abundance comes from translation-inhibition experiments (Figure 31B).

If protein synthesis is inhibited by antibiotics or genetic mutation, the exponential growth rate can be reduced up to 20-fold with a corresponding change in the metabolic protein abundance. Nevertheless, in a given nutrient environment, the empirical parameter $\kappa_N$ appears to be independent of the metabolic protein abundance, suggesting that the effective catalytic rates $\kappa_{E_i}$ are likewise independent of the metabolic protein abundance over the range of translational inhibition surveyed.

More direct evidence comes from the perturbation experiments of Hwa and coworkers [642, 643]. Like the translational inhibition experiments discussed in this section, growth is inhibited in a variety of ways, including carbon-limitation, nitrogen-limitation and protein over-expression. Using mass spectrometry, the abundance of hundreds of proteins are measured in each growth condition. Consistent with the 'equivalent circuit' coarse-graining, enzymes appear to change their mass fraction abundance proportional to the proteome sector of which they are a part. That is, it appears that most of the metabolic flux is regulated by enzyme abundance (fixed $\kappa_{E_i}$) rather than by fine-tuning the catalytic rates via cofactors and substrate abundance.

Finally, the proteomic work by the Hwa lab suggests a general rule for quantifying proteome-partitioning constraints in other organisms. By focusing on a given strategy of growth inhibition (*eg.* translational inhibition), proteins can be grouped into those whose fraction correlates inversely with the growth rate ('regulated' response) and all others whose fraction necessarily has the inverse growth rate dependence. For *E. coli*, those proteins that increased their mass fraction with translational inhibition were translation-associated proteins such as the ribosomal proteins and elongation factors. There is no reason that the identity of the individual proteins in any given sector should be conserved in other bacterial species or single-cell eukaryotes; that of course will depend upon the precise details of the regulation.

## 6. Modern bacterial physiology, Part II: The fundamental unit of cell size and the general growth law

In the last section, we discussed the sector model of proteome partitioning, which is useful because it has quantitative predictive power. Historically, the nutrient growth law (see Box 6 below) introduced in Section 2 was one of the first laws that allowed quantitative prediction in bacterial physiology. For example, for *E. coli*, it is sufficient to know the nutrient-imposed growth rate to predict the average cell size in steady-state growth (Figures 1B and 2C;

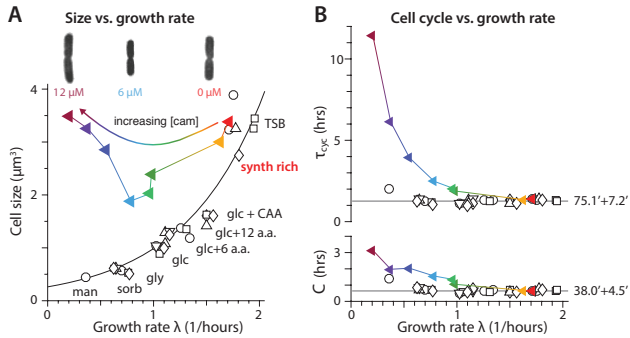

Figure 37: **Changes in cell size and cell cycle under translational inhibition. A.** The nutrient growth law for normal growth conditions under different nutrient conditions of an *E. coli* K12 NCM3722 strain. Each data point represents approximately $10^4$ cells. Solid line is an exponential fit to data (empty symbols). **B.** The duration of replication (**C** period) and one complete round of cell cycle ($\tau_{\text{cyc}}$) both increased with increasing dose of chloramphenicol.

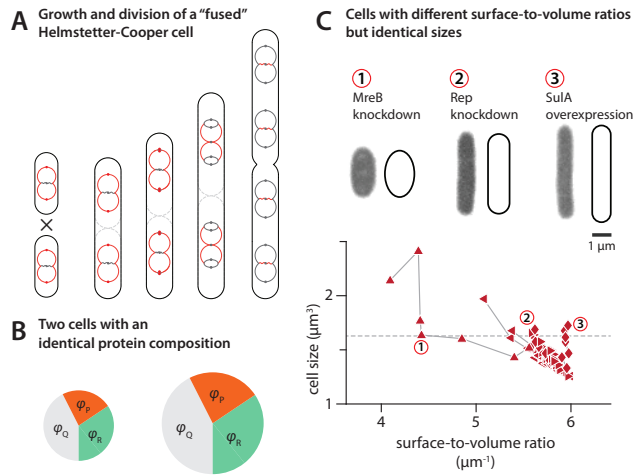

Figure 38: **Why models based on intensive parameters need additional constraints to determine the cell size. A.** Fusion of two synchronized cells still follow the Helmstetter-Cooper model. **B.** Two cells with an identical protein composition. **C.** Cells with different surface-to-volume ratios can have identical cell size (replotted from data of [252]).

[2, 4, 5]). The nutrient growth law (and the two back-to-back papers from whence it came [2, 3]) thus established a basic practice in bacterial physiology of plotting any physiological parameter of interest against the growth rate. In that spirit, the proteome sector model was based upon the analysis of the ribosome protein mass fraction $\phi_R$ vs. growth rate under different growth conditions.

One may wonder how general and robust the nutrient growth law is. A physiologist's approach to answer this question is to perturb the physiological state of the cells, and measure the changes in average cell size vs. growth rate under the new steady-state growth condition. One such perturbation is systematic growth inhibition using a pharmacological method, e.g., inhibition of protein synthesis as we discussed in Section 5.

Shown in Figure 37 is an example of how the cell size and cell cycle respond to a sublethal dosage of chloramphenicol, an antibiotic that inhibits translation. In this example, the cell size decreases first as growth slows with increasing dosage of chloramphenicol. However, cell size starts to increase at some point despite slowing growth. The data clearly shows that steady-state cells under translational inhibition deviate from the prediction of the nutrient growth law. In other words, the nutrient growth law is not robust to growth inhibition, and knowing the growth rate is not sufficient to predict the cell size once growth is inhibited. This is perhaps not surprising given that the growth rate is not sufficient to predict the ribosome mass fraction $\phi_R$ either – under conditions of translational inhibition, the cell has significantly higher ribosome mass fraction than a cell growing at the same rate, but in a poorer nutrient medium.

Given the predictive power of the proteome sector model of the previous section, one may be tempted to model how cell size changes under growth inhibition. The answer is that we need a different approach to understand the principle of cell size control. As we shall explain in this section, the key is to identify what remains invariant, rather than what changes under growth inhibition. The invariance will lead us to the "general growth law," which provides a complete description of cell size control for any steady-state growth condition.

### 6.1. General challenges in approaching cell size control with intensive parameters

Consider two *E. coli* cells of an identical size with synchronized replication cycles (Figure 38A). If we could fuse the two cells into one, the fused cell still would follow the Helmstetter-Cooper model, although the fused cell would be two times bigger than the cells before fusion. In other words, Helmstetter-Cooper model cannot determine the absolute size of the cell.

The above thought experiment illustrates why models that only depend on intensive parameters, such as a concentration of a protein, are unable to determine the absolute size of the cell. This limitation applies to the proteome sector model, as well. Two cells can have exactly the protein composition, but their size can be different (Figure. 38B). Therefore, the cell can growth indefinitely without changing its protein composition, and the sector model does not provide any instruction on when to divide. It is also clear that two cells can have two different sizes without changing the ratio of their rates of surface and volume synthesis [252] (Figure 38C).

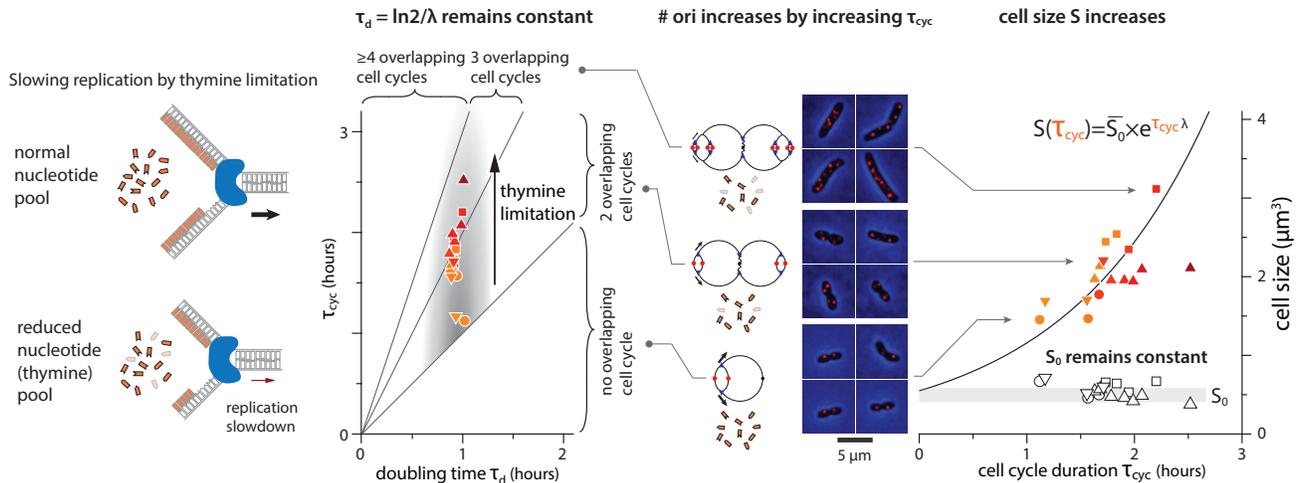

Figure 39: **Cell size increase by increasing C period** [252]. Left: Thymine limitation reduces the nucleotide pool and, as a consequence, DNA replication slows down. Middle: $\tau_{cyc}$ increases in thymine limitation while $\tau_d$ remains unchanged, increasing the number of overlapping cell cycles. Chromosome schematics and cell images with foci qualitatively show increasing number of replication origins as a result of multifork replication. Right: Cell size increases exponentially with the cell-cycle time $\tau_{cyc}$ in thymine limitation, as predicted by Eq. 99 (solid line, no free parameters). The empty symbols are $S_0$, and the thickness of the grey band denotes ±SD. Symbol shapes reflect biological replicates and the symbol colors indicate the level of thymine limitation.

Because of the basic limitations associated with intensive parameters, models based on them must employ additional biological constraints. For example, an earlier sector-based model imposed a constraint on the absolute size of one of the sectors to determine cell division [9]. The surface-to-volume synthesis rates model is based on a hypothesis that cells in a specific growth condition have a precisely tuned amount of excess cell wall materials to control the timing of cell division [473]. So far, no direct experimental evidence is available to test these hypotheses. It is thus important to identify canonical processes that are necessary and sufficient to determine the cell size in both population and single-cell levels.

### 6.2. Size tautology and the origin of the nutrient growth law

Consider a steady-state population of *E. coli*. The average cell size can be expressed in the following tautological manner.

$$S = S/\#ori \times \#ori$$
$$= S_0 2^{\tau_{cyc}/\tau_d},$$

where $S$ is the average cell size of the cell population, $S_0 = S/\#ori$ is the average cell size per replication origin, and $\#ori$ is the average number of replication origins per cell. The relationship $\#ori = 2^{\tau_{cyc}/\tau_d}$ can be obtained straightforwardly from the age distribution (Section 2.2.1) or graphically as was done by Bremer & Churchward (Section 2.2.4). The power of 2 implies that the total number of origins per cell doubles at replication initiation. $S_0$ was known to be proportional to the initiation mass [295], so $S_0$ is used to denote initiation mass or 'unit cell' interchangeably throughout this section (also see Eq. 10 in Section 2.2.5).

We have already seen something similar before (Figure 1B). If $S_0$ and $\tau_{cyc}$ are constant, the above equation is identical to the nutrient growth law. In other words, the original nutrient growth law was a special case of the more general relationship

$$S(S_0, \tau_{cyc}, \lambda) = S_0 e^{\tau_{cyc}\lambda}, \tag{99}$$

where the average generation time $\tau_d$ is converted to the average growth rate $\lambda = \ln 2/\tau_d$. The importance of Eq. 99 is that it provides a complete description of cell size for any steady-state growth condition using three canonical variables $S_0, \tau_{cyc}, \lambda$.

Biologically, the three canonical variables in Eq. 99 represent the initiation control ($S_0$), the progression of combined replication and division cycle ($\tau_{cyc} = \mathbf{C}+\mathbf{D}$), and the global biosynthesis rate ($\lambda$). Considering a major question in biology is how growth and the cell cycle are coordinated, the pioneering experiments that led to the nutrient growth law can be understood as an independence between the control of the cell cycle ($S_0$ and $\tau_{cyc}$) and the global biosynthesis ($\lambda$) under nutrient-modulated growth.

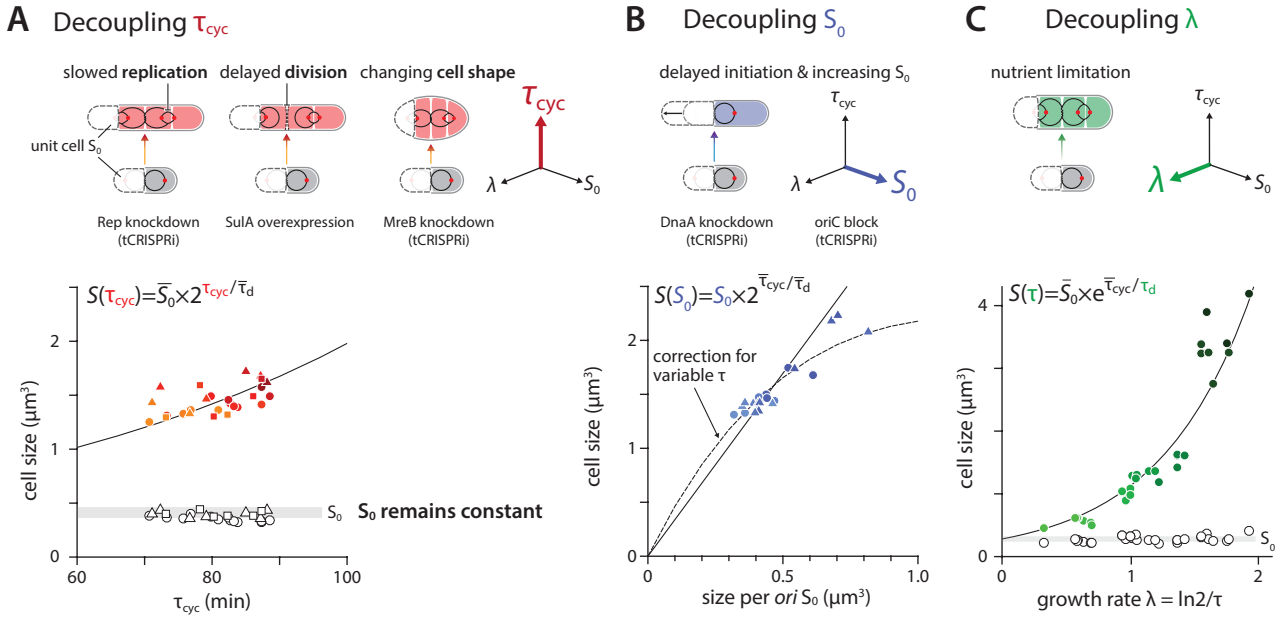

Figure 40: **Decoupling three canonical processes from one another** [252] **A.** Top: $\tau_{cyc}$ can be decoupled from $S_0$ and $\lambda$ by three orthogonal methods: slowing replication, slowing cell division, or changing cell shape. The symbol colors represent the degree of knockdown or overexpression (same for **B** and **C**). Bottom: Cell size increases exponentially as predicted by the general growth law (solid line, no adjustable parameters; same for **B** and **C**). $S_0$ remains unchanged (open symbols). Grey band indicates average $S_0$ from no-induction controls and its thickness indicates ±SD (same for **B** and **C**). **B.** Top: $S_0$ can be decoupled from $\tau_{cyc}$ and $\lambda$ using two orthogonal methods: repression of DnaA to delay DNA replication initiation or sequestration of oriC. Bottom: The solid line is Eq. 99 with constant $\lambda$. The dashed line is Eq. 99, assuming a linear dependence of $\lambda$ on $S_0$ (fitted separately) to account for the slight decrease in growth rate in the $S_0$ vs. $\lambda$ data. **C.** Top: Decoupling $\lambda$ from $\tau_{cyc}$ and $S_0$ by nutrient limitation. Bottom: The nutrient growth law is a special case of Eq. 99, where $S_0$ and $\tau_{cyc}$ are constant. $S_0$ is constant over all growth conditions.

In the 1970s, Zaritsky and Pritchard [219] were the first to explicitly examine the relationship among these canonical processes. They noticed from Eq. 99 that cell size will increase exponentially with respect to $\tau_{cyc}$, if the initiation mass and the growth rate can be decoupled from the duration of DNA replication and remain constant. To test their hypothesis, they used an *E. coli* mutant that is deficient of thymine synthesis. By supplying a controlled amount of thymine in the growth media, they were indeed able to exclusively change the DNA replication time (**C** period), which resulted in an increase in cell size. A modern version of Zaritsky and Pritchard's experiment is shown in Figure. 39.

### 6.3. Decoupling the canonical processes in E. coli

The thymine limitation experiments are an example of how DNA replication and global biosynthesis are coordinated yet can be modulated independently. However, it is also well known in bacterial physiology that both the **C** period and cell cycle time $\tau_{cyc} = \mathbf{C} + \mathbf{D}$ become prolonged if the cells grow in nutrient poor media [631, 632]. In fact, for a typical *E. coli* cells, **C** and $\tau_{cyc}$ increase approximately in proportion to the doubling time $\tau_d$ when $\tau_d$ is larger than 1 hour. In other words, the independence of the cell cycle time $\tau_{cyc}$ from the population growth rate $\lambda$ is not general even for nutrient modulated growth. It is an important physiological issue to understand the extent to which the three canonical processes can be decoupled from one another.

Fortunately, this issue has been settled conclusively by the recent work of Si *et al.*. It appears that over a wide range of growth conditions, the three canonical processes can be modulated independently. To show this, Si *et al.* used genetic and biochemical methods to selectively inhibit key genes or steps for each canonical process. In particular, they developed tunable CRISPR interference (tCRISPRi) system, which allows precise and systematic repression of a specific gene of interest (Box 5) [666]. A good example is knockdown of Rep, a DNA helicase that facilitates replication and loss of which causes increase in the DNA replication period [667]. Thus, knockdown of Rep should have the same effect on cell size as thymine limitation by increasing $\tau_{cyc}$, and this indeed was the case (Figure 39A, left panel).

A major lesson from these decoupling experiments

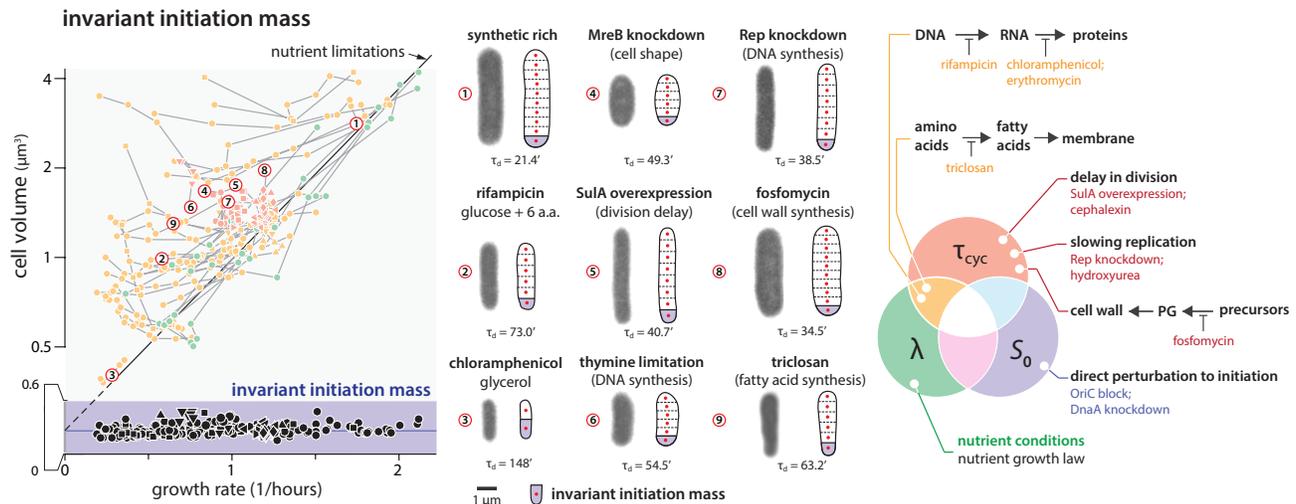

Figure 41: **Invariance of the initiation mass** $S_0$ [252]. Left: The measured $\tau_{\text{cyc}}$ vs. $\tau_d$ shows a linear relationship under growth inhibition. Empty circles represent pooled single-cell data from [631]. Coloring reflects which core biosynthetic process is perturbed. Right: An 'inhibition diagram' mapping perturbations to the three core biosynthetic processes underlying the general growth law.

[252] is that, in *E. coli*, it is possible to selectively inhibit biosynthesis underlying each of the three canonical processes, yet the inhibition does not feedback to the other canonical processes.

A broader implication here is that *E. coli* on average does not actively modulate the rate of biosynthesis (e.g. synthesis rate of DNA, RNA, ribosome etc.) except that of global biosynthesis (quantified as the growth rate), consistent with balanced growth discussed in Section 2. Under selective inhibition, the resulting cell size, which is the most downstream consequence, follows Eq. 99.

### 6.4. How do the canonical processes respond to general physiological perturbations?

Historically, bacterial physiology has been largely focused on understanding the effect nutrient limitations [1–3, 58, 86, 88, 111, 169, 208, 226, 283, 351, 467, 628, 654, 658, 673, 674], though antibiotic inhibition studies have played an important role in the elucidation of DNA replication kinetics [144] and ribosome biosynthesis [675–678]. A natural question is then how growth inhibition will affect the canonical processes, and the extent of their decoupling. We have already seen two examples that growth inhibition simultaneously affects both the growth rate $\lambda$ and the cell cycle time $\tau_{\text{cyc}}$ in Figure 37.

Si *et al.* also investigated this issue systematically and discovered an unexpected invariance principle. Shown in Figures 41 are changes in $\tau_{\text{cyc}}$, $\lambda = \ln 2/\tau_d$, ribosomal fraction $\phi_{\text{R}}$, and cell size $S(S_0, \tau_{\text{cyc}}, \lambda)$ under perturbations to translation, transcription, ribosome content, fatty acids synthesis, cell wall synthesis, and surface-to-volume ratio by genetic or pharmacological methods with different nutritional limitations (these perturbations can be regarded as "growth inhibition" since the $\lambda$ decreases in these cases). Clearly, each of these perturbations can cause changes in more than one canonical process simultaneously (Figure 41). What is

**The law of a fundamental unit of cell size:**
**cell size is the sum of all (invariant) unit cells**

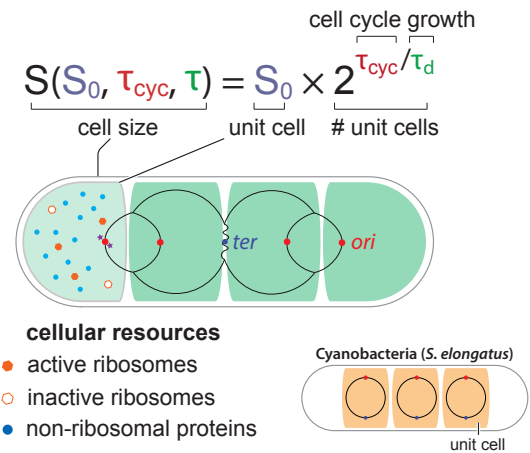

Figure 42: **The law of a fundamental unit of cell size.** The general growth law states that cell size $S$ is the sum of all unit cells $S_0$, each unit cell containing the minimal resource for self-replication from a single replication origin, for both *E. coli* and the Cyanobacteria *S. elongatus* [252, 679].

remarkable, however, is that the initiation mass $S_0$ is invariant.

Another recent study by Wallden *et al.*revealed the invariance of the initiation mass $S_0$ at the single-cell level under nutrient-mediated slow growth [631]. Under these growth conditions, the cell cycle time $\tau_{\mathrm{cyc}}$ changes in direct proportion to the doubling time $\tau_d$ at both the single-cell and population level. This invariance of the unit cell size $S_0$ in slow growth was previously unknown, and underscores the robustness of initiation control.

### 6.5. Unit cell: the fundamental unit of cell size in bacteria

Why is the initiation mass $S_0$ invariant? While we could approach this from several different perspectives (*e.g.*, evolutionary considerations), Si *et al.*has provided an interpretation in the context of cell size control. To see this, note that $2^{\tau_{\mathrm{cyc}}/\tau_d}$ in Eq. 99 is the average number of replication origins per cell in steady-state growth. Because $S_0$ is the cell size per replication origin, Eq. 99 states that **cell size is the sum of all "unit cells"** (Donachie and colleagues also introduced the notion of unit cell for a different reason; Section 2.3.1)

$$\text{(Cell size) } S = S_0 \times 2^{\frac{\tau_{\mathrm{cyc}}}{\tau_d}} = \sum\{\text{unit cells } S_0\}. \quad (100)$$

This 'general growth law' has a universal property. If we rescale the cell size and the growth rate by their respective physiological parameters, unit cell size $S_0$ and cell cycle time $\tau_{\mathrm{cyc}}$, all data from the perturbation experiments in Figure 41 collapse onto a single exponential master curve (Figure 44). In other words, the general growth law clarifies the origin of the nutrient growth law by Schaechter, Maaløe, and Kjeldgaard (1958) [2], whose exponential relationship is a special case with $\lambda = \ln 2/\tau_d$ being the only variable in the experiments. (see Box 6 for a summary of the "growth laws" established in the history of bacterial physiology).

From its tautological construction, the general growth law is applicable beyond *E. coli*. For example, evolutionary divergent cyanobacteria appear to follow the same principle [679, 680]. Cyanobacteria are photosynthetic prokaryotes and their growth rate depends on the intensity of illumination [681–687]. Surprisingly, data shows that neither the cell size distributions nor the chromosome copy number distributions are affected by the illumination-imposed growth rate in their experimental conditions. Most newborn cells contain on average three chromosomes, and double their number by the time they divide. The average newborn size is independent of the growth rate. Furthermore, previous work suggests that replication initiation is asynchronous and, at any given time, only one of the chromosome copies undergoes DNA replication [688–690]. Taken together, growth and the chromosome replication cycle are coupled such that the average cell size added per replication cycle of one chromosome is invariant, identical the the general growth law that the cell size is the sum of all unit cells (Figure 42).

### 6.6. Remaining challenges

The general growth law, that cell size is the sum of all unit cells, provides a simple and straightforward interpretation of bacterial cell size control via coordinated biosynthesis for steady-state growth conditions. However, there are several outstanding issues that need to be resolved by future work.

First, there is still a significant gap in our understanding of how the adder principle at the single-cell level and the general growth law at the population level are connected. A simple reason is that correlations between the three canonical processes are difficult to measure at the single-cell level, particularly during multifork replication. Nevertheless, these correlations are important in understanding how much individual cells grow between birth and division with respect to the birth size and other reference points during the cell cycle. As such, this is a question that new experimental breakthroughs are required to answer, and this in turn will guide further theoretical developments.

**Initiator threshold model**

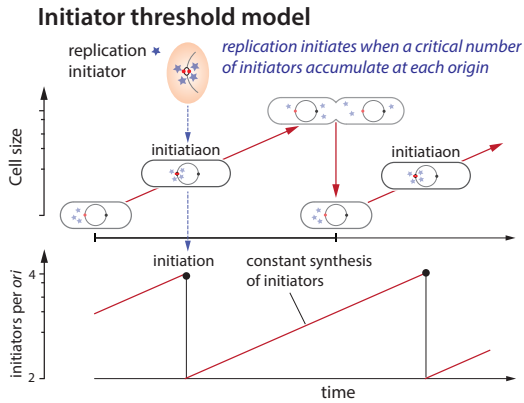

Figure 43: **The initiator threshold model.** Initiation-competent initiators (stars in purple) accumulate at the same rate as the growth rate $\lambda$ and trigger initiation at a critical number per *ori* (four in this illustration).

Another important issue is the mechanistic origin of the robustness of the initiation control. Current experimental data are consistent with the combination of the "initiator threshold" and the "autorepression" hypotheses that we discussed in Section 2.2.7 [56, 261, 360]. The essence of these hypotheses is that the expression of initiator proteins is autoregulated so that their concentrations are maintained at a constant level independent of the growth condition. Under these hypotheses, initiators are synthesized at the same rate as the growth rate, accumulating at each replication origin to a threshold level. Once the growth-rate independent threshold is reached, initiation is triggered, thus the name "initiator threshold." (see the illustration in Figure 43) This threshold is the constraint required by size control models based on intensive parameters, which we explained in Section 6.1.

While similar ideas go back to the Helmstetter-Cooper model in the 1960s [72, 111], no direct experimental evidence is currently available. The closest evidence is biochemical data from the 1990s by Hansen and Kohiyama who showed that the level of DnaA is approximately constant under nutrient limitation [351]. Therefore, this issue also awaits new experimental data, with deeper theoretical understanding of the control of initiation mass fluctuations.

*6.7. Summary*

Sixty years ago, the discovery of the nutrient growth law signaled the arrival of the first golden era of bacterial physiology. With the new progress in the field since the 2010s, we understand the origin of the nutrient growth law at the phenomenological level as summarized in Figure 44. We also summarize various quantitative principles and laws discovered in the past decade in Box 6.

## Box 6 – Various "growth laws" in bacterial physiology

**The growth law** by Schaechter, Maaløe, and Kjeldgaard was the foundational principle in bacterial physiology (Figure 1) [2]. It states that the the average cell size - the most apparent extensive property - has an exponential dependence on the nutrient-imposed growth rate $\lambda$ in steady-state growth (see Section 2.2.2).

$$\text{(cell size) } S \propto e^{\beta\lambda},$$

where $\beta$ is a constant.

**The growth law of exponential growth** was contributed to by many bacterial physiologists and is well-summarized by Arthur Koch [69]. This law states that the total mass and the cell number of a microbial population increase exponentially in steady-state growth (see Section 2.3.1).

$$\text{mass} \propto e^{\lambda t}$$

**The growth law of ribosome synthesis** is the mechanistic foundation underlying the former two laws. It originated from Neidhardt, Magasanik and Harvey's works [58, 677] and characterizes the positive linear relationship between RNA or ribosome content and the growth rate $\lambda$ under moderate-to-fast nutrient modulated growth (also see Section 2.2.3).

$$\frac{\text{RNA}}{\text{protein}} \quad \text{or} \quad \frac{\text{ribosome}}{\text{protein}} = a + b \times \lambda,$$

where $a$ and $b$ are constants.

**The growth law of proteome partitioning** is about constraints imposed on protein composition, and was first described in Scott *et al.* in 2010 (see Section 5) [251]. Under nutrient-modulated growth, or translational inhibition, the ribosome mass fraction $\phi_R$ is given by,

$$\phi_R = \frac{\lambda}{\kappa_T} + \phi_R^{\min} \quad \text{or} \quad \phi_R = -\frac{\lambda}{\kappa_N} + \phi_R^{\max}$$

where $\kappa_T$ and $\kappa_N$ are constants related to the rates of protein synthesis and nutrient assimilation, respectively. All other proteins in the cell are constrained such that their mass fraction $\phi_P$ is given by $\phi_P = \phi_{\max} - \phi_R$ ($\phi_{\max} \approx 0.45$ for *E. coli* MG1655 K12 strains under moderately-rich nutrient conditions; see Section 5).

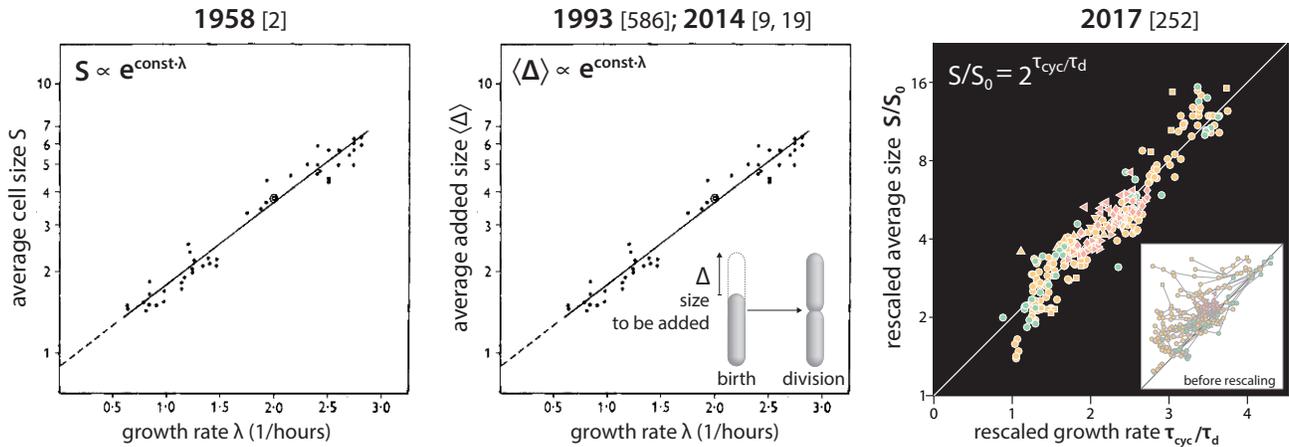

Figure 44: **From the nutrient growth law to the adder principle, and to the general growth law.** Left: The nutrient growth law discovered in 1958 revealed a quantitative relationship between the average cell size and the nutrient-imposed growth rate. Middle: The adder principle explained the origin of the y-axis of the nutrient growth law, linking to cell size homeostasis of individual cells under nutrient limitation. Right: The general growth law, or the growth law of a fundamental unit of cell size, extends the nutrient growth law to any steady-state growth, by stating that the cell size is the sum of all unit cells, where the size of unit cell is determined by the control of replication initiation.

---

**The growth law of a fundamental unit of cell size** states that the cell size is the sum of all (invariant) unit cells, and it was reported in Si *et al.* 2017 [252]. This law is based on the coordination of three canonical, decouplable physiological processes represented by the initiation mass or the unit cell size ($S_0$), the duration of replication-division cycle $\tau_{\mathrm{cyc}}$, and the global biosynthesis rate ($\lambda$). This work shows that the growth law by Schaechter *et al.* is a special case of

$$\text{(cell size)}\ S = S_0 e^{\tau_{\mathrm{cyc}}\lambda} = \sum \{\text{unit cells}\ S_0\},$$

where $\lambda$ was the only experimental variable. The measurement of the three canonical variables $S_0$, $\tau_{\mathrm{cyc}}$, $\lambda$ is sufficient to predict the average cell size for any steady-state growth condition.

## 7. Modern bacterial physiology, Part III: The cell as a factory

The previous two sections revisited different 'growth laws' (Box 6). The 'growth law of proteome partitioning' in Section 5 focused on how the ribosomal fraction of the total proteome changes under different growth conditions, whereas the 'growth law of a fundamental unit of cell size' in Section 6 reviewed physiological properties that largely remain invariant under different steady-state physiological conditions.

In this section we discuss 'self-replicating factory' proposed by the mathematician von Neumann, who developed the concept to describe a machine that replicates itself in a non-trivial manner. Following [691], we mathematically formulate this model and show it has rich theoretical results. There are several parallels between the self-replicating machine model and the process of global biosynthesis during growth of the cell. As such, the model of self-replicating machine suggests a new research avenue to navigate quantitative microbial physiology.

### 7.1. History of the self-replicating factory concept

Although the writer Samuel Butler was probably the first to write about the idea of self-replicating machines that can evolve and even develop intelligence (beginning in his essay "Darwin among the machines" which later evolved to "the book of the machines" as part of his novel "Erewhon"), it was John von Neumann who considered this idea from a scientific standpoint. His pioneering study of self-replication in the early late 40's of the previous century was summarized posthumously in [692]. The main goal of von Neumann's analysis was to understand how a physical system could become more complex over time. As a mathematician, he also searched for a definition that would make self-replication "non-trivial," and distinguishable from inanimate self-replication like crystal growth. Motivated by the successful introduction of the universal Turing machine by Alan Turing to the theory of computation, and by its successful physical implementation, led by von Neumann himself in what is arguably the first general purpose computer, von Neumann suggested an intriguing definition for a non-trivial self-replicating factory.

In his first model, the kinematic self-replicator, he envisioned a room full of parts and a self-replicating factory that assembles a copy of itself by consuming these parts as substrates. He defined the factory as "non-trivial" if it contains a machine, the universal constructor (U) that can read instructions and translate them into physical assembly actions that result in the construction of any machine in the factory, self included, provided all the substrates are available.

To keep things simple, von Neumann only considered serial scheduling of the self-replication process [692]. The universal constructor first reads the instructions and subsequently constructs a copy of all the machines in the factory including itself. However, to obtain a fully functional independent copy, the instruction set should also be replicated. Here, von Neumann recognized a potential problem. Instructing the U machine to copy the instructions requires a separate set of instructions. But then this set also requires a set of instructions for it to be copied, and this appears to lead to an infinite recursion, and an infinite instruction set, which is clearly not physically realizable.

## Box 7 – the first universal constructor

*"Ten things were created at twilight of Saturday's eve, between the suns ... And some say also ... the original tongs, for tongs are made with tongs." Avot, chapter 5, paragraph 6.*

Ancient mishnaic-period Rabbis had pondered on the following paradox – Tongs are used used by toolmakers to create new tools. How then were the first set of tongs created? Not surprisingly, some offered a resolution of this apparent paradox by using a *Deus ex machina*-type argument, claiming the almighty had created the first forging tongs, just before resting, after a very busy week. ‡

Rosen suggested a modern version of this paradox, as a critique on von Neumann's model [694], by noting that in order to replicate a universal-constructor you need a universal constructor; hence it cannot be created from scratch. He then concluded that the concept is biologically irrelevant, as it cannot form naturally. A simple resolution of the paradox actually invokes random mutations as a simple mechanism for creating variations in the instruction set. Because one approximately universal construction machine can create a slightly better U-machine by chance due to random mutations in their instructions, the paradox is removed [695].

‡ The mathematician Mark Kac (1914-1984) once gave a lecture at Caltech, with Feynman in the audience. When Kac finished, Feynman stood up and loudly proclaimed, "If all mathematics disappeared, it would set physics back precisely one week." To that outrageous comment, Kac shot back with that yes, he knew of that week; it was "Precisely the week in which God created the world." [693]

There are a few known tricks to solve this problem in computer science, and von Neumann identified the simplest solution. The U machine should be instructed to construct a "Copying machine" (R) and this machine will template-copy the instructions without translating them. Although von Neumann eventually abandoned this line of research, in retrospect what is remarkable about it is its ability to correctly predict the existence of a special type of machinery to template copy the DNA. Furthermore, it offered a functional schematics of all cellular processes that is consistent with the modern view (reflected e.g. in gene ontologies such as GO), but based entirely on abstract reasoning. Indeed, the universal constructor rather than being one machine, is in fact the translation-transcription machinery (with its key players — ribosomes and RNA-polymerase), the instruction set is of course the DNA molecule, and the "copying machine" is the replisome machinery (with its key players — the DNA polymerases).

Despite this retrospective success, von Neumann's model did not influence the development of molecular cell biology and was ignored for many years. Revisiting von Neumann's original model today, it is rather striking how a few logical and simplifying considerations can lead to so much biological relevance.

In fact, von Neumann's architecture seems to apply without exception to all known cells and thus presents itself as a universal model for cellular self-replication. There are, however, several important aspects that were not studied originally by von Neumann that are nevertheless interesting from a modern perspective. And *vice versa*, aspects that von Neumann originally thought to be important, turned out to be uninteresting from a modern standpoint. In regard of the latter, von Neumann put a lot of effort on designing the transport of goods from one place to another, however, in bacteria, diffusion in a limited volume will effortlessly take care of this issue.

There are three missing theoretical concepts in von Neumann's model. These are essential if we wish to describe actual biological cells rather than a fictitious machine. The first is the temporal organization (scheduling) of the self-replication process and the dynamical rules determining cellular resource allocation "policies." Second, physical and chemical constraints on performance, especially those originating from non-equilibrium thermodynamics and biochemistry. Last is the control module which process both internal and external cues.

## 7.2. The transcription-translation machinery— a realization of the universal constructor

It is tempting to identify the universal constructor with the ribosome, yet on a closer inspection, this cannot be correct since the catalytic core of a ribosome is made of rRNA, but the ribosome cannot synthesize rRNA. Indeed, rRNA is transcribed by RNA-polymerase. Hence a ribosome cannot self-replicate; a requirement for being certified as a universal constructor. Indeed, the universal constructor is not composed of a single molecular machine, but rather from a conglomerate of molecular machines comprising the transcription-translation machinery with the forefront players being the ribosome and RNA-polymerase. To those, a team of chaperons and helper molecular machines join, to form the universal constructor that can collectively both self-replicate itself, as well as produce all the other machines in the factory, machines required to sustain the universal constructor by supplying it with substrate (metabolism), membrane bound volume (membrane synthesis), DNA instructions (replisome machinery) and control and regulation. This is done collectively, and asynchronously by reading ribosomal DNA (rDNA) as template as well as by transcribing and subsequently translating coded DNA instructions while consuming substrate (nucleotides and amino-acids) and free energy.

The Universal constructor can construct any protein, including all the proteins needed for its own operation. Among them we name initiation, elongation, and release factors, aminoacyl-tRNA-synthetase proteins, chaperons, restriction enzymes, essential promoters and so on.

The self-replication of the ribosome and RNA-polymerase is done as follows — RNA polymerases transcribes mRNA for the subunits from which they are composed. These mRNAs are translated by ribosomes to form new RNA-polymerase subunits (mainly $\alpha, \alpha', \beta, \beta'$ and $\sigma_{70}$) these then self-assemble to form a new RNA-polymerase. RNA polymerases transcribe both rDNA to form rRNA, and tRNA (which is further processed with the help of dedicated proteins and restriction enzymes that modify the nascent strands to form a mature functional tRNA). They also transcribe mRNAs of all the ribosomal proteins. These mRNAs are translated by ribosomes to form ribosomal proteins. Ribosomal proteins and rRNA self-assemble to form the two ribosomal subunits in a self-assembly process with a predefined partial temporal order discovered by Nomura for the small sub unit and later by Nierhaus and others [508, 696,

697] for the large sub-unit (for a recent interesting discussion on why ribosomes have so many small ribosomal proteins see [698]). Thus, overall these two molecular machines are self-replicating themselves and all the helper molecular machines that comprise the universal constructor. $U$ of course also performs its other duties to make the remainder of the proteome, including the part of the proteome required to template copy the instructions themselves (replisome machinery), and also synthesize the membrane, thus expanding the membrane bound protected volume required to sustain these processes, which would quickly dilute and cease working in the absence of a membrane.

## 7.3. Concurrency in self-replication

A standard factory is often modeled as a network of queues [699]. The factory consumes raw materials that lines up in queues before a set of dedicated processing units. The materials are processed in a predetermined partial temporal order. The average rate of production is the ratio between two key system parameters: the work in process $\eta_{\mathrm{WIP}}$ — how many products are in production concurrently on average, and the cycle time $\tau_{\mathrm{C}}$ — the average duration to complete one product, which is determined by the average critical path duration [691].

In many circumstances, decreasing the cycle time in order to increase the production rate can compromise performance. For example, the average time required to synthesize a given protein results from a fine balance between speed and accuracy [700, 701]. In light of this, it is tempting to conclude that the maximal completion rate of a task with a cycle time $\tau_{\mathrm{C}}$ is given by its reciprocal $\tau_{\mathrm{C}}^{-1}$. However this is not the case. There are two fundamentally distinct ways to increase the production rate or throughput, $\kappa_{\mathrm{TH}}$, beyond the apparent limit set by the reciprocal of the cycle time, (i) *parallelization* and (ii) *pipelining*. Parallelization refers to the simultaneous production of several products and is obtained by having multiple production lines running in parallel. Pipelining is defined as *starting a new task before the previous task has been completed*, using the same production line. See Figure 45 for an illustration from protein synthesis.

An important relation known as Little's law [699, 702] exists between the production rate (or throughput) $\kappa_{\mathrm{TH}}$, the work-in-process $\eta_{\mathrm{WIP}}$ — the number of processing units (e.g. ribosomes) concurrently active in making the product, and the cycle time $\tau_{\mathrm{C}}$,

$$\kappa_{\mathrm{TH}} = \frac{\eta_{\mathrm{WIP}}}{\tau_{\mathrm{C}}}. \tag{101}$$

The rate of production $\kappa_{\mathrm{TH}}$, the latency $\tau_{\mathrm{C}}$ and the level of concurrency $\eta_{\mathrm{WIP}}$ are *system parameters*

**A**  Parallel production of proteins, TH = 3/CT

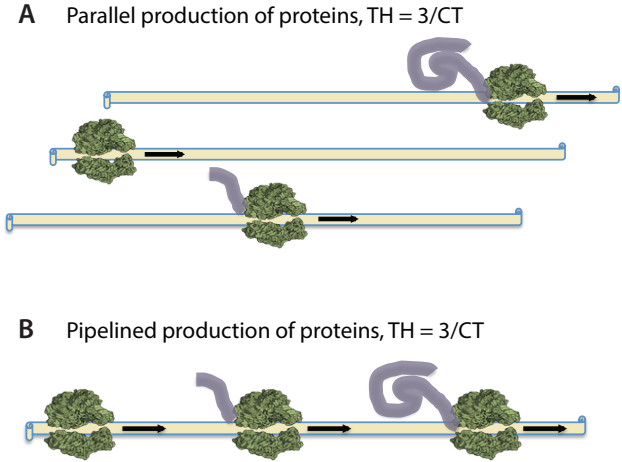

**B**  Pipelined production of proteins, TH = 3/CT

Figure 45: **Illustration of Little's law in protein production.** Panel **A** shows $m = 3$ parallel production lines producing an enzyme. The elongating peptide chain is represented by a purple solid line. The average rate of production (the throughput —$\kappa_{TH}$) of the protein is $3\tau_{C,a}^{-1}$, where $\tau_{C,a}$ is the latency of a single ribosome i.e. the average delay from initiation to complete translation by a single ribosome (neglecting the excess folding time). The work-in-process is $\eta_{WIP} = m = 3$. The average number of ribosomes on mRNA is equal to 1. Panel **B** shows pipelined protein production. A new ribosome start to translate prior to the completion of the translation by the previous ribosome. The level of concurrency — $\eta_{WIP}$ is the average number of ribosomes concurrently translating the mRNA and is equal to $\eta_{WIP} = n = 3$, hence the throughput is $\kappa_{TH} = 3\tau_{C,b}^{-1}$. The latency $\tau_{C,b} = \tau_{C,b}(\eta_{WIP})$ is the average latency of a single ribosome, in the presence of $\eta_{WIP} - 1$ ribosomes on the same mRNA and is typically larger than $\tau_{C,a}$ when $\eta_{WIP}$ is high. The two methods depicted in Panel **A** and Panel **B** can be combined by having $m$ mRNA's with $n$ ribosomes on each, resulting in a protein production rate $\kappa_{TH}$ that is $\eta_{WIP} = nm$ times larger than $\tau_C^{-1}$ if $\tau_{C,a} = \tau_{C,b}$.

and as such they typically depend on external parameters as well as on each other.

It should be emphasized that in general cycletime depends on the concurrency $\eta_{WIP}$, and if the factory is congested above some critical concurrency $\eta_{WIP}$ value, delays occur and the average cycle time increases. In protein synthesis a single ribosome that halts in a specific location on the mRNA can cause a "traffic jam" since other ribosomes cannot bypass it due to the one-dimensionality of the mRNA and the self-exclusion interaction between the ribosomes. Thus, the cycle time - average time a single ribosome translates an mRNA of length $L_{mRNA}$ bases, is a function of the ribosome density $\rho_R$ or $\eta_{WIP}$ (= $\rho_R \times L_{mRNA}$), see Figure 45. It should be evident then that in order to increase the rate of production, it is not always beneficial to increase the level of concurrency $\eta_{WIP}$, because cycle time is typically monotonically increasing with concurrency, unless the system is highly non-random and synchronized (e.g. if all ribosome travel back to front at a constant speed).

The queuing network modeling scheme facilitates the study of non-Markovian stochastic dynamics. This is achieved by taking into account the noise in the servicing time and in the products inter-arrival and departure times without assuming that the related distributions are exponential. Correlation in arrival or departure times is also readily implementable. Importantly, this scheme also facilitates the discovery of hidden relations between the "microscopic" production floor rules and practices, and their "macroscopic" system-wide outcome e.g. of revenue flow rate. For example, the allowed maximal size of intermediate buffers, the scheduling of work with different priorities or due dates, implementation of quality control procedures, utilization of critical processing units, authorization of production, general production policy ("push" vs. "pull" [699]) are all affecting the global outcome, which is a measure of how well the entire endeavor operates in light of its objective.

A cell can also be viewed as a factory whose product is another factory. This circularity is the core difference from the standard analysis of factories using the tools of systems engineering and operations research. While in a standard factory the processing units are transforming substrates to products, here the processing units transform the substrate to form new processing units, leading to an overall doubling of the factory. This leads to a novel type of pipelining, since if the temporal ordering is properly chosen, the newly formed processing units can start working on a new generation, prior to the completion of the previous one, leading to an overall pipelining of self-replication (Figure 46).

Thus, because the factory's product is another factory, there are only two factors that affect the doubling time — the critical path duration $\tau_{crit}$, which is the duration of the longest serial process that is bound to occur, and the level of concurrency $\eta_{WIP}$, which is a measure of overlap between production of different generations — it equals $2\ln 2$ if the newly formed factory is triggered to start replication only after it is fully completed. If it is larger than $2\ln 2$, then there is pipelining of self-replication.

To make this observation quantitative, consider a self-replicating factory for which the critical path (latency) is equals $\tau_{crit}$. For example, $\tau_{crit}$ could be the time it takes to replicate DNA. Indeed, we already encountered Cooper-Helmstetter model that predicts pipelining of DNA replication with great success. We will show that other forms e.g. asynchronous pipelining are also possible, in particular in the process of biomass synthesis. For the purpose of this discussion we assume that all external resources are readily available.

When there is concurrency in the self-replication process, the equation for the overall doubling rate should take into account the fact that newly formed machinery, as well as existing machinery that is free, can start producing machinery for the next round of replication, prior to the completion of the previous round of replication. This pipelining of self-replication can be *phenomenologically* captured by the following equation:

$$\frac{dB}{dt} = \frac{\eta_{\text{WIP}}}{\tau_{\text{crit}}} B(t - \tau_{\text{crit}}), \qquad (102)$$

where $B$ represents the dry biomass, $\kappa_{TH} = \frac{\eta_{\text{WIP}}}{\tau_{\text{crit}}}$ is the throughput of the bottleneck process, i.e. $\eta_{\text{WIP}}$ is the level of concurrency on the critical path. When $\eta_{\text{WIP}} \leq 2\ln 2$ there is no pipelining, and the doubling time is equal to $\tau_{\text{crit}}$, when $\eta_{\text{WIP}} > 2\ln 2$ there is pipelining of self-replication and a new generation initiates prior to the completion of the previous generation.

This equation admits an exponential solution, which can be seen by inserting a trial function $B(t) = B_0 e^{\lambda t}$ into (102). The exponential growth rate $\lambda$ is given by the transcendental equation $(\lambda \tau_{\text{crit}}) e^{\lambda \tau_{\text{crit}}} = \eta_{\text{WIP}}$ which can be solved analytically using the Lambert-W function [703]:

$$\lambda = \frac{\mathcal{L}_w(\eta_{\text{WIP}})}{\tau_{\text{crit}}}. \qquad (103)$$

The Lambert function satisfies that $\mathcal{L}_w(x) \approx x$, for $x \ll 1$, and $\mathcal{L}_w(x) \approx \ln x$, for $x \gg 1$. Thus, if $\eta_{\text{WIP}} \ll 1$ $\lambda \sim \frac{\eta_{\text{WIP}}}{\tau_{\text{crit}}}$. The case $\eta_{\text{WIP}} \ll 1$ represents a scenario where due e.g. to low utilization the doubling time is longer than the critical path duration.

If the concurrency $\eta_{\text{WIP}}$ is large, and the critical latency time $\tau_{\text{crit}}$ is fixed, then doubling at a rate that is $n$ times larger than $(\ln 2)\tau_{\text{crit}}^{-1}$ requires an increase in the level of concurrency (as measured by $\eta_{\text{WIP}}$) by a factor $n2^{n-1}$ i.e. exponentially for large $n$'s.

To illustrate a situation with asynchronous pipelining of self-replication, consider a simplified universal constructor $U$ that is composed of a single subunits $U_1$, which on average requires $U$ to work for $\tau$ units of time in order to make it. Upon completion, $U_1$ requires an extra $\tau_{SA}$ time units to self-assemble into a mature $U$ e.g. by acquiring further modifications. After this time, $U_1$ is transformed to a new functional universal constructor unit $U$.

The equations that describe the process of making a new $U$ are $\frac{dU_1}{dt} = \frac{\alpha}{\tau} U(t - \tau)$, and $\frac{dU}{dt} = \frac{1}{\tau_{SA}} U_1(t - \tau_{SA})$. Inserting an exponential ansatz and solving for the growth rate we obtain: $\lambda = \frac{\mathcal{L}_w(\frac{a}{g}\sqrt{\alpha})}{a}$ where $a = \frac{\tau + \tau_{SA}}{2}$ is the arithmetic mean between subunit production and assembly times, and $g = \sqrt{\tau \tau_{SA}}$ is the geometric mean. If the time to make each subunit ($\tau$) and the self-assembly time ($\tau_{SA}$) are sufficiently far

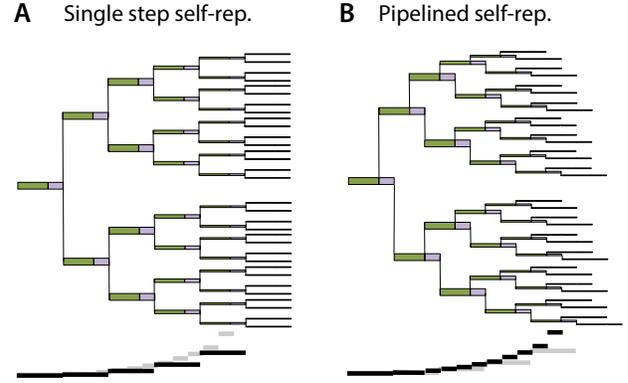

**A** Single step self-rep. **B** Pipelined self-rep.

Figure 46: **Two models for self-replication of the universal constructor** ($U$). Model in inset **A** — each existing U is making a single copy non-preemptively by reading the instructions to build a copy encoded in $DNA(U)$ and consuming raw materials which we assume are abundant. The average duration a single $U$ machine replicates a single copy is $\tau_U$, after which two $U$'s are immediately made available - the old and the new. Model in inset **B** represent a $U$ machinery composed a single subunits $U_1$ which has to self-assemble (mature) after being made. The time it takes for $U$ to synthesize $U_1$ is $\tau$. The maturation time is $\tau_{SA}$. We asume that $\tau + \tau_{SA} = \tau_U$. Upon maturation, the subunits $U_1$ is transformed into a new $U$ which is added to the pool of available $U$'s. Under these assumptions we show in the text that the doubling rate of model **B** will be faster than the doubling rate of model **A** because model **B** is pipelining self-replication, i.e. each $U$ that completes the production of a subunit $U_1$, can start making another subunits, prior to the maturation of the previous subunit.

apart, then the doubling time $\mu$ is shorter than $\frac{\ln 2}{\tau_{SA} + \tau}$. This can be modeled more coarsely by writing an effective equation like Eq. 102 and using $\eta_{WIP} = \frac{\sqrt{\alpha} a}{g}$ and latency $CT = a$. For example, if $\tau = 30 \ min$, $\tau_{SA} = 2 \ min$, and $\alpha = 0.5$ then $\lambda \sim 0.045 \ min^{-1}$ while $\frac{\ln 2}{\tau + \tau_a} \sim 0.022 \ min^{-1}$. The pipelining occurs because any U that finished making one subunit, can immediately start making a new subunit for the next round of $U$-production, prior to the completion of the previous round.

We can also model exponential growth using the standard equation $\frac{dB}{dt} = \lambda_0 B(t)$ but this equation does not explain how $\lambda_0$ depends on the critical path, and whether there is pipelining. For that purpose we need a model that either coarsely or specifically accounts for $\eta_{WIP}$ and critical path duration.

Equation 103 can be seen as a non rigorous generalization of Little's law to concurrently self-replicating factories. It corrects the phenomenological equation $\frac{dB}{dt} = \lambda_0 B$ by explicitly showing that the growth rate of biomass in the present, is due to initiation of replication that occurred $\tau_{\text{crit}}$ units of time ago, i.e. growth rate is history dependent. Thus

Eq. 103 only describes the asymptotic (balanced) growth rate. Any typical history of initiations (e.g. via nutrient shifts up/down) will present decay and oscillations towards the state of balanced exponential growth. Interestingly, the infinitely many complex solutions of Eq. 103 gives the spectrum of oscillations and decay with the first complex solution being the leading order of both the oscillation periods and the decay rate [704].

Note that in practice, $\tau_{\mathrm{crit}}$ is not fixed, deterministic quantity, but rather a stochastic variable, requiring the development of more sophisticated mathematical machinery to capture the statistics of the growth rate.

Interestingly, an increase in the growth rate is not possible via synchronized parallelization if the latency is kept fixed. This is because doubling the number of machines dedicated to self-replication also doubles the number of machines in need of replication. If however, the machines are unsynchronized then the fast machines can restart before the slow machines finished, thus effectively pipelining the process.

### 7.4. The cell as a self-replicating queuing network

Simulating complex assembly lines using a queuing network modeling scheme facilitates the study of non-Markovian stochastic dynamics. This is achieved by taking into account noisy production times and product inter-arrival and departure times without assuming that the distributions are exponential.

Consider a standard parallel queue with substrates as jobs to be processed and several catalysts as the processing units. Two general probability distributions characterize this queue: (i) the distribution of substrate inter-arrival durations; (ii) the service time distribution, the duration of time the substrate is being processed before it exit as product. To make the queue self-replicating, we add the following rule that when a processing unit completes its work, it releases a new processing unit. This processing unit immediately joins the pool of available processing units and will start working (i.e. replicating) as soon as it finds substrate and a free-to-read instruction set.

In Figure 48 a simulation of the self-replicating queue with unlimited resources is shown. Note that contrary to naïve expectations, the distribution of doubling times is not the same as the distribution of service times, because the doubling time is the difference between two random first passage events on the causal tree: from 1 to 2n, minus from 1 to n. Since these two first passage times are also correlated for small n (because of the partial overlap in the trajectories), an analytic calculation is not readily obtained.

Using the concept of a self-replicating queue we can also construct self-replicating queuing networks.

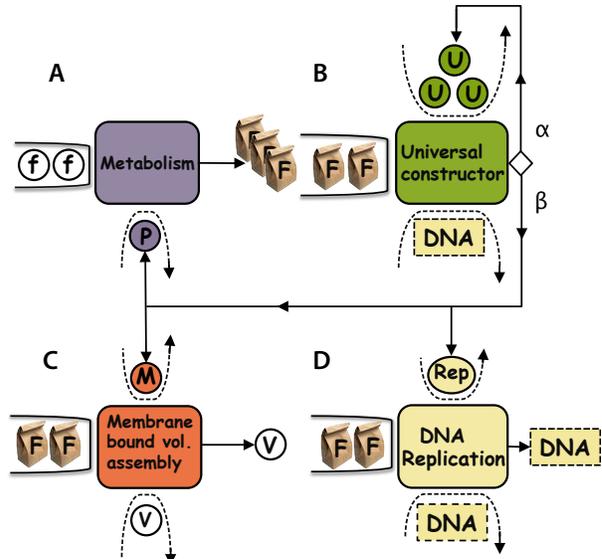

Figure 47: **A graphical model of cell growth as a self-replicating network of queues.** All material components reside in queues. Complex *de-novo* synthesis reaction networks are grouped by their functionality and are depicted as squares: **A** Metabolism (purple square) — production of substrates ($F$) by metabolic proteins ($P$); **B** Universal constructor (green square) — the transcription and translation machinery $U$; **C** Production of membrane bound volume (orange square) — synthesis of additional membrane bound volume ($V$) by dedicated $M$ proteins; Although all units require this protected volume as an essential resource, we did not represented it graphically. **D** Replication of $DNA$ (off-white square) by the replisome proteins (*Rep.*). Each of these processes doubles its product during the doubling time $T_\lambda$. Material inputs that are consumed by a reaction (substrates) are placed in a solid queue. Material inputs that are used as processing units or "servers" e.g. essential catalysts or templates that are required for a certain duration to perform their task and are subsequently released back to the general pool and can serve again, are located in a dashed line arrowed queues. Material inputs that are *de-novo* synthesized by a reaction are marked by a solid arrow emanating from the square towards them. The overall work in progress of a particular reaction is the minimum among all the input materials, divided by the stoichiometric demand for inputs by one indivisible bio-synthesis task. Due to the dual role of the universal constructor, a fraction $\alpha \in [0,1]$ units are allocated to self-replication while a fraction of $\beta \leq \phi_{\mathrm{max}} - \alpha$ units are allocated to protein production. *Rep* represents proteins involved in copying the DNA. $M$ proteins are involved in the assembly of new membrane bound volume. $P$ proteins are metabolic proteins that import and convert external substrates lower case $f$, to internal "Foods" (marked by capital F in a "take-away" bag) — internally consumed substrates e.g. amino acids and nucleotides.

Here, we focus on a specific self-replicating queuing network with the von Neumann architecture (Figure

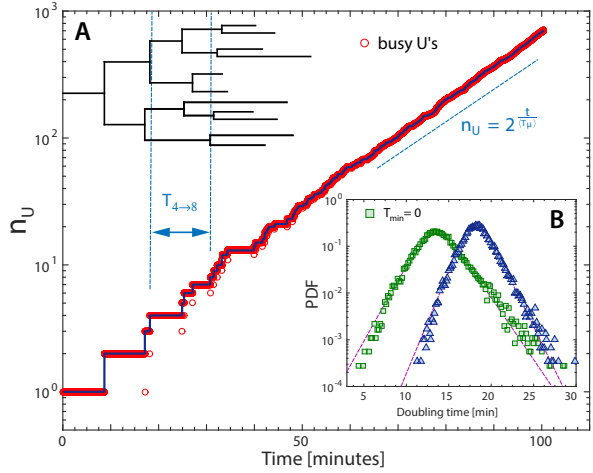

Figure 48: **Single realization of a self-replicating queue in a rich environment.** Number of self-replicating servers $n_U$ as a function of time. Busy $U$'s marked in red circles closely follow. Asymptotic exponential growth is obtained after a few doublings. Panel **A** shows the causal tree structure of the process. Doubling from 4 to 8 servers is marked with light blue dashed lines. Inverted black triangles mark the position where $n_U$ equals a power of two for the first time. Panel **B** statistics of the doubling times with arbitrary initial size. The service time distribution for a single $U$ to complete replication is distributed with a non-Markovian distribution of the form $\Theta(t - T_{min})e^{-\frac{(t - T_{min})}{\tau}}$, where $\Theta(\cdot)$ is the Heaviside step function, with $T_{min} = \tau = 6$ minutes. Green squares are for $T_{min} = 0$, $\tau = 12$ minutes Markovian case.

47). This architecture was first described by von Neumann, albeit not in the context of cellular dynamics and queuing theory, in [692]. von Neumann was interested in understanding whether an artificial self-replicating factory could be constructed, and if so, what is its structure.

As already mentioned, the most prominent machine in von Neumann's factory is the universal constructor, the "machine that makes machines" which reads the instruction set (I) and builds all the processing units in the network, including a copy of itself. In our framework, the universal constructor is a self-replicating queue. The analogue in cells to the universal constructor is not a single machine but rather a set of molecular machines — the transcription-translation machinery, that is capable of transcribing and translating DNA, thus producing all the cellular machinery in the cell, including new functional copies of itself. The other processing units — "supportive machinery" have three major roles: to convert external substrate $f$ to internal substrate $F$ (by the action of the metabolic proteins and transporters contained in the $P$ protein class), to replicate the instruction set I (by the DNA replisome machinery contained in the $D$ protein class), and to double the volume V (by cell-wall synthesizing enzymes in the $M$ protein class), to make room for all the newly formed units.

## 7.5. A simplified model for biomass growth

Consider the following equations that describe the growth of biomass which is grossly composed of the universal constructor machinery ($U$) and all other metabolic proteins ($P$) (see green and purple boxes in Figure 47).

$$\dot{U} = \alpha \, \frac{U\left(t - \frac{\tau_U}{u_U}\right)}{\frac{\tau_U}{u_U}}, \tag{104}$$

$$\dot{P} = (\phi_{max} - \alpha) \, \frac{U\left(t - \frac{\tau_P}{u_U}\right)}{\frac{\tau_P}{u_U}}, \tag{105}$$

$$\dot{F}_{new} = \frac{P(t - \frac{\tau_F}{u_P})}{\frac{\tau_F}{u_P}}, \tag{106}$$

$$u_U = \frac{1}{T} \int_{t-T}^{t} \min\left(1, \frac{\dot{F}_{new}(t')}{\kappa U(t')}\right) dt', \tag{107}$$

$$\kappa \equiv \alpha \frac{F_U}{\tau_U} + (\phi_{max} - \alpha)\frac{F_P}{\tau_P}. \tag{108}$$

The integration time scale $T$ drops out from any steady-state calculation, but may be relevant for transients. The first equation describes the self-replication of the universal constructor by a subset $\alpha$ of the universal constructors. The allocation parameter $\alpha$ represent the probability that one $U$ will be involved in duties related to self-replication upon completing its previous task. The utilization parameter $u_U \in [0,1]$ measures the average fraction of time $U$ is "busy." When the substrate is in excess, the utilization equals 1, and the latency is $\tau_U$; when substrate is limiting, the utilization drops below one, indicating that some $U$'s are starving for resources. This causes them to halt mid-process and wait for the necessary resources to arrive. Hence the latency grows above the nominal completion time $\tau_U$ and becomes $u_U$. To illustrate, if on average, 50% of the time $U$ is starving for resources, then $u_U = 0.5$ and the latency will grow by a factor of two $\tau_U' = \tau_U/u_U = 2\tau_U$. Similarly, the rate will decrease by the same factor.

The second equation describes the fraction of $U$'s that are busy making the rest of the machinery (which in this simplified version is simply the metabolic proteins). The third equation describes the rate of production of *new* substrate $F_{new}$ by the metabolic proteins $P$. The overall change on the number of free substrate $F$ is given by,

$$\dot{F} = \dot{F}_{new} - (\alpha F_U + \beta F_P)\dot{U},$$

where $F_U$ and $F_P$ are the fraction of substrate consumed in the synthesis of $U$ and $P$, respectively, and $\beta = \phi_{max} - \alpha$.

The total protein mass $M = U + P$ reads,

$$\dot{M} = (\dot{U} + \dot{P}) =$$

$$= u_U \left[ \frac{\alpha}{\tau_U} U \left( t - \frac{\tau_U}{u_U} \right) + \frac{(\phi_{max} - \alpha)}{\tau_P} U \left( t - \frac{\tau_P}{u_U} \right) \right]$$

The idling proteins that correspond to the growth independent sector $Q$ are given by $Q = (1 - u_P)P$.

The demand for the substrate $F$ per unit time per mass of the replicator $U$ is given by $\kappa = \alpha F_U / \tau_U + (\phi_{max} - \alpha) F_P / \tau_P$, thus the utilization is a finite moving time average of the level of resource starvation — i.e. the average over the instantaneous supply relative to the instantaneous demand (note that if the supply is above the demand then the utilization will be locked to 1).

Inserting an exponential ansatz $U(t) = U_0 e^{\lambda t}$, $P(t) = P_0 e^{\lambda t}$, $F(t) = F_0 e^{\lambda t}$ we obtain

$$\lambda U_0 = \alpha \, \frac{u_U}{\tau_U} \, U_0 \, e^{-\frac{\lambda \tau_U}{u_U}}, \tag{109}$$

$$\lambda P_0 = (\phi_{max} - \alpha) \, \frac{u_U}{\tau_P} \, U_0 e^{-\frac{\lambda \tau_P}{u_U}}, \tag{110}$$

$$\lambda F_0 = \frac{1}{\tau_P} \, P_0 \, e^{-\lambda \tau_F}, \tag{111}$$

$$\bar{u}_U = \min \left( 1, \frac{e^{-\frac{\lambda(\tau_P - \tau_U)}{u_U}} e^{-\lambda \tau_F}}{\kappa \tau_F} \frac{(\phi_{max} - \alpha)}{\alpha} \frac{\tau_U}{\tau_P} \right). \tag{112}$$

Where we assumed that the utilization of the metabolic proteins is maximal $u_P = 1$ since external substrate is abundant. Using $\lambda = \frac{\mathcal{L}_w(\alpha) u_U}{\tau_U}$ we obtain after some algebra that

$$\bar{u}_U = \min \left( 1, \frac{\mathcal{L}_w \left( e^{\mathcal{L}_w(\alpha) \left( 1 - \frac{\tau_P}{\tau_U} \right) \frac{(\phi_{max} - \alpha) \mathcal{L}_w(\alpha)}{\kappa \alpha \tau_P}} \right)}{\mathcal{L}_w(\alpha) \frac{\tau_F}{\tau_U}} \right). \tag{113}$$

If $\tau_P = \tau_U$, then this expression simplifies to $\bar{u}_U(\tau_P = \tau_U) = \min \left( 1, \frac{\mathcal{L}_w \left( \frac{(\phi_{max})}{\alpha} \mathcal{L}_w(\alpha) \right)}{\mathcal{L}_w(\alpha) \frac{\tau_F}{\tau_U}} \right)$. The growth rate is given by

$$\lambda = \min \left( \frac{\bar{u}_U \mathcal{L}_w(\alpha)}{\tau_U}, \frac{\mathcal{L}_w(\alpha)}{\tau_U} \right) \approx \frac{\bar{u}_U \alpha}{\tau_U}. \tag{114}$$

There are two growth regimes in this model. If $\alpha \leq \alpha_{opt}$ then $U$'s never starve for resource, the utilization is always $u_U = 1$, and what limits growth is the internal duration for $U$ to make $U$, and the level of allocation for self-replication $\alpha$. If however $\alpha > \alpha_{opt}$ then $U$ is attempting to produce too much of itself at the expense of making the $P$'s that are required to supply $U$ with substrate $F$, hence the $U$'s will be underutilized, and the growth rate will be nutrient limited.

This model uses five parameters: the three latencies: $\tau_U, \tau_P$ and $\tau_F$, and the two allocation parameters $\alpha$ and $(\phi_{max} - \alpha)$. In general $\alpha$ is not the same as the mass fraction of $U$ (relative to the entire biomass) $\alpha_U(t) = \frac{U(t)}{(U(t) + P(t) + F(t))} \simeq \frac{U(t)}{U(t) + P(t)}$. This is only the case if $\tau_U = \tau_P$. This is because in balanced growth conditions,

$$\alpha_U = \frac{U_0}{U_0 + P_0 + F_0} \simeq \frac{U_0}{U_0 + P_0}$$

$$= \frac{\alpha}{\alpha + \frac{\tau_P}{\tau_U}(\phi_{max} - \alpha)e^{\mathcal{L}_w(\alpha) \left( 1 - \frac{\tau_P}{\tau_U} \right)}}.$$

If $\tau_U = \tau_P$ we obtain that $\alpha = \phi_{max} \alpha_U$ i.e. that the fraction of active $U$'s times the overall mass fraction of all the $U$'s equals the fraction of active $U$'s involved in self-replication.

## 7.6. Coordinating DNA replication with biomass growth

Copying DNA in E. coli has to start from the origin of replication (in other organisms - eukaryote and certain archea, there are multiple origins that can be initiated concurrently). The need to perform error corrections, and the existence of an error threshold below which the cell's ultimate survival will be hampered, dictate a minimal completion time of the order of 40 minutes for an *E. coli* genome (4.64 Mbps). Further complications arise because of the need to finish segregating the two copies to both sides of the septum, which consumes more time. How do these processes coordinate with the overall biomass doubling process?

We will show that applying Helmstetter's initiator model [705] in the context of the present model, successfully captures two key features observed in experiments: (i) if the doubling time is faster than the cycle time for DNA replication and segregation then DNA replication is pipelined (Cooper-Helmstetter model). (ii) if growth is slow, then DNA replication is initiated after an 'idle' period known as the B phase. The model only applies to bacteria that have a single origin of replication. Let's assume that in order to initiate replication, each origin requires that a certain number of replisome machines (designated as $I$ in fig x) will accumulate on it. In other words, each origin has a threshold $T_I$ and only after the number of $I$'s attached to the origin crosses this threshold, will they initiate the replication process. Finally, we will assume that the initiators $I$ are constitutively expressed with an allocation parameter $\gamma > 0$. We summarize these assumptions by the following equations that supplement Equations 20-24 above.

$$\dot{I} = \gamma \frac{u_U}{\tau_1} U \left( t - \frac{\tau_{rmI}}{u_U} \right), \tag{115}$$

$$\dot{\text{DNA}} = \frac{u_D}{\tau_D} \left( O(t) - O \left( t - \frac{\tau_D}{u_D} \right) \right), \tag{116}$$

$$\dot{O} = O(t - \tau_I) \dot{I} \delta \left( I - T_I O(t - \tau_o) \right). \tag{117}$$

These equation recreates the Cooper-Helmstetter model without accounting for the dilution due to division. The first equation accounts for the production of the replisome machinery (including initiator proteins) by the universal constructor ($\tau_1$ is the time to produce it, $\gamma$ the allocation of $U$ towards this tasks — of the order of a few percent [706]). The second equation accounts for DNA synthesis, $\tau_D$ is the time to replicate a single DNA genome, $u_D$ is the utilization of the replisome (since the DNA replisome also requires internal substrates from metabolism, which have so far been collectively represented by a single variable $F$. If the required substrates for DNA replication are missing, then the time to replicate increases accordingly), and $O(t)$ is the number of origins at time $t$. Finally, the last equation accounts for initiation of replication (marked by a discrete event — the doubling of the origins of replication). The variable $T_I$ is the threshold number of initiators $I$ that are required for initiation, and $\tau_o$ is the short duration it takes to replicate the origin of replication ($\tau_o \ll \tau_I$). To see how this model accounts for the coupling between cell replication and DNA synthesis, consider the case where $\tau_D > \tau_U$. Then in order to have at least one fully copied DNA per doubling time, we need to initiate one DNA replication every doubling time. This will ensure (after a settling period of length $\tau_D$ ) that there is one DNA copy per biomass doubled. Since the number of initiators $I$ grows exponentially at the same rate as biomass (because $\gamma > 0$), they will cross the threshold and simultaneously initiate all the existing origins once per doubling time in steady growth conditions. If the biomass doubling time, (which equals the doubling time of the $U$ machinery) is shorter than the DNA replication time, then DNA will be replicated in a pipeline — a new round will be initiated prior to the completion of the previous round. If on the other hand, the biomass doubling time is longer than the DNA doubling time, then it will take time for the initiator $I$ to accumulate above the threshold, resulting in the emergence of the 'idling' or $B$ period.

Note that since all constitutive proteins are guaranteed to grow at the same rate in this model, there is no need to assume it as in Helmstetter's original model.

## 7.7. Hinshelwood and then Koch's model of the cell as a simple autocatalytic cycle

In 1946 Hinshelwood offered a simple model that accounts for self-replication in a mathematically appealing manner [566]. In his model, a set of $n$ enzymes catalyzes each other in a cyclical manner. Assuming all the substrates are abundant, the rate of catalysis is a function of the individual rates at substrate saturation, times the concentration of enzymes:

$$\frac{dE_i}{dt} = k_i E_{i+1}. \tag{118}$$

Where the addition is modulo $n$ i.e. $n+1 = 1$. To solve this equation, take $n$ successive derivatives of one of the equations. This yields

$$\frac{d^n E_i}{dt^n} = \Pi_{i=1}^n k_i E_i \tag{119}$$

The characteristic equation is given by $z^n = \Pi_{i=1}^n k_i$ , thus the asymptotic growth rate is $\lambda = (\Pi_{i=1}^n k_i)^{\frac{1}{n}}$ — the geometric mean of all the rates of the individual reactions when saturated. If we assume these rates to be stochastic and independently drawn from the same distribution, then the logarithm of the growth rate is a sum of independent, and identically-distributed random variables. This means that the distribution of the growth rate is log-normal, which further implies that the distribution of doubling times is also log-normal, because the log-normal distribution is stable with respect to reciprocation. Perhaps due to its simplicity, many authors use this distribution as the default distribution for doubling rates. However, a careful study of its ability to account for measured distributions of doubling times from 'mother-machine' experiments [9, 384] shows that even after fitting, it fails to account correctly for the observed distributions (see e.g. SI in [691]).

Koch suggested that the Hinshelwood cycle could be a good coarse-grained model for cellular self-replication [69]. He somewhat crudely argued that that 'ribosomes make proteins' and 'proteins make ribosomes' i.e. $\frac{dR}{dt} = k_R P$ and $\frac{dP}{dt} = k_P R$. Hence the overall biomass growth rate obeys $\lambda = \sqrt{k_R k_P}$ where $k_R$ and $k_P$ are the associated kinetic rates. One should contrast this result with a Markovian version of the von Neumann model that claims instead that the universal constructor makes copies of itself and the rest of the proteome hence (again assuming abundance of substrate and the presence of many copies of U-machines) $\frac{dU}{dt} = \alpha k_U U$ and $\frac{dP}{dt} = (1 - \alpha) k_U U$, which instead suggest that $\lambda = \alpha k_U$. Koch's model can be easily refuted by the observation that increasing the rate of production of a protein in the cell does not always increase its growth rate.

Recently, Biswas *et al.*[385] employed the Hinshelwood model to explain distribution of cell size and division times in *C. cresentus* grown in various temperatures. In their model, the stochasticity in division times is attributed to stochasticity in the rates of enzymes. They used an additive noise model for the rates of individual enzymes in an Hinshelwood cycle, and calculated the first passage time distribution, claiming it to be a good model of the division time distribution. However, as described by Jun and others (*e.g.*, Ref. [20]), the division time cannot be a simple first passage process, since it is inconsistent with a stable size distribution. Furthermore, there is no reason to believe that rates have additive noise, unless the noise is very weak and the correction to the average rate is taken only to first order.

### 7.8. *Non-Markovian model for the Hinshelwood cycle*

Interestingly, we can solve analytically for the growth rate of a non-Markovian generalization to the Hinshelwood cycle. Once again we insert latencies to account for the fact that the synthesis process has a finite non-zero duration. The equations are then

$$\frac{dE_i}{dt} = k_i E_{i+1}(t - \tau_{i+1}), \tag{120}$$

multiplying all the equations together and rearranging we obtain $\lambda^n e^{\lambda(\tau_1 + \tau_2 + \ldots + \tau_n)} = k_1 \times k_2 \times \ldots \times k_n$ or $\lambda e^{\lambda \frac{\tau_1 + \ldots \tau_n}{n}} = \sqrt[n]{k_1 \times k_2 \times \ldots \times k_n}$ Thus

$$\lambda \bar{\tau} e^{\lambda \bar{\tau}} = \langle k \rangle \bar{\tau}, \tag{121}$$

with $\bar{\tau} = \frac{\tau_1 + \ldots + \tau_n}{n}$ and $\langle k \rangle = \sqrt[n]{k_1 \times \ldots \times k_n}$. The growth rate is then

$$\lambda = \frac{\mathcal{L}_w(\langle k \rangle \bar{\tau})}{\bar{\tau}}, \tag{122}$$

with $\mathcal{L}_w(\cdot)$ being the Lambert-W function [703]. Note that this solution equals to the previous solution if $\langle k \rangle \bar{\tau} \ll 1$ since $\mathcal{L}_w(x) \sim x, x \ll 1$. If however, $\langle k \rangle \bar{\tau} \gg 1$ then $\lambda \sim \frac{\ln(\langle k \rangle \bar{\tau})}{\bar{\tau}}$. Using Little's law we can set each $k_i$ to be equal to $k_i = \frac{\eta_{\text{VIP}}^i}{\tau_i}$ with $\eta_{\text{VIP}}^i$ being the number of enzymes of type $i$ concurrently catalyzing (in a unit volume). This leads to the following equation for the growth rate $\lambda = \frac{\mathcal{L}_w(\langle \eta_{WIP} \rangle)}{\bar{\tau}}$.

### 7.9. *Summary*

The present model is based on von Neumann's kinematic self-replicator concept. It assumes each cell contains within it a set of molecular machines that self-replicate themselves, and also make all the other machines in the cell, machines that support the universal constructors by supplying material inputs i.e. substrates, and energy (metabolism), membrane bound volume, and the required DNA instructions. Since there are many copies, and they work concurrently, the doubling time is not simply $\ln 2$ divided by the time to make a single copy. The equations that describe the average behavior of this model are delay-differential equations, with state dependent delay. Nevertheless, they predict exponential growth at steady-growth conditions. The delay accounts for the fact that no major de-novo synthesis process of cellular complexes (e.g. ribosomes, DNA) can be made arbitrarily fast, as opposed to the what is expected from an exponential distribution. In this model the system will stabilize to balanced growth condition where all constitutive proteins grow exponentially locked to the rate of growth of the universal constructor. A more elaborate model would also account for the stochasticity of the delay and is beyond the scope of this discussion.

## 8. **Future**

Bacterial physiology is a subject that is integrative by nature. True understanding of its key observables — growth rate and cell size – requires deep insights into all the major cellular processes and their interrelations, including DNA replication, the transcription-translation machinery with its feedback and feed-forward controls, and membrane growth and division. The appeal of the subject to physicists is the presence of general rules that seem to apply to diverse types of bacteria. These rules allow such integration and suggest that despite the complexity, there is a universal *logica-ex-machina*.

### 8.1. *Towards "Control laws"*

To develop a full understanding of bacteria and further uncover new universal laws it is important to remember that "no bacterium is an island" to paraphrase John Donne, as bacteria typically live in the company of different species of microorganisms. Furthermore, bacteria do not typically live in an artificial environment in which nutrients are constantly replenished. Thus, it is necessary to develop a better understanding of how bacteria dynamically allocate their resources between all major *de-novo* synthesis processes as the external conditions change, or in the presence of other bacteria. The response to starvation for amino-acids, in *E.coli* called the 'the stringent response' [506, 507], appears to be well-conserved among different types of bacteria. A big challenge in the future, will be to develop 'control laws' analogous to the 'growth laws' that will generalize our understanding of steady-state behavior to a dynamic regime [707]. For example, while the biochemical details of how to inhibit overproduction of amino-acids might vary, the presence of such negative product feedback seems to be very broad, suggesting a general rule [708]. Why should we expect such rules to exist at all? E. coli contains more than 4000 genes, if indeed regulation of gene expression was totally random, we would need more than 4000 bits to describe the on/off control logic. Yet many genes are expressed constitutively, and other genes are jointly expressed (reflecting, for instance, the fact that they are all genes of proteins in the same metabolic pathway, or subunits of the same complex end-product, like the ATP-synthethase molecule). The fact that all the key molecular machinery in the cell is common to all known life-forms gives us hope that there are also universal control modules and that the design logic is simple and can be unraveled in the forthcoming decade.

## 8.2. Non-equilibrium thermodynamics of living systems

Apart from growth rate another interesting observable which is the focus of interest in many biotechnological applications is the growth yield. For example, the carbon growth yield measures how many carbon molecules are consumed per average cell produced. From a theoretical perspective, an interesting parameter that is currently difficult to measure is the thermodynamic growth yield - how much free energy is dissipated per unit-cell produced. Note that this yield can vary in time because not all of the entropy produced during growth is in the form of heat, as the cell dynamically exchanges substrates with the environment and these substrates have inherent molar entropy as well. If all the available free energy a cell has in a given environment was used for growth, the growth yield would be maximal. It is tempting to conjecture however, that this will also happen at the price of slow growth. Indeed, syntropic bacteria that live very close to the thermodynamic limit in terms of energy-conversion efficiency, are also notoriously slow growers [709]. Is there a trade-off between growth and thermodynamic yield?

Intriguingly there seem to be a close relation between thermodynamic yield, growth rate and the manner in which the cell schedules its reproduction. Growing quickly requires pipelining of self-replication, *i.e.* increasing the level of concurrency in the process of biomass production. This requires energy, and as we explained in Section 7.3, any linear increase in doubling rate results in an *exponential* increases in the demand for substrate and energy. For organisms that live in nutrient-poor environments, on the other hand, energetic efficiency is vital, and pipelining is riskier – what if resources unexpectedly and rapidly deplete?

In many real engines, a trade-off between energetic efficiency and power exist, so perhaps there is a similar trade-off between growth rate and energetic efficiency for bacteria. Is there a minimal entropy tax for producing a unit cell? It is possible to measure the heat released during growth, but total entropy increase is also composed of the molar entropy of the all the molecules exchanged. Nevertheless, some differential calorimetry measurements hint that the amount metabolic heat production normalized to cell mass released during growth is constant [710]. Given that heat is the 'last stop' of energy down the ladder of utility, perhaps cells try to minimize lossy dissipation when energy is scarce? Finally, obligatory phototrophic cyanobacteria have subordinated their gene expression to the diurnal cycle. They seem to avoid, for example, the production of ribosomes during nighttime when energy is scarce [711]. This serial scheduling results in slower growth as opposed to pipelining, but is also more efficient. Can we devise a model that can explain the discrepancy between the life-style of these cyanobacteria and enteric bacteria based upon their habitat, or on general thermodynamic and control theory arguments?

Developing new methods to measure the thermodynamic efficiency of growth, even on a population level, can facilitate interesting comparisons between different types of bacteria. For example, all slow growers have a single rDNA operon, but despite the many interesting theoretical and empirical works [712–715], there is no theory that can predict when a bacteria will require more than one operon, how many, and where to locate them with respect to the origin and terminus sites, given the ecological niches it occupies and evolutionary history. This too remains a challenge for the future.

## 8.3. Issues on the variability and causality of physiological controls

In steady-state growth all biomolecules on average double their copy numbers in each division cycle; however, the kinetics of their biosynthesis is stochastic so that there is significant cell-to-cell variability in biomolecule abundance and subsequent physiological parameters. Despite the stochasticity at the molecular level, every daughter cell still must inherit the proteins, the chromosome, and the cell envelope to be viable. If the synthesis of each member of this 'trinity' is independent from one another, then the cell will face a serious threat. To see this, let us assume 10% of variability for the timing of replication initiation and cell division. In each generation, initiation and division timings will perform a 'random walk.' After 100 generations the cell will inevitably reverse with 100% probability its order of replication initiation and cell division ($10\% \times \sqrt{100} = 100\%$). It is currently unknown how *E. coli* coordinates replication initiation and cell division, and more generally all biosynthesis, to avoid this catastrophe. This is a major question that needs to be answered in the future.

## 8.4. Evolution of physiological controls

Last but not least, evolution of physiological controls is a profound yet difficult theme to study. For example, *E. coli* and *B. subtilis* are one billion years divergent, and they are the textbook examples of how Gram-negative and Gram-positive bacteria are fundamentally different at the molecular level in their cell cycle control. Nevertheless, both organisms appear to follow the adder principle for size homeostasis. If so, what is the hierarchy of physiological controls associated with growth, the cell cycle, and cell size, and how did each control emerge during the course of evolution? Bacteria thus offer outstanding opportunities to bring physiology and genetics together to deepen our understanding of the overarching quantitative principles of cellular reproduction and their evolution.

### Acknowledgments

For over a decade, we have been fortunate to have known some of the founders of bacterial physiology and those who were closely related to them, including Tove Atlung, Stuart Austin, Hans Bremer, Stephen Cooper, Willie Donachie, Pat Dennis, Flemming Hansen, Nanne Nanninga, the late Fred Neidhardt, the late Kurt Nortdstrom, Steen Pedersen, Elio Schaechter, Conrad Woldringh, Andrew Wright, and Arieh Zaritsky. Many of them shared their inspirational anecdotes with us and taught us – with patience – their beautiful science that is largely forgotten. But above all, their stories, much of which we couldn't share in this article, reminded us what should drive scientists. We dedicate this article to them. We thank Serena Bradde, Johan Paulsson, Petra Levin, Massimo Vergassola, and the former and current members of Hwa and Jun labs for many years of collaborations and interactions, and Stephen Cooper, Willie Donachie, Petra Levin, JT Sauls, and Elio Schaechter for critical reading. This work was supported by the Paul G. Allen Frontiers Group, Pew Charitable Trust, NSF CAREER grant MCB-1253843, NIH grant R01 GM118565-01 (to S.J.) and NSERC Discovery grant (to M.S.).

### List of symbols

$\lambda$, exponential growth rate

$\mu$, doubling rate, growth rate in base 2

$B$, latency time between birth and the initiation of DNA replication

$C$, time for DNA replication

$D$, time between termination of DNA replication and cell division

$\varphi(a)$, age distribution (PDF a cell has age $(a, a + da)$)

$\rho_l(l)$, length distribution (PDF a cell has length $(l, l + dl)$)

$\tau_{\text{cyc}}$, cell cycle time $(C + D)$

$\tau_{\text{d}}$, division (doubling) time $(1/\mu)$

$V_x$, elongation rate

$\rho_{\tau_{\text{d}}}(\tau)$, doubling time distribution (PDF the division time for a given cell is $(\tau, \tau + d\tau)$)

$\rho_{l_d}(l)$, division length distribution (PDF a cell has length $(l, l + dl)$ at division)

$\rho_{\Delta_d}(\Delta)$, added-mass-at-division distribution (PDF a cell has added mass $(\Delta, \Delta + d\Delta)$ at division)

$\rho_Z(z)$, PDF for a random variable $Z$

$\Psi(l)$, birth size distribution (PDF a cell has length $(l, l + dl)$ at birth)

$\eta(l, t)$, number of cells with length $l$ at time $t$

$\eta(l, \xi, t)$, number of cells with length $l$ and age $\xi$ at time $t$

$p$, number of generation from the mother $(p = 0)$

$l_b$, birth size

$l_d$, division size

$\Delta_d$, added mass between birth and division $(l_d - l_b)$

$\Delta(t)$, incremental added mass $(l(t) - l_b)$

$\gamma(l)$, division rate function (PDF a cell of size $l$ will divide $(t, t + dt)$)

$S_0$, unit cell size (also known as initiation mass, or mass/origin of DNA replication)